# The dynamics of internal electric field screening in hybrid perovskite solar cells probed using electroabsorption

Davide Moia,[1,2,*] Ilario Gelmetti,[3] Philip Calado,[1] Yinghong Hu,[4] Xiaoe Li,[5] Pablo Docampo,[4,6] John de Mello,[5] Joachim Maier,[2] Jenny Nelson,[1] Piers R.F. Barnes[1,*]

[1] Department of Physics, Imperial College London, London SW7 2AZ, UK

[2] Max Planck Institute for Solid State Research, Stuttgart, Germany

[3] Institute of Chemical Research of Catalonia (ICIQ), Barcelona Institute of Science and Technology (BIST), Tarragona, Spain

[4] Department of Chemistry and Center for NanoScience (CeNS), LMU München, München, Germany

[5] Department of Chemistry, Imperial College London, London SW7 2AZ, UK

[6] Department of Chemistry, University of Glasgow, Glasgow, G12 8QQ, UK

*d.moia@fkf.mpg.de; piers.barnes@imperial.ac.uk

## Abstract

Electric fields arising from the distribution of charge in metal halide perovskite solar cells are critical for understanding the many weird and wonderful optoelectronic properties displayed by these devices. Mobile ionic defects are thought to accumulate at interfaces to screen electric fields within the bulk of the perovskite semiconductor on application of external bias, but tools are needed to directly probe the dynamics of the electric field in this process. Here we show that electroabsorption measurements allow the electric field within the active layer to be tracked as a function of frequency or time. The magnitude of the electroabsorption signal, corresponding to the strength of the electric field in the perovskite layer, falls off for externally applied low frequency voltages or at long times following voltage steps. Our observations are consistent with drift-diffusion simulations, impedance spectroscopy, and transient photocurrent measurements. They indicate charge screening/redistribution on time-scales ranging from 10 ms to 100 s depending on the device interlayer material, perovskite composition, dominant charged defect, and illumination conditions. The method can be performed on typical solar cell structures and has potential to become a routine characterization tool for optimizing hybrid perovskite devices.

## I. Introduction

Metal halide perovskites combine broad light absorption spectra, good electronic properties and solution processability, making them promising materials for solar cells. [1] Possibly the most peculiar feature of hybrid perovskite solar cells is the observation of electronic and optoelectronic dynamics that extend to very long time-scales (milliseconds to hundreds of seconds). [2,3] This includes the presence of hysteresis in current-voltage measurements, transient photocurrent measurements, switchable photovoltaic behavior and huge values of apparent capacitance. [4,5]

Among the explanations for the slow dynamics, migration of ionic defects within the perovskite layer is now the generally accepted cause. [6–9] While hysteresis effects in hybrid perovskites can be exploited in some other fields, [10,11] they remain a concern for solar cell applications where a stable and reliable power output upon illumination is desirable. In particular, the degradation of perovskite solar cells is believed to correlate with slow time-scale processes caused by ion migration. [12] Recent advances in interlayer engineering, perovskite composition and crystal growth superficially appear to have solved the problem of hysteresis and, in some cases, improved the stability of devices. However, measurements and simulations indicate that an absence of hysteresis does not rule out ionic migration in the cell. [13] Understanding and controlling ion migration in perovskites remains crucial to achieving stable and reliable photovoltaic modules.

An equivalent circuit model recently proposed by some of us, based on ionically gated transistors describing the hybrid perovskite-contact layer interfaces, can describe most of the optoelectronic properties shown by hybrid perovskite devices. [14] The model assumes that, when no external bias is applied to the device, the ionic charge carriers (which are also assumed to be the majority carriers) in a perovskite solar cell accumulate at the interfaces to screen the built-in potential due to the difference in work functions of the contact materials (Figure 1b). Thus, assuming sufficiently large concentrations of mobile ionic defects, only the narrow (relative to the thickness of the active layer) space-charge regions close to the interfaces with the contacts experience significant electric field. On the other hand, the bulk of the perovskite layer remains almost free of electric field. If the electrical potential across the device is changed (through application of a bias or generation of a photovoltage) this results in an instantaneous electric field within the perovskite bulk (Figure 1c or 1d) that is subsequently screened with a time constant corresponding to the time-scale for ionic charge redistribution (Figure 1e or 1f). Ionic redistribution changes the electrostatic potential at the interface and this 'gates' the transfer of electronic charge across the interface (recombination or injection processes) in a manner similar to how charge transfer through a bipolar transistor is governed by its base voltage. [11] Although a coherent picture of the underlying device physics is emerging, independent methods to directly interrogate the processes at work are needed. In particular, probes of the electrostatics in the perovskite active layer will help elucidate the physical process of ionic migration and identify the parameters controlling its dynamics.

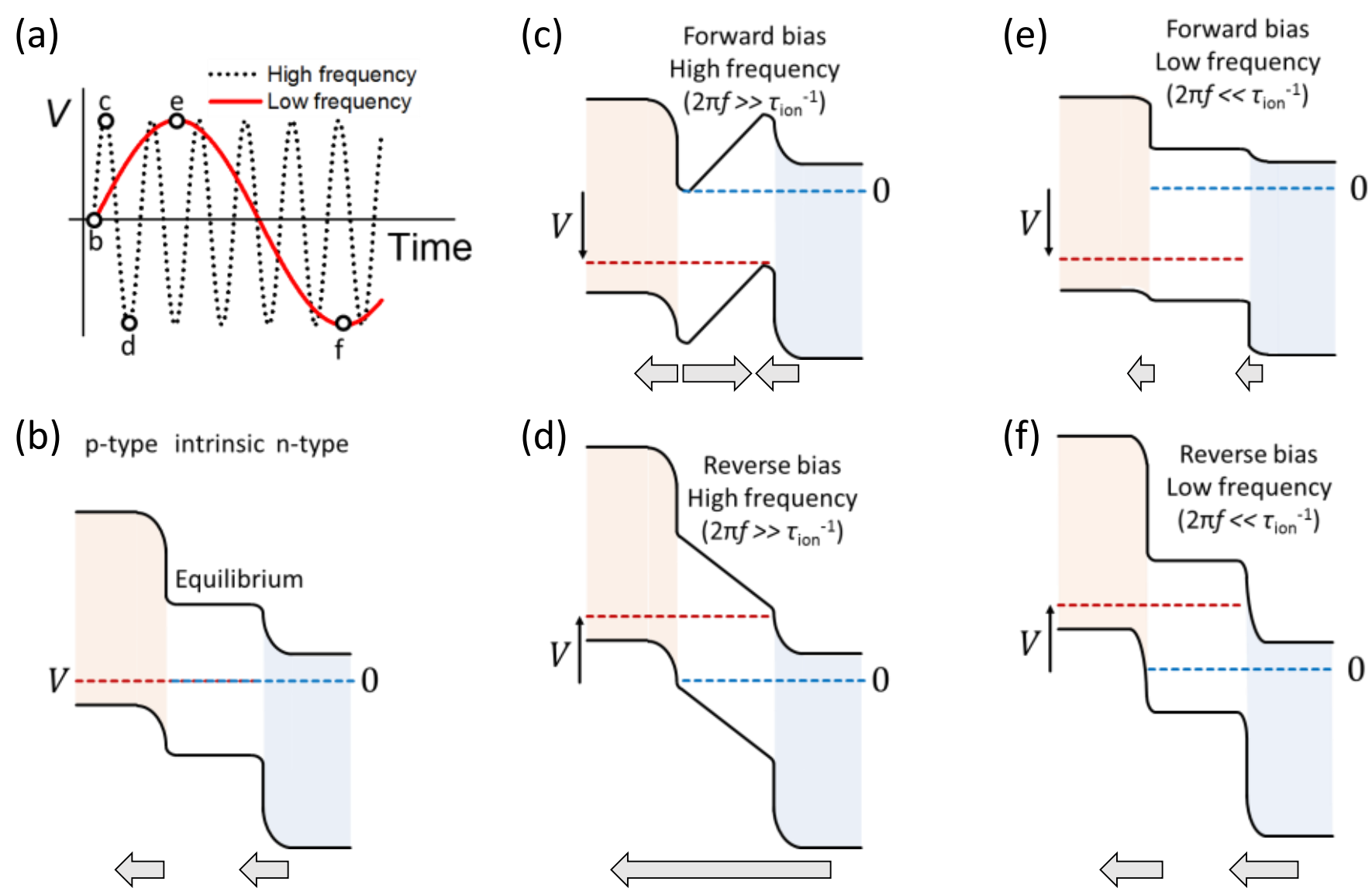

**Figure 1. Schematic energy level diagrams and electric field direction of a p-i-n perovskite solar cell.** The schematics show the effect of mobile ions redistribution comparing the behavior under applied voltage bias with high or low frequency as illustrated in (**a**). (**b**) The device at zero bias in the dark. The p-type, intrinsic, and n-type layers are indicated by pink, white and blue regions respectively. The hole and electron (quasi) Fermi levels are indicated by the respective red and blue dashed lines. Resulting energy level diagrams upon the application of a high frequency voltage for the case of (**c**) forward bias and (**d**) reverse bias, or of a low frequency voltage for (**e**) forward bias and (**f**) reverse bias. The time constant $\tau_{ion}$ is associated with the ionic redistribution in the perovskite and internal electric field screening. Below each energy level diagram, arrows indicating the direction of the electric field in the contacts and in the active layer are included. The schematics show approximately flat quasi-Fermi levels in the perovskite layer, corresponding to the high electronic mobility limit (i.e. where the rate of electronic transport is much greater than the rate of recombination and thermal generation in the bulk of the active layer).

Measurements of electric field at the surface of exposed device cross sections have been reported using Kelvin Probe Microscopy, or indirectly using Electron Beam Induced Current measurements. [15–18] While these techniques can give access to field distribution in devices, their surface sensitivity and resolution can pose challenges to data interpretation. [19] Spectroscopic, contactless techniques have also been presented. Electroabsorption (EA) enables changes in optical properties of the materials induced by an electric field to be monitored. Electroabsorption (referred to also as Stark spectroscopy) has previously been applied to hybrid perovskite layers processed in architectures which include one or two insulating layers (e.g., poly (methyl methacrylate), PMMA), in lateral devices, or in solar cells. [20–25] The technique offers insights into the photophysics of charge generation in perovskite solar cells at steady state, as well as the ability to quantify the exciton binding energy and effective mass. [21] The measured EA signal for methylammonium lead iodide (MAPI) based devices has been interpreted in terms of the Stark effect [20,26–29] as well as Franz-Keldysh-Aspnes theory. [21] Measurements of EA on complete perovskite solar cells have only been explored to a limited extent, with an observed hysteresis in the measured built-in potential value extracted from reflection mode EA during solar cell potential scans being attributed to ionic migration by Li *et al*. [28] EA measurements applied with time or frequency dependence have the potential to give useful information with respect to the dynamics of ion migration and charge distribution in devices. While frequency dependent EA has been recently used to investigate the effect of light intensity on the shape and magnitude of the EA spectrum in the 40 Hz–1kHz range, for MAPI based devices that included a thin insulating PMMA layer, [30] the method has not been reported to date for solar cells.

In the low-field limit, Franz-Keldysh-Aspnes theory and the quadratic Stark effect predict that an electric field applied to a material will modulate its optical transmittance in proportion to the (energy-dependent) imaginary part of the third order nonlinear optical susceptibility of the material, $\chi^{(3)}(h\nu)$, and can be approximated using superposition of different order derivatives of the material absorption coefficient. [21,25,31–33] If an electric field, $F$, with a static and oscillating component is applied ($F = \bar{F} + \Delta F \sin[2\pi f t]$ where $\bar{F}$ is the steady state electric field, $\Delta F$ is the amplitude of the oscillation with frequency $f$, and *t* is the time), then the transmittance is modulated (due to a periodic variation in the material's dielectric constant): $T = \bar{T} + \tilde{T}$, where $\bar{T}$ and $\tilde{T}$ are the steady state and oscillating parts of the transmittance signal. The quantity $\tilde{T}$ can be split into components varying at the frequency of the applied field, $\tilde{T}_f$ (the first harmonic), at twice the frequency, $\tilde{T}_{2f}$ (the second harmonic), as well as higher order components (see Figure 2a and more details in the methods). The amplitude of the first harmonic signal, $\Delta T_f$, is proportional to the product of the static field and the amplitude of the oscillating component:

$$\frac{\Delta T_f}{\bar{T}}(h\nu) \propto \bar{F}\,\Delta F\, Im[\chi^{(3)}(h\nu)]. \quad (1)$$

The amplitude of the second harmonic signal, $\Delta T_{2f}$, is proportional to the square of the amplitude of the oscillating field:

$$\frac{\Delta T_{2f}}{\bar{T}}(h\nu) \propto \Delta F^2 \mathrm{Im}[\chi^{(3)}(h\nu)]. \quad (2)$$

Regardless of whether Franz-Keldysh-Aspnes theory and/or the quadratic Stark effect represent the underlying physical mechanism relating $T$ to $F$, the form of the relationship described by Equations 1 and 2 is expected to be the same. Thus, the relative electric field strength within a perovskite solar cell can be probed by measuring the intensity of the first and second harmonic EA signals.

Here, we use frequency domain and time resolved EA to investigate the dynamics of the internal electric field, and we observe internal field screening at low modulation frequencies for all perovskite solar cells we measured. The cut-off frequency below which the screening is detected ranges between 100 Hz – 1 kHz depending on the active layer material and device architecture measured. We verified the measurements by comparing the frequency of this transition, which corresponds to a time-scale of electric field screening, with additional independent experimental techniques: step-dwell-probe (SDP) photocurrent measurements and impedance spectroscopy. We finally show examples of how these techniques can be used to investigate the influence of perovskite morphology, composition and thickness, as well as the effect of applied DC voltage or light bias. Using these combined spectroscopic and optoelectronic methods, we show a self-consistent multi-method approach that can be directly applied to solar cell devices and integrated to other routine measurements such as current voltage and impedance spectroscopy.

## II. Methods

*Device fabrication*: solar cell devices were fabricated on FTO substrates and included an oxide ($TiO_2$ or $SnO_2$) interlayer, a hybrid perovskite active layer, a Spiro OMeTAD hole transporting materials and a semitransparent (~40 nm thick) gold top contact. The details for the fabrication are described in section 1 of the Supporting Information.

*Electroabsorption measurements*: transmission-mode electroabsorption (EA) spectroscopy was performed on solar cell devices (Figure 2a). A xenon lamp was used for continuous probe light. The light beam was passed through a monochromator before being focused onto the active area of the device. The sample was oriented so that the probe light was incident on the metal contact to reduce the intensity of the probe reaching the active layer of the solar cell. This was estimated to be in the order of 0.01 sun (~300 µW cm$^{-2}$ at 760 nm for measurements of MAPI devices) by measuring the photocurrent of the device and by accounting for a beam size of approximately 2 mm$^2$. The probe beam was then passed through a second monochromator, using filters at the output slit to remove higher order diffraction components. Finally, the beam was focused on a silicon photodiode connected to the input of a SR830 lock-in amplifier or of a HF2LI Zurich instrument lock-in amplifier (the latter was used for frequency dependent EA measurements). The baseline transmitted signal, $\bar{T}$, was recorded before performing the EA measurements by using a chopper at 560 Hz placed between the probe light and the first monochromator and recording the signal from the silicon photodetector through the lock-in amplifier referenced to the external trigger of the chopper. Fractional changes in optical transmission were then calculated as $\Delta T/\bar{T}$ as explained below. Two different photodiodes were used for the measurement: a low noise slow (~10 kHz) photodetector was used for measuring EA spectra at 1 kHz; a second, high speed (5 MHz) photodetector was used for the frequency dependent measurements from 1 Hz up to 1 MHz. An oscillating voltage with angular frequency $2\pi f$ applied to the cell was taken from an auxiliary output of the lock-in amplifier ($\tilde{V}$) and combined with a DC voltage ($\bar{V}$) using a custom-made unity gain amplifier circuit to give an output voltage, $V = \bar{V} + \tilde{V} = \bar{V} + \Delta V \sin[2\pi f t]$, applied to the solar cell. Values of $\Delta V \leq$0.8 V were used to limit injection of electronic charge carriers in the solar cell (which can result in additional charge-induced electromodulation features, see e.g. Ref. [34]) and to limit degradation of the perovskite mixed conductor. The resulting amplitude of the averaged change in electric field across the solar cells was in the range $\Delta F \approx$ 0.01 – 0.06 MV cm$^{-1}$ depending on the device and the type of measurement. Oscillations in optical transmission $\tilde{T}$ inferred from the signal produced by the photodetector and measured by the lock-in amplifier were recorded using a LabView program. The amplitude ($\Delta T$) and phase ($\Phi$) of $\tilde{T}$ were recorded at the frequency of $\tilde{V}$ (giving the 1$^{st}$ harmonic signal), and at twice the frequency of $\tilde{V}$ (giving the 2$^{nd}$ harmonic signal). The DC voltage $\bar{V}$ was kept at 0 V unless stated otherwise. We briefly address the effect of varying $\bar{V}$ below.

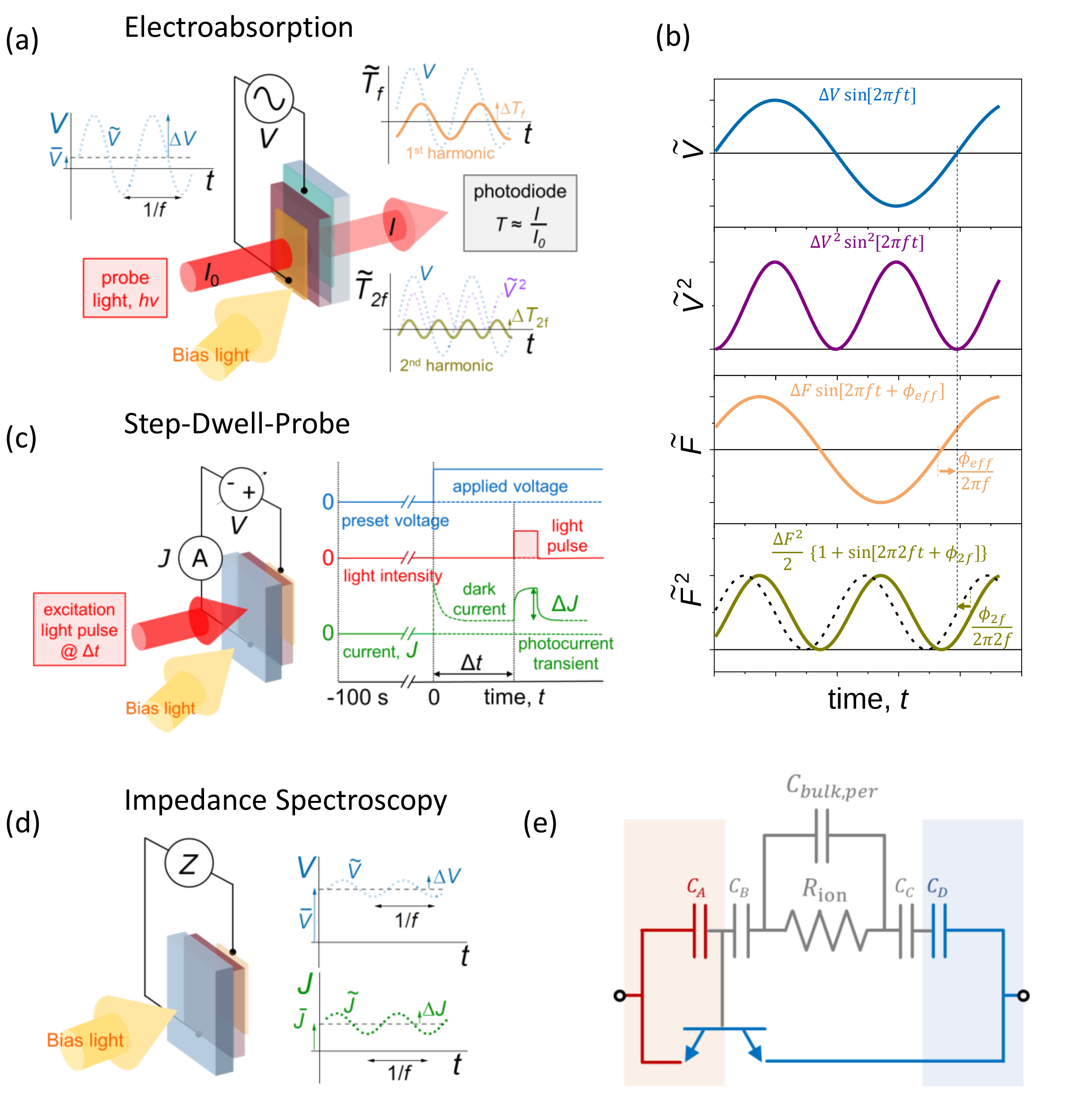

**Figure 2. Schematics for the electroabsorption, the step-dwell-probe and the impedance spectroscopy setup.** (**a**) Schematic of the experimental setup and measured photodiode signals in response to an applied voltage ($V = \bar{V} + \tilde{V}$) across the device composed of a static DC component ($\bar{V}$) and a periodic component ($\tilde{V}$) with amplitude Δ*V* oscillating with frequency *f*. The intensity of the probe light ($I_0$) transmitted through the sample (*I*) oscillates in response to *V*. Two components of the oscillating transmittance ($\tilde{T}$) signal were monitored: the 1st harmonic which varies at frequency *f* with an amplitude $\Delta T_f$, and the 2nd harmonic which varies at frequency 2*f* with an amplitude $\Delta T_{2f}$. The measurement can be run while the device is illuminated by a bias light. (**b**) Relative phase between the applied voltage in an EA experiment and the other relevant components: square of the applied voltage $V^2$, change in electric field $\tilde{F}$ averaged through the perovskite layer resulting from the applied voltage but reduced by any screening in the device, average square of the change in electric field $\tilde{F}^2$. The diagram shows the definition of $\phi_{eff}$ with respect to the measured phase $\phi_{2f}$ during the 2nd harmonic measurement. $\phi_{eff}$ is representative of the phase lag between the change in field and the applied voltage. (**c**) For step-dwell-probe experiments, the device is left at a preset voltage for $t < 0$. The voltage is then stepped to a probe voltage at $t = 0$, and the photocurrent $\Delta J$ induced by a light pulse applied to the solar cell after a dwell time $\Delta t$ at the probe voltage is measured. A bias light can optionally be used during the measurement. In this study the bias light (white LED) was switched on at $t = 0$. (**d**) Impedance spectroscopy measurements were run on devices under dark or under light. (**e**) Equivalent circuit model used in the analysis of the data.

*Analysis of electroabsorption data*: as discussed in the introduction, $\tilde{T}$ can oscillate at the fundamental voltage perturbation frequency, *f*, and also at higher order harmonic frequencies, *nf*, where *n* is an integer. The time lag of internal screening processes means that changes in electric field can be out-of-phase with the applied voltage, which in turn is reflected in the phase of $\tilde{T}$. Consequently, a phase term must be included when considering the full expression of $\tilde{T}$, such that $\tilde{T} = \sum_{n=1}^{\infty} \Delta T_{nf} \sin[2\pi nft + \phi_{nf}]$. For the *n*-th harmonic of the signal at angular frequency $2\pi nf$, we can define in-phase and out-of-phase components, $\Delta T_{nf} \sin\left[2\pi nft + \phi_{nf}\right] = \Delta T'_{nf} \sin[2\pi nft] + \Delta T''_{nf} \cos[2\pi nft]$ with amplitudes $\Delta T'_{nf} = \Delta T_{nf} \cos \phi_{nf}$ and $\Delta T''_{nf} = \Delta T_{nf} \sin \phi_{nf}$, respectively. On this basis, for the 1st harmonic measurements, $\Delta T'_f$ is in-phase and $\Delta T''_f$ is out-of-phase with the applied voltage. For the 2nd harmonic measurements, $-\Delta T''_{2f}$ is in-phase and $\Delta T'_{2f}$ is out-of-phase with the square of the applied voltage (since $V^2 \sin^2[2\pi ft] = V^2(1 - \sin[2\pi 2ft + \pi/2])/2$). Therefore, we use $-\Delta T''_{2f}/\bar{T}$ to illustrate the component of the 2nd harmonic EA spectrum which is in-phase with the square of the applied voltage. Furthermore, we will present the magnitude and phase of the 2nd harmonic signal as a function of frequency. To give a more intuitive interpretation of the measured 2nd harmonic signal as it varies with the frequency of the applied potential, we have plotted the square root of the EA signal which is proportional to the change in field $\Delta F$, as indicated by Equation 2. The magnitude corresponds to $\left(\Delta T_{2f}/\bar{T}\right)^{1/2}$ while the resulting effective phase is $\phi_{eff} = \phi_{2f}/2 + \pi/4$ (in radians). The phase shift, $\phi_{eff}$, is 0° when the change in absorption (and therefore, the square of the electric field in the active layer) is in-phase with the square of the applied potential. If $\phi_{eff}$ is nonzero, it refers to a phase shift with respect to the applied voltage. Therefore, $\phi_{eff}$ is representative of the phase-lag of the electric field in the absorber layer, which oscillates at the frequency of the applied perturbation as shown in Figure 2b. Note that $\phi_{eff}$ as defined above corresponds to $\phi_f$ in the simplest case where $\tilde{F}$ has a sinusoidal profile. This is not expected to be necessarily the case for our experiments, where large values of $\Delta V$ are used (see section 2 of the Supporting Information for more details). Nevertheless, defining $\phi_{eff}$ still provides insights into the charge screening effects occurring in the perovskite layer. For all the $\Delta T/\bar{T}$ results, we accounted for the fact that the data obtained from the lock-in is the root mean square of the signal and we also corrected the amplitude of the signal for the trapezoidal shape of $T$ during the measurement of $\bar{T}$ with chopped light.

*Optoelectronic transient measurements*: step-dwell-probe (SDP) measurements (see Figure 2c) were performed using an automated transient optoelectronic measurement rig (TRACER) described in ref. [35]. The measurement has been described in ref. [36] and consists of a photocurrent measurement performed with a specific protocol for the voltage bias of the device: the cell is left at a preset voltage (0 V in this work) in the dark for long enough to reach an approximately steady-state (between 1 s and 100 s depending on previous measurement). The device is then stepped to a probe voltage (a positive voltage below the open circuit voltage, $V_{oc}$, in this work) for a particular dwell time which is long enough to allow complete electronic capacitive charging of the device leaving only the residual dark current. Following this dark dwell time, a LED light-pulse of 6 µs is applied to the cell and the resulting photocurrent during the pulse is recorded 4 µs after the application of the pulse (see Figure 2c). $\Delta J$ is the amplitude and direction (positive or negative) of the corresponding photocurrent pulse. The sign of $\Delta J$ is in part related to the direction that the internal electric field in the perovskite layer carries the photogenerated charge. The measurement is repeated with the same conditions but with different dwell times, such that a $\Delta J$ *vs* dwell time plot can be obtained. The measurement can be performed with or without an additional bias light (in our case, white LEDs were used). A green LED was used for the pulsed source with an intensity of 2 sun equivalent for devices using $MAPbBr_3$ as the active layer. A red LED with 1 sun equivalent intensity was used for all other cells. To make sure that the cell was in a comparable state when probing at different dwell times, a control measurement at the shortest dwell time was performed before every measurement to check that the inverted photocurrent (negative $\Delta J$) had not changed for this shortest dwell time condition.

*Impedance spectroscopy*: impedance measurements were performed following the method described in reference [14]. In short, a voltage amplitude of 20 mV was used for all measurements and a 1 MHz–0.1 Hz frequency range was used in all cases. For measurements under light bias at open circuit, a chronopotentiometry measurement was performed for 100 seconds to obtain an estimate of the steady state open-circuit voltage. The device was then left equilibrating under light (or dark if applicable) by applying the final voltage measured during the chronopotentiometry measurement for a further 100 second period. The impedance measurement was then performed. The impedance spectra were interpreted using the equivalent circuit model presented in ref. [14].

The model (see Figure 2e) includes an ionic and dielectric branch (gray elements) and an electronic branch (blue and red elements). The former involves a resistor $R_{ion}$ representing ionic transport in the perovskite layer. The resistor $R_{ion}$ is connected on either side with a capacitor associated to the space-charge layer in the contact ($C_A$ and $C_D$) and a capacitor associated with the space charge layer or Debye layer in the perovskite ($C_B$ and $C_C$). The series of the two capacitors at either interface represents the total interfacial capacitance and is referred to as $C_{ion}^{\perp}$. The resulting circuit determines the characteristic time constant $\tau_{ion}$ (e.g., assuming symmetrical interfaces $C_{ion}^{\perp} = C_{ion,1}^{\perp} = C_{ion,2}^{\perp}$ one obtains $\tau_{ion} = R_{ion}C_{ion}^{\perp}/2$ ). Bipolar transistors are used to approximate the electronic branch. These elements reproduce the gating of injection and recombination currents at the interfaces by the changes in electrostatic potential. In this work, we consider a single transistor associated with the surface recombination process occurring at one of the interfaces to dominate the electronic component of the measured impedance. Finally a geometric capacitor, $C_{bulk,per}$, is connected in parallel to $R_{ion}$. The overall geometric capacitance, $C_g$, is determined by the series of all the capacitors described above.

*Simulations*: drift-diffusion simulations were performed using the open-source software package Driftfusion. [37] Driftfusion solves Poisson's equation and the continuity equations for electrons, holes, mobile ions and the electrostatic potential for a one-dimensional device and their evolution over time. Further details of the simulation methods can be found in section 3 of the Supporting Information. We performed simulations of the EA, SDP and impedance measurements.

## III. Results

### A. Electroabsorption of perovskite solar cells

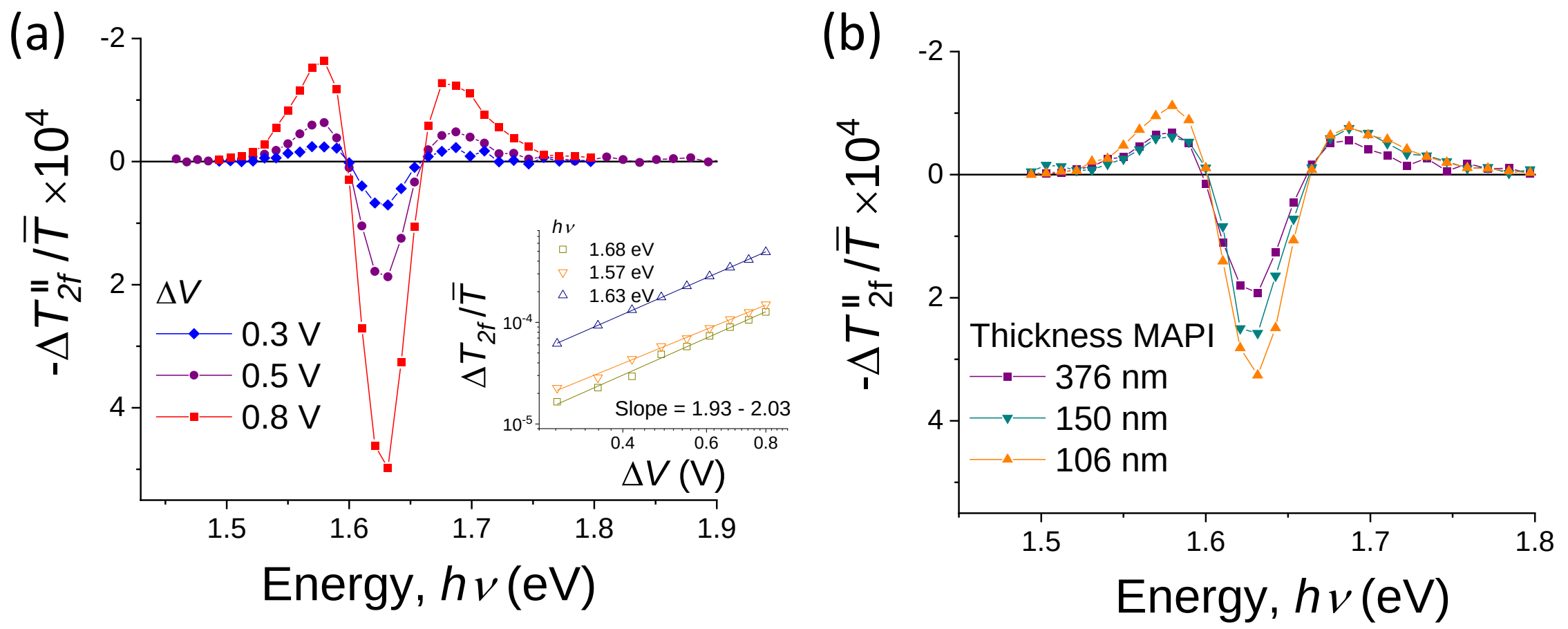

**Figure 3. Second harmonic electroabsorption of $TiO_2$/MAPI/Spiro OMeTAD solar cells measured by applying a sinusoidal voltage at 1 kHz with $\overline{V}$ = 0 V.** (**a**) The effect of different applied voltage amplitudes, $\Delta V$, on the 2nd harmonic $-\Delta T''_{2f}/\overline{T}$, plotted as function of probe light photon energy, $h\nu$, for a solar cell with MAPI thickness of about 380 nm. The inset in (**a**) shows the signal amplitude, $\Delta T_{2f}/\overline{T}$, plotted as function of $\Delta V$ for photon energies corresponding to the minima ($h\nu$ = 1.57 eV and 1.68 eV) and maxima ($h\nu$ = 1.63 eV). (**b**) Effect of the MAPI active layer thickness on the 2nd harmonic EA for $\Delta V$ = 0.5 V.

Figure 3a shows EA spectra measured by applying a 1 kHz sinusoidal voltage of varying amplitude to a MAPI solar cell and probing the changes in optical transmission at the 2nd harmonic of the frequency of the applied potential. The data correspond to the component of the signal that is in-phase with the square of the applied voltage. The EA spectra obtained by applying voltage perturbations $\tilde{V}$ with amplitudes $\Delta V$ = 0.3, 0.5 and 0.8 V show the same profile. We observe a quadratic dependence of the EA signal with the applied voltage in this voltage range (see inset in Figure 3a). Figure 3b shows a similar measurement performed using $\Delta V = 0.5\ V$ on devices with the same structure but different active layer thicknesses. The magnitude of the signal increases for thinner active layers suggesting a correlation of the signal with the average electric field present in the bulk of the material for these devices. The sublinear dependence of the square root of the EA signal *vs* perovskite thickness suggest a significant drop in field through the contacts and the possibility that most electric field variations occur at the interface with the contacts (see discussion section). We also observe a slight variation of the EA spectral profile for the three devices, which could be ascribed to optical interference effects due to the different thickness of the active layer or to inherent variations in spectral shape due to the changes in average photo-generated charge density in the films due to the probe beam. [30] The quadratic dependence of the EA signal on the applied voltage is consistent with both Franz-Keldysh-Aspnes and Stark theories, as discussed in previous reports [21,25,38] and in the introduction. It also suggests that the electric field in the MAPI layer varies in proportion to the applied potential. In next section, we proceed to investigate the dynamics of the absorber layer's electric field by varying the frequency of the applied voltage ($\tilde{V}$).

### B. Frequency dependent electroabsorption

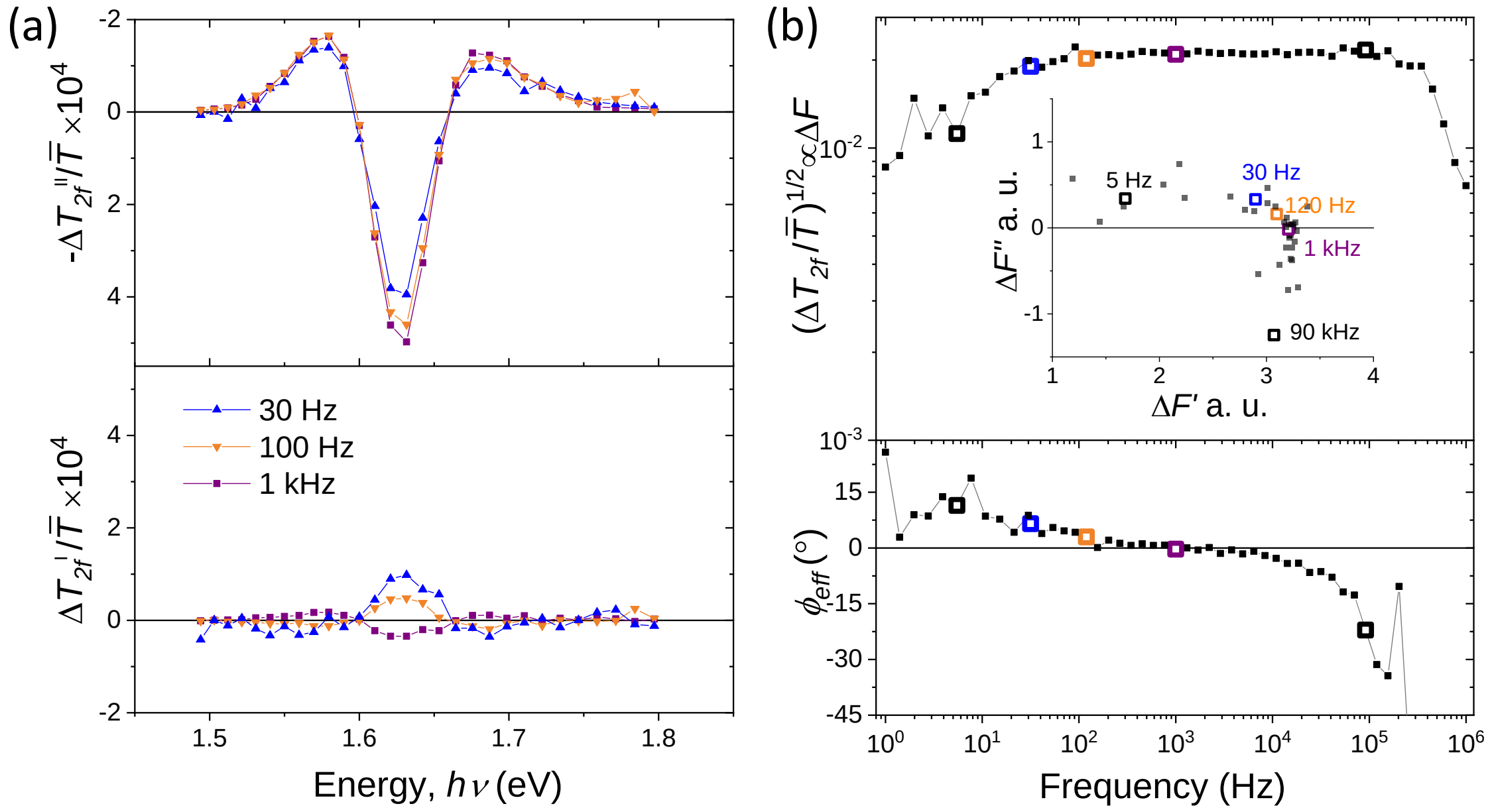

**Figure 4. Frequency dependent electroabsorption measurement on a solar cell with $TiO_2$/MAPI/Spiro OMeTAD architecture.** (**a**) 2nd harmonic EA spectra recorded at different frequencies of the applied voltage ($\Delta V$ = 0.8 V). The top and the bottom panels refer to the components in-phase and out-of-phase with the square of the applied potential, respectively (see Figure 2b). (**b**) Bode plot of the square root of the EA peak at 1.63 eV photon energy (760 nm) measured at different frequencies. The inset in (**b**) shows the equivalent Nyquist representation of the change in electric field in the perovskite using frequency as an implicit parameter. The phase $\phi_{eff}$ has been calculated as $\phi_{eff} = \phi_{2f}/2 + \pi/4$ (see method section).

Figure 4a shows the 2nd harmonic EA spectra of a MAPI solar cell measured at different frequencies of applied voltage, using a constant amplitude ($\Delta V$ = 0.8 V) and no DC bias ($\bar{V} = 0\ V$). When decreasing the frequency of the voltage perturbation from 1 kHz to 30 Hz, we observe a decrease (with approximately unchanged spectral shape) in the in-phase EA signal and an increase in the out-of-phase EA signal with respect to $\tilde{V}^2$ (Figure 4a). By measuring the EA signal at the peak wavelength of 760 nm (1.63 eV photon energy) as a function of applied frequency, we obtain the frequency response of EA signal in the device. Since we expect the EA signal to be related to the square of the change in electric field in the perovskite, in Figure 4b we show the Bode plot of the square root of the EA signal. The equivalent Nyquist representation of the change in electric field is also displayed as inset of Figure 4b, where we plot $\Delta F'$ versus $\Delta F''$ which are, respectively, in-phase and out phase with the applied potential. The frequency spectrum of $(\Delta T/\bar{T})^{1/2}$ in Figure 4b shows a plateau region at intermediate frequencies (between 100Hz and 200kHz) where the effective phase $\phi_{eff}$ is close to 0°. This is consistent with the field in the perovskite layer giving rise to the EA signal being in-phase with the applied voltage perturbation. The signal shows a decrease in magnitude and a phase shift when either high or low frequencies of the applied voltage are used. At high frequencies, we expect the intrinsic RC time constant of the cell (in the order of 1 μs for the device under consideration) to limit the maximum rate at which the applied potential changes across the solar cell stack. We therefore ascribe the decrease in magnitude above ~200 kHz to an increasing drop of potential on the series resistance ($R_s$) of the cell and the capacitive charging of the geometric capacitance associated with the high frequency dielectric properties of the whole device stack ($C_g$). In the low frequency region (for $f$ < 100 Hz), a process with slow dynamics appears to screen the internal electric field in the active layer, thus decreasing the EA signal, also consistent with the data in Figure 4a. In the discussion section we discuss the nature of this screening process, which appears to be due to ionic charge migration. In the next section we validate this observation by comparing the frequency dependent EA results with two other independent techniques.

### C. Measuring electric field screening using electroabsorption, step-dwell-probe and impedance spectroscopy

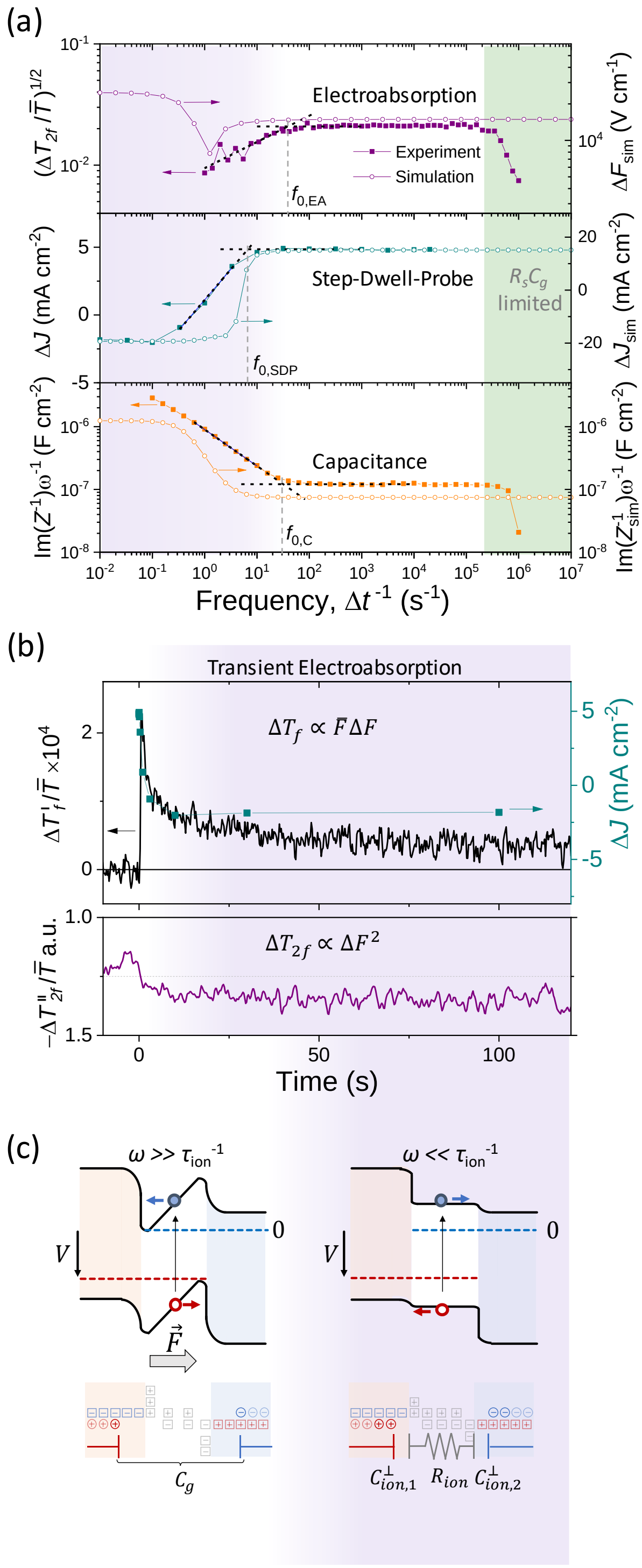

**Figure 5. Comparison of different frequency and time resolved techniques measuring the dynamics of electric field screening for a MAPI solar cell in the dark.** (**a**) Frequency dependent EA (top), SDP photocurrent plotted against inverse dwell time $\Delta t^{-1}$ (middle) and capacitance extracted from impedance spectroscopy (bottom). The light green shaded area corresponds

to the frequency range where we expect the limitations of series resistance and geometric capacitance $R_sC_g$ to become significant. Experimental data (solid squares, left axis) as well as simulated results using drift-diffusion simulations (empty circles, right axis) are displayed. The values of the characteristic frequency $f_0$ for each technique is obtained based on the intersection of fitting lines (black dashed lines) to the high frequency region and to the transition to low frequency region. (**b**) Time resolved transient EA measured for the in-phase 1st harmonic (top) and 2nd harmonic (bottom) signal at 760 nm (1.63 eV) upon the application of a voltage step $\bar{V} = 0\ V\ \rightarrow 0.5\ V$ at $t = 0\ s$ superimposed to $\tilde{V}$ ($\Delta V$ = 0.5 V and 1 kHz). The dwell time dependent photocurrent response from SDP shown in (**a**) is superimposed to the 1st harmonic transient measurement. (**c**) Schematics of the energy level diagrams for the early time-scale (high frequency, white background) and long time-scale (low frequency, shaded purple background) response to a voltage perturbation. The implications in terms of electric field, photocurrent direction, and charge distribution contributing to the observed capacitance are also illustrated.

Figure 5 shows five different measurements of a $TiO_2$/MAPI/Spiro OMeTAD solar cell under dark (for EA experiments a weak probe light is applied, see methods section). Figure 5a compares the frequency dependent EA data from Figure 4b, with SDP data where the photocurrent is plotted against the reciprocal of the dwell time, $\Delta t$. This is done to compare the estimated time-scales of electric field screening on the frequency scale. Finally, the capacitance extracted from the impedance of the cell is shown. The three measurements highlight the spectroscopic, optoelectronic and electrochemical response of the device. The data from each measurement show two frequency regions with different behavior, a high frequency region (fast time-scale, white background in Figure 5) and a low frequency region (long time-scale, purple background in Figure 5):

- Electroabsorption: an approximately constant EA signal is observed in the $100\ Hz - 100\ kHz$ region, with a drop in signal magnitude for $f < 10 - 100\ Hz$. This is consistent with the (partial) screening of the electric field (see $\vec{F}$ in Figure 5c) by ionic charges occurring at low frequencies.
- Step-dwell-probe: the photocurrent induced by a light pulse incident on the cell at different times after the application of a step potential (see direction of electron transport and hole transport in Figure 5c) is positive (charges moving away from their respective selective contact) for $\Delta t < 0.1\ s$ and changes sign for light pulses applied at later times. This is consistent with changes in direction for the driving force (affected by the electric field in the perovskite layer) of the photocurrent. [36,39]

- Impedance spectroscopy: for high frequencies the capacitance of the solar cell remains constant, which we attribute to the geometric capacitance contribution. Its value increases for $f < 10 - 100$ Hz due to ion redistribution within the perovskite active layer (see electronic charge accumulating at the contacts and change in ionic distribution in Figure 5c).

For each technique we highlight the transition in behavior described above by defining a parameter $f_0$ to be the frequency (or inverse dwell time for SDP measurements) that divides the high frequency "plateau" from the low frequency electric field screening regime. We extract $f_0$ as the frequency value obtained from the intersection between a (constant) line fitted to the high frequency plateau and a line fitted to the transition to the low frequency behavior (see black dashed lines in Figure 5). While the type and degree of perturbation is different among the three techniques, we observe a very similar trend and similar estimates of $f_0$ at which we expect the electric field screening to happen. Note that $f_{0,C}$ is not strictly associated with the low frequency impedance time constant. However, it is easily accessible and seems to correlate qualitatively with the results from other techniques. In Figure 5a, we also include results from simulations obtained using the Driftfusion software [37], where we simulate the EA, the SDP and the impedance measurement on a solar cell stack with representative parameters (see full discussion in section 3 of the Supporting Information). The general trend discussed above is well reproduced, with a high frequency response that can be attributed to the dielectric properties of the simulated device without the contribution of ion transport and a low frequency behavior that matches the electric field screening effect due to the redistribution of mobile ionic defects. Discrepancies observed between the experimental and simulated data are due to the simplified system considered for the simulations, which assumed non dispersive transport and a single mobile ionic species. For $\Delta F_{sim}$, which is representative of the simulated EA signal, we observe an increase in magnitude for frequencies below 1 Hz, a frequency region that cannot be resolved experimentally. We provide more detail on this as well as further simulated results in section 3 of the Supporting Information, where we show that the increase in the low frequency simulated signal is obtained when using similar (or larger) values of the interfacial capacitance on the contact side as the one on the perovskite side (e.g., $C_C \approx C_D$ for the simulations in Figure 5).

In order to investigate the very low frequency behavior of devices further, we developed a time resolved transient EA technique. Figure 5b shows an experiment performed on the same type of solar cell as described in Figure 5a, where we apply a voltage $V = \bar{V} + \tilde{V}$ and measure the 1st or the 2nd harmonic signal at 760 nm as a function of time following a step change in $\bar{V}$. While in the previous EA experiments we kept $\bar{V} = 0\ V$ throughout the measurement, here we apply a step in the DC voltage such that $\bar{V} = 0\ V$ for $t < 0\ s$ and $\bar{V} = 0.5\ V$ for $t > 0\ s$. An oscillating voltage ($\tilde{V}$) with $\Delta V = 0.5\ V$ at 1 kHz is superimposed to the step potential to allow the evolution of field to be monitored using the lock-in technique. As shown in Figure 5b (top graph), the change in absorption related to the 1st harmonic due to the step potential presents a sharp peak at early times ($t < 1\ s$) and tends to stabilize over a time-scale in the order of 10 to 100 s. In the same figure we also include the SDP data from Figure 5a for comparison of the long time-scale response between the two techniques. This result, combined with the observations above, shows that the process of electric field screening occurs over a large range of time-scales. After the decay, the baseline in the measurement also shifts slightly, suggesting a detectable change in $\bar{F}$ at long time scales (after the screening process) on application of the forward bias (see Equation 1).

When the $\bar{V}$ step was reversed, the time resolved signal is inverted (see Figure S9) as would be expected since Equation 1 predicts that the EA signal depends on the sign of $\bar{F}$. We note that the 1st harmonic EA spectrum for MAPI solar cells shows a similar shape as the 2nd harmonic spectrum (as expected from Eqs 1 and 2). A similar behavior is observed for different active layer compositions (see 1st harmonic EA characterization of a $MAPbBr_3$ cells in section 4 of the Supporting Information). On the other hand, a much larger change in baseline at long time-scale is observed for MAPI solar cells with $SnO_2$ contacts (Figure S10). Following a peak in the data at early time-scales similar to the one in Figure 5b, these cells showed a long time baseline change that was opposite to the case with $TiO_2$. We have checked that this measurement is not affected by electroluminescence contributions (Figure S11). This result could not be explained on the basis of the interpretation presented in this study. As a general note, the analysis of 1st harmonic EA signal is complex, because of changes in spectral shape over several measurements, possibly due to additional contributions from other induced absorption features. We also find that changes in the magnitude and sign of the 1st harmonic signal vary over very long time-scales. This made it difficult to perform, for example, frequency dependent 1st harmonic EA measurements.

The transient EA measurement for the 2nd harmonic signal (bottom graph of Figure 5b) does not present a clear peak at $t$ = 0 s like the one observed for the 1st harmonic measurement. Instead, the data show a small step, followed by a very slow variation occurring over similar time-scales as the slow transient of the 1st harmonic signal. The absence of a peak is consistent with Equation 2 (2nd harmonic signal is independent of $\bar{F}$, see Equation 2). The increase in signal magnitude upon application of the forward bias is likely to be due to the increase in the capacitance of the contacts at forward bias, which increases the fraction of applied $\tilde{V}$ dropping across the absorber layer during the measurement. On the basis of these considerations, the dependence of the EA signal on the applied DC voltage (Figure S19) can provide information on the voltage dependence of the capacitances involved in the cell.

The consistency between the results obtained via these different techniques shows that each can be used to probe the time-scale of electric field screening within the active layer of perovskite solar cells. These tools are therefore suitable to test the effect of the different optical bias and device parameters on this process, as we will show in the next section.

### D. Effect of bias light, interlayers and absorber film thickness, morphology and composition

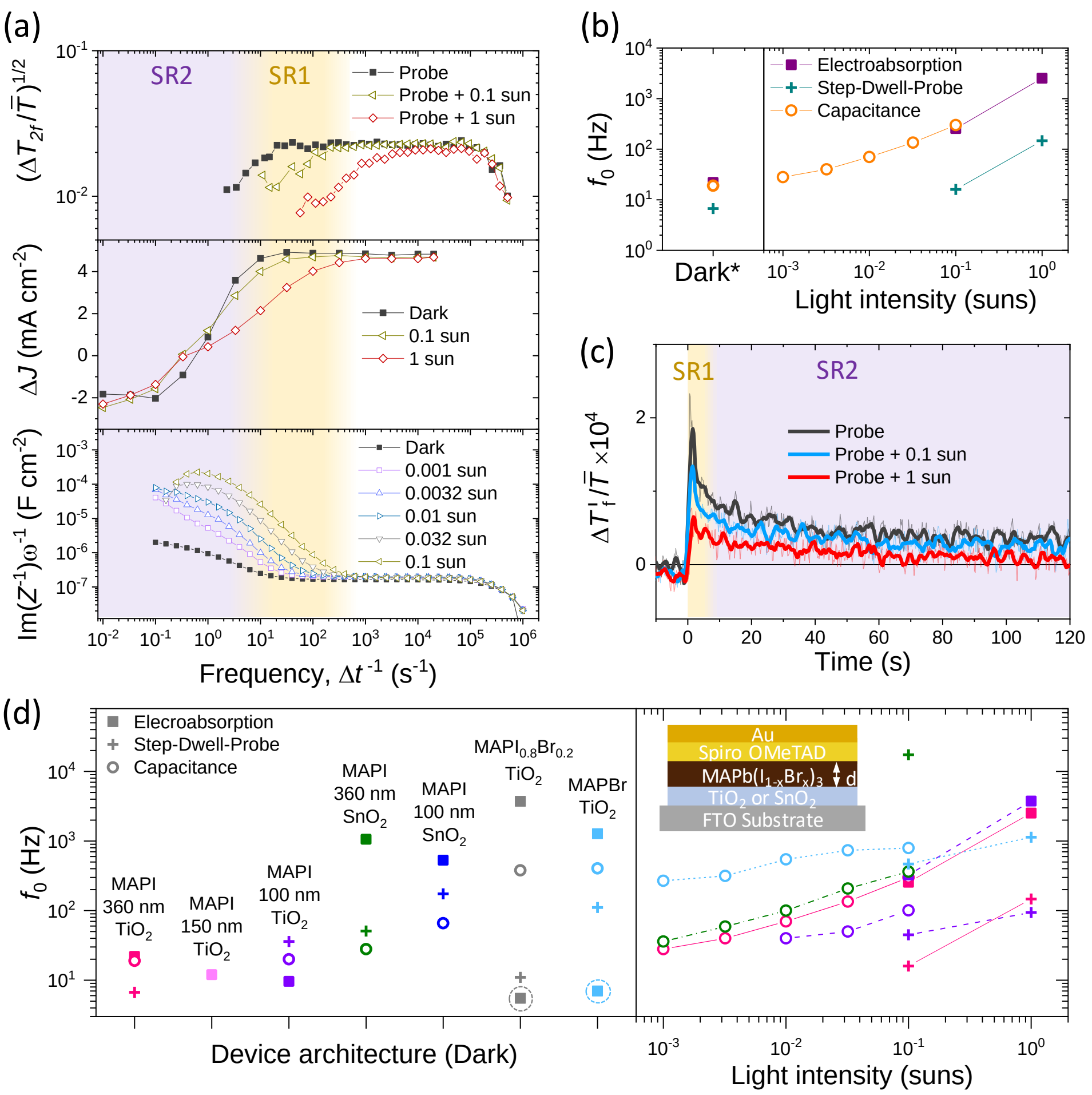

**Figure 6. Effect of different solar cell parameters on electroabsorption, step-dwell-probe and capacitance measurements.** (**a**) (**c**) Measurements on $TiO_2$/MAPI/Spiro OMeTAD solar cells as in Figure 5a and b, also showing the effect of light on $f_0$ (MAPI thickness 360 nm). The purple and the yellow background indicate the approximate frequency range of screening regime 1 (SR1, observed only under illumination) and screening regime 2 (SR2, observed under dark and light conditions). In (**c**) smoothed data (15 points, ~ 3 s window) are shown superimposed to the raw data. (**b**) Values of $f_0$ extracted from the data plotted in (**a**). (**d**) Summary of the data obtained in the dark (left) and under light (right) for solar cells with different architectures. The inset of (**d**) shows the device structure and the parameters investigated in this study: interlayer material, hybrid perovskite composition and thickness, *d*. The circled data points indicate an additional characteristic frequency extracted from the EA of some devices. *For EA data measured without bias light, the cell was exposed to a weak probe light (see methods).

We performed EA, SDP, and (apparent) capacitance measurements for a number of different device architectures under both dark and a range of bias light conditions. Figure 6 shows a summary of these measurements and the frequencies at which the field screening transition $f_0$ occurs. First, we describe the observed changes in dynamics for each measurement technique when the device is exposed to a bias light with respect to measurements performed in the dark. Figure 6a shows that for all measurements on the reference solar cell ($TiO_2$/MAPI-360 nm thick/Spiro OMeTAD), the field screening transition shifts to higher frequencies as bias light intensity increases. The corresponding values of $f_0$ are shown in Figure 6b where the increasing trend with light can be observed (ranging between 1 – 10 $^4$ Hz). The values of $f_{0,EA}$ and $f_{0,C}$ show close agreement while the values for $f_{0,SDP}$ show a similar trend with light intensity but with an offset to lower frequencies – this offset is likely to be intrinsic to the technique, resulting from a time-lag required before sufficient ionic redistribution has occurred to start influencing photocurrent direction. The latter is sensitive to the direction of the quasi-Fermi levels gradient rather than the field only. Figure 6a also shows that the abruptness of this transition for the SDP data is reduced by bias light, with the appearance of a high frequency decay in photocurrent to about 0 mA cm$^{-2}$, followed by the, roughly unchanged, low frequency transition to negative photocurrent values (see more examples for other device structures in section 5 of the Supporting Information). When the first harmonic of the EA was monitored following a potential step we observed that the initial, rapidly changing, component of the signal could no longer be resolved as bias light intensity was increased (Figure 6c). Consistent with the observations in Figure 6a, the slow time-scale component of the first harmonic signal remained at all light intensities.

Figure 6d shows the values of $f_0$ extracted from measurements on other device architectures and active layer thicknesses (impedance and current-voltage curves of the devices can be found in section 8 and 9 of the Supporting Information). In all cases $f_0$ increases with increased light intensity implying more rapid field screening. The perovskite layer thickness appears to have a relatively weak influence on $f_0$ or on the dynamics of the long time-scale measurements (see Figure S22 and S28). Only values of $f_{0,SDP}$ show an appreciable increase with decreasing layer thickness. In section 6 of the Supporting Information we also show that $f_{0,EA}$ is not influenced by very different MAPI layer morphologies.

In contrast, the electron transporting material (ETM) appears to have a significant influence on $f_0$. When $TiO_2$ was replaced by $SnO_2$, the values of $f_0$ increased by up to 50 times depending on the measurement and device (Figure 6d). A lower capacitance at high frequencies for $SnO_2$ compared to $TiO_2$, consistent with the trend in $f_{0,C}$ (Figure S17), could explain this effect (see also next section). Changing the hole transporting material – Spiro OMeTAD doped with lithium ions – for undoped

Spiro OMeTAD did not significantly influence $f_{0,EA}$ (see Supporting Information section 7). The observation suggests that lithium ions are not directly involved in the screening behavior at the frequencies monitored in this experiment.

Finally, Figure 6d also shows the influence of changing the perovskite layer composition. For the frequency dependent EA spectrum of $MAPbBr_3$ or $MAPb(I_{0.8}Br_{0.2})_3$ (Figure S21), we identify a mild decrease in signal for frequencies below about 1 kHz, followed by a sharp drop below about 10 Hz. In Figure 6d we indicate both these frequencies (the latter is highlighted by a dashed circle). Higher values of $f_0$ for $MAPbBr_3$ compared to MAPI were also recorded using SDP and capacitance measurements. We also recorded an increase in $f_{0,SDP}$ with light for this device, similar to MAPI cells, while we observed a drop in signal magnitude but no clear shift for $f_{0,EA}$ to higher frequencies when illuminating the $MAPbBr_3$ cell. A more complex behavior for $MAPb(I_{0.8}Br_{0.2})_3$ was observed under light with a non-monotonic drop in signal with decreasing frequencies (see Figure S21b). In section 7 of the Supporting Information, we also include measurements on mixed cation hybrid perovskite solar cells. We show that for all compositions containing between 2 and 4 different A-cations, the value of $f_{0,EA}$ in the dark increases to about 1 kHz compared to the case of MAPI on $TiO_2$ (100 – 200 Hz). Figure S26 shows that this effect is present regardless of using $TiO_2$ or $SnO_2$ as interlayer.

## IV. Discussion

### A. Interpretation of measurements in the dark

The frequency and time dependent EA techniques that we have developed allow a direct measurement of the time-scale for electric field screening in the metal halide perovskite layers of solar cells. The results show good qualitative agreement with indirect measurements of this field screening process inferred from SDP photocurrent measurements and impedance spectroscopy measurements. These screening dynamics depend on the details of the device architecture, composition and operation. We now discuss possible interpretations of the screening mechanisms based on the experimental observations reported in this work.

The field screening that we observe is likely to be due to redistribution of ionic charge in the hybrid perovskite layer. This is consistent with the many previous reports suggesting that ion transport in these photovoltaic materials is responsible for the slow optoelectronic dynamics observed in solar cell devices resulting in behavior such as current-voltage hysteresis. [6,7,13,14,40] Our observations are consistent with this interpretation, also discussed by Rana *et al.*, [30] and provide a direct, non-destructive measurement of field screening within the active layer of complete devices.

Device architecture can influence the screening time-scale. When device contact materials are changed there is an associated change in the capacitance of the interfaces ($C_{ion}^{\perp}$, see also Figure 5c) which leads to a change in the ionic screening time of the device ($\sim R_{ion}C_{ion}^{\perp}/2$ for the symmetrical case), assuming that electrode polarization dominates such dynamics. For example, a lower doping density in the $SnO_2$ relative to $TiO_2$ would lead to a smaller capacitance and thus shorter screening time. Active layer composition could similarly influence both the capacitance of the interfaces (due to defect density, energetic offset, and dielectric constant, all of which would influence space-charge layer widths and thus capacitance) as well as the ionic resistance (due to the different mobile defect concentration and mobility). Further contribution from electronic charge carriers cannot be ruled out, as we discuss below.

### B. Electric field screening under light

We now consider the effect of light on the measured characteristic frequency of electric field screening. Clearly the techniques used in this study offer different probes with different sensitivity to the electric field screening and, especially for the case of measurements under light, they point towards a complex, possibly multi-process, behavior. When light is applied to the reference device structure $TiO_2$/MAPI/Spiro OMeTAD, we observe a pronounced shift to higher values of the frequency $f_0$, below which the field screening is "activated" (Figure 6a, b). We refer to this behavior under light and the frequency range where it occurs as screening regime 1 (SR1). This effect is also visible from time resolved transient EA measurements (smaller initial amplitude of the signal at the early time-scales for cells under illumination, Figure 6c). At longer times following a voltage step for the time resolved EA and the SDP measurements, a second slower component to the change in electric field is also detected, the rate of which is relatively uninfluenced by light. We refer to this as screening regime 2 (SR2). The light independence of SR2 is also displayed by the similar transient dynamics observed for time resolved EA measurements where the bias light is switched on or off, while keeping $\bar{V}$ unchanged (see Figure S18). For the impedance measurements, the appearance of SR1 can be associated with the increase in apparent capacitance under light at low frequencies, which increases the value of $f_{0,C}$. [14] As mentioned above, $f_{0,C}$ does not have an intuitive physical meaning, and it is used here to track the changes in magnitude and/or dynamics of the low frequency capacitance component (see discussion in section 8 of the Supporting Information). We note that, because the use of a probe light is necessary to carry out the EA measurements, it is not possible to evaluate $f_{0,EA}$ for the completely dark case. Based on the trend discussed above, the value of $f_{0,EA}$ extracted from measurements without a bias light may be affected by the probe light intensity, as also reported by Rana *et al*. [30] Different average charge densities induced by the probe light in different devices could potentially contribute to some of the observed trends in $f_{0,EA}$ as a function of active layer composition and ETM, as discussed below (see also section 9 of the Supporting Information). In this context, the comparison in Figures 5 and 6 of EA data with the results obtained from the other techniques, for which measurements in the dark are possible (here SDP and impedance), is particularly useful.

The two regimes of electric field screening SR1 and SR2 detected by these techniques highlight the possibility for two processes with different dynamics and different dependence on illumination to occur. These processes can be electronic or ionic in nature. Regarding electronic charge carriers, an additional contribution to electric field screening from photo-generated charges and injected charges needs be considered. As for the ionic contribution, it is well established that iodide vacancies are the majority carriers in MAPI. [6,7] Lower diffusion coefficients have been evaluated for methylammonium cations and other defects, which may anyway provide further screening ability. [41–43]

In a first scenario, SR1 could be due to injected and photo-generated electronic charges redistributing in the cell to screen the electric field induced by the externally applied potential. SR2, which is similar between measurements on the same device irrespective of the illumination level, may be related to the ionic relaxation, e.g., due to redistribution of iodide vacancies. Analysis of simulations suggest that electronic screening, while active at faster time-scales compared to experiments, is present and increases in magnitude under light because of the contribution from the photo-generated electronic charge carriers (see section 3 in the Supporting Information). Furthermore, we tested the frequency dependent EA for a $TiO_2$/MAPI/Spiro OMeTAD device by including a DC offset $\overline{V} = 0.5\ V$ to the applied AC voltage. We observe a slight drop in the signal at frequencies below about 500 Hz (Figure S20), consistent with an increase in the background electronic concentration in the device. This would suggest that electronic currents are playing some role in the fast component of the screening process. This effect may also be related to the accumulation of photo-generated charge in trap states, which is not accounted for explicitly in our simulations at present. We note that ionic redistribution may also contribute to the faster screening at high bias light intensity (SR1) based on the possible increase in ionic conductivity under light which was reported to be relevant for iodide defects. [44–46] The long time-scale behavior (SR2) could then also be related to a second ionic species, which is comparatively unaffected by the background light intensity. [47] A detailed investigation of these screening effects deserves future work.

### C. Spatial distribution of the electric field

We now comment on the spatial distribution of the changes in electric field that we measure. The fact that the magnitude of the 2nd harmonic EA signal scales with the square of the applied potential (with some fluctuation, this was the case for all devices, see Figure S27) suggests that the square root might be used to evaluate the electric field strength in the active layer. However, the weak dependence of both signal magnitude and some of the time constants extracted from the measurements on active layer thickness, indicates a more complex situation. Regarding the thickness-dependent magnitude of the EA signal at high frequencies, we note that the contribution to the EA signal would be more significant from the regions of the device where the active layer is the thinnest (because of the large electric field in those regions for a certain applied high frequency voltage). It follows that if surface roughness of the active layer were to increase superlinearly with thickness, this could also contribute to the trend shown in Figure 3b. The thickness dependence of the EA signal magnitude could also be consistent with only a small fraction of the changes in potential dropping in the active layer, while a large fraction of it drops across the contact materials (see section 2 and 10 of the Supporting Information). It is also possible that, if screening were fast enough within the active layer bulk, most of the potential drop would be confined to its interfaces (e.g., across the depletion layer at the MAPI/$TiO_2$ interface [48]). A mixed situation between the cases described above is likely.

## V. Conclusions

In this work, we developed frequency-dependent and time-domain transient electroabsorption spectroscopy techniques that can be applied to solar cell devices to probe their internal electric field dynamics. By varying the frequency of the applied potential, we investigated the time-scale of the electric field screening for different solar cell architectures and tested the effect of experimentally accessible parameters on the internal electrostatics of the device. We compared the results with data from other techniques, such as optoelectronic transient step-dwell-probe and impedance spectroscopy, to extract independent (although indirect) evaluations of the electric field screening dynamics in devices. This set of techniques provides a number of characteristic frequencies ($f_0$) defining the transition in behavior from the high frequency electronic and dielectric response of the hybrid perovskite, to a low frequency response where change in applied electric field becomes screened in the perovskite. For dark conditions, we ascribe the latter to ionic defect transport and redistribution in the active layer. Under light, we find that, in addition to the process occurring at long time-scales, a faster process is responsible for part of the screening. The process can be attributed to screening by photo-generated electronic charge or to a photo-enhanced ionic conductivity effect. Our analysis shows the potential of using time resolved and frequency domain electroabsorption to evaluate the internal electric field dynamics in hybrid perovskite solar cells and to validate the interpretation of data obtained using other optoelectronic and electrochemical techniques.

## Acknowledgements

We thank Jiachen Gu and James Bannock for their help in testing the electroabsorption setup and assistance in the lab. We acknowledge funding from the UK Engineering and Physical Sciences Research Council (grants EP/M025020/1, EP/R020574/1, EP/T028513/1, EP/R023581/1, EP/M014797/1). DM is grateful to the Alexander von Humboldt foundation for funding.

Supporting information

# The dynamics of internal electric field screening in hybrid perovskite solar cells probed using electroabsorption

Davide Moia,[1,2,*] Ilario Gelmetti,[3] Philip Calado,[1] Yinghong Hu,[4] Xiaoe Li,[5] Pablo Docampo,[4,6] John de Mello,[5] Joachim Maier,[2] Jenny Nelson,[1] Piers R.F. Barnes[1,*]

[1] Department of Physics, Imperial College London, London SW7 2AZ, UK

[2] Max Planck Institute for Solid State Research, Stuttgart, Germany

[3] Institute of Chemical Research of Catalonia (ICIQ), Barcelona Institute of Science and Technology (BIST), Tarragona, Spain

[4] Department of Chemistry and Center for NanoScience (CeNS), LMU München, München, Germany

[5] Department of Chemistry, Imperial College London, London SW7 2AZ, UK

[6] Department of Chemistry, University of Glasgow, Glasgow, G12 8QQ

*d.moia@fkf.mpg.de; piers.barnes@imperial.ac.uk

**Table of contents**

# 1. Device fabrication procedure

**Perovskite precursor solutions:**

**MAPI:** Stoichiometric amounts of $PbI_2$ (576 mg, 1.25 mmol, TCI Co., Ltd.) and methylammonium iodide (MAI, 199 mg, 1.25 mmol, Greatcell Solar) were dissolved in 1 mL of a 4:1 (v:v) solvent mixture of dimethylformamide (DMF, anhydrous, Sigma-Aldrich) and dimethyl sulfoxide (DMSO, anhydrous, Sigma-Aldrich) by heating up to 100 °C in a closed vial. This ~1.25 M MAPI precursor solution was diluted in appropriate amounts of the DMF:DMSO solvent mixture to obtain ~0.75 M and ~0.625 M MAPI solutions. All solutions were filtered through a 0.45 µm syringe filter before use. According to SEM cross-sections, the MAPI solutions with the concentrations 1.25 M, 0.75 M and 0.625 M resulted in average perovskite film thicknesses of 367 nm, 150 nm and 106 nm, respectively.

**MAPBrI:** Methylammonium bromide (MABr, Greatcell Solar, 56 mg, 0.25 mmol), MAI (119 mg, 1 mmol) and $PbI_2$ (576 mg, 1.25 mmol) were dissolved in 1 mL DMF:DMSO (4:1, v:v) at 100°C to obtain the nominal composition of $MAPb(Br_{0.2}I_{0.8})_3$ in the mixed halide perovskite precursor solution.

**MAPBr:** MABr (140 mg, 1.25 mmol) and $PbBr_2$ (TCI Co., Ltd., 459 mg, 1.25 mmol) were dissolved in 1 mL DMF:DMSO (4:1, v:v) at 100°C.

**FAMAPI:** Formamidinium iodide (FAI, 183 mg, 1.06 mmol), MAI (30 mg, 0.19 mmol) and $PbI_2$ (576 mg, 1.25 mmol) were dissolved in 1 mL DMF:DMSO (4:1, v:v) at 100°C to obtain the nominal composition of $FA_{0.85}MA_{0.15}PbI_3$.

**FAMA:** Following the protocol reported by Saliba *et al.*,[1] a multiple cation mixed-halide perovskite solution was prepared by dissolving $PbI_2$ (508 mg, 1.1 mmol), $PbBr_2$ (80.7 mg, 0.22 mmol), FAI (171.97 mg, 1 mmol) and MABr (22.4 mg, 0.2 mmol) in 1 mL of a 4 : 1 (v/v) mixture of anhydrous DMF and DMSO. This nonstoichiometric precursor solution for $(FA_{0.83}MA_{0.17})Pb(I_{0.83}Br_{0.17})_3$ contains a 10 mol% excess of both $PbI_2$ and $PbBr_2$, which was introduced to enhance device performance. The FAMA solution was filtrated through a 0.45 µm syringe filter before use. We would like to note that volume changes upon dissolving the salts are expected.

**FAMAC:** CsI (389.7 mg, 1.5 mmol) was dissolved in 1 mL DMSO and filtrated through a 0.45 µm syringe filter. To obtain the desired triple cation perovskite composition of ~5% Cs, 42 µL of the ~1.5 M CsI stock solution was added to 1 mL FAMA solution, yielding a nominal composition of $Cs_{0.05}[(FA_{0.83}MA_{0.17})]_{0.95}Pb(I_{0.83}Br_{0.17})_3$. The change in the I : Br ratio was neglected and we note that volume changes upon dissolving CsI in DMSO were not taken into consideration.

**FAMAR:** RbI (318.5 mg, 1.5 mmol) was dissolved in 1 mL of a 4 : 1 (v/v) DMF : DMSO mixture and filtrated through a 0.45 µm syringe filter. To obtain the desired triple cation perovskite composition of ~5% Rb, 42 µL of the ~1.5 M RbI stock solution was added to 1 mL FAMA solution, yielding a nominal composition of $Rb_{0.05}[(FA_{0.83}MA_{0.17})]_{0.95}Pb(I_{0.83}Br_{0.17})_3$. However, it is likely that the Rb is effectively not (fully) incorporated into the perovskite structure. The change in the I : Br ratio was neglected and we note that volume changes upon dissolving RbI in the DMF : DMSO mixture were not taken into consideration.

**FAMARC:** To obtain the quadruple cation perovskite composition of ~5% Rb and ~5% Cs, 42 µL of the RbI solution and 42 µL of the CsI solution were added to 1 mL FAMA solution, yielding a nominal composition of $Rb_{0.05}Cs_{0.05}[(FA_{0.83}MA_{0.17})]_{0.9}Pb(I_{0.83}Br_{0.17})_3$. The change in the I : Br ratio was neglected and we note that volume changes upon dissolving the Cs and Rb salts were not taken into consideration to calculate the additive concentration.

**Perovskite film deposition:**

For the MAPI, MAPBrI and MAPBr films, 75 µL of the perovskite precursor solution was spin-coated inside a nitrogen-filled glovebox at 1000 rpm and 5000 rpm for 10 s and 30 s, respectively. Approximately 15 s before the end of spinning, 500 µL of chlorobenzene was added to the film. The MAPI films were first annealed at 40 °C for 40 min and subsequently annealed at 100 °C for 10 min. The MAPBrI and MAPBr films were annealed at 100 °C for 10 min. For the FAMA, FAMAC, FAMAR and FAMARC films, the perovskite solution was spin-coated at 1000 rpm and 4000 rpm for 10 s and 30 s, respectively. Chlorobenzene was added as an anti-solvent 20 s before the end of spinning. The perovskite film formation was completed by annealing at 100 °C for 60 min on a hotplate in nitrogen atmosphere.

**Device Fabrication:**

Fluorine-doped tin oxide (FTO) coated glass substrates (7 Ω/sq) were patterned by etching with zinc powder and 3 M HCl solution and successively cleaned with deionized water, a 2% Hellmanex detergent solution, ethanol and finally treated with oxygen plasma for 5 min. A compact $TiO_2$ layer was deposited as hole blocking layer on the substrate *via* a sol-gel approach. A mixture of 2 M HCl (35 µL) and anhydrous isopropanol (2.53 mL) was added dropwise to a solution of 370 µL titanium(IV) isopropoxide (Sigma-Aldrich) in isopropanol (2.53 mL) under vigorous stirring. The $TiO_x$ solution was spin-coated dynamically onto the FTO substrates at 2000 rpm for 45 s, followed by annealing in air at 150 °C for 10 min and subsequently at 500 °C for 45 min. Alternatively, 10–15 nm compact $SnO_x$ electron transport layers were prepared by atomic layer deposition (ALD) on FTO-coated glass substrates which were patterned and cleaned as in the $TiO_2$ preparation. Tetrakis (dimethylaminotin) (IV) (TDMASn, Strem, 99.99%) was used as a tin precursor. The deposition was conducted at 118 °C with a base pressure of 5 mbar in a Picosun R-200 Advanced ALD reactor. The tin precursor was held at 75 °C during depsitions. Ozone gas was produced by an ozone generator (INUSA AC2025). Nitrogen (99.999%, Air Liquide) was used as the carrier and purge gas with a flow rate of 50 sccm per precursor line. The growth rate was determined via spectroscopic ellipsometry on Si(100) witness substrates. A Cauchy model was used for the tin oxide layer and the growth rate was 0.69 Å per cycle. After the deposition of the electron transporting layer and the perovskite layer, a 2,2',7,7'-tetrakis-(*N,N*-di-*p*-methoxyphenyl-amine)-9,9'-Spirobifluorene (Spiro-OMeTAD) hole transporter layer was applied. 1 mL of a solution of Spiro-OMeTAD (Borun Chemicals, 99.8%) in anhydrous chlorobenzene (75 mg/mL) was doped with 10 µL 4-*tert*-butylpyridine (Sigma-Aldrich, 96%) and 30 µL of a 170 mg/mL lithium bis(trifluoromethane)sulfonimidate (Li-TFSI) (Sigma-Aldrich, 99.95%) solution in acetonitrile (Sigma-Aldrich, anhydrous) and deposited by spin-coating at 1500 rpm for 40 s and then 2000 rpm for 5 s. After storing the samples overnight in air at 25% relative humidity, 40 nm Au was deposited through a patterned shadow mask by thermal evaporation at $8 \times 10^{-7}$ mbar to form the back electrode.

## 2. Analysis of frequency dependent 2nd harmonic electroabsorption and interpretation of electric field screening

***On the evaluation of $\phi_{eff}$***

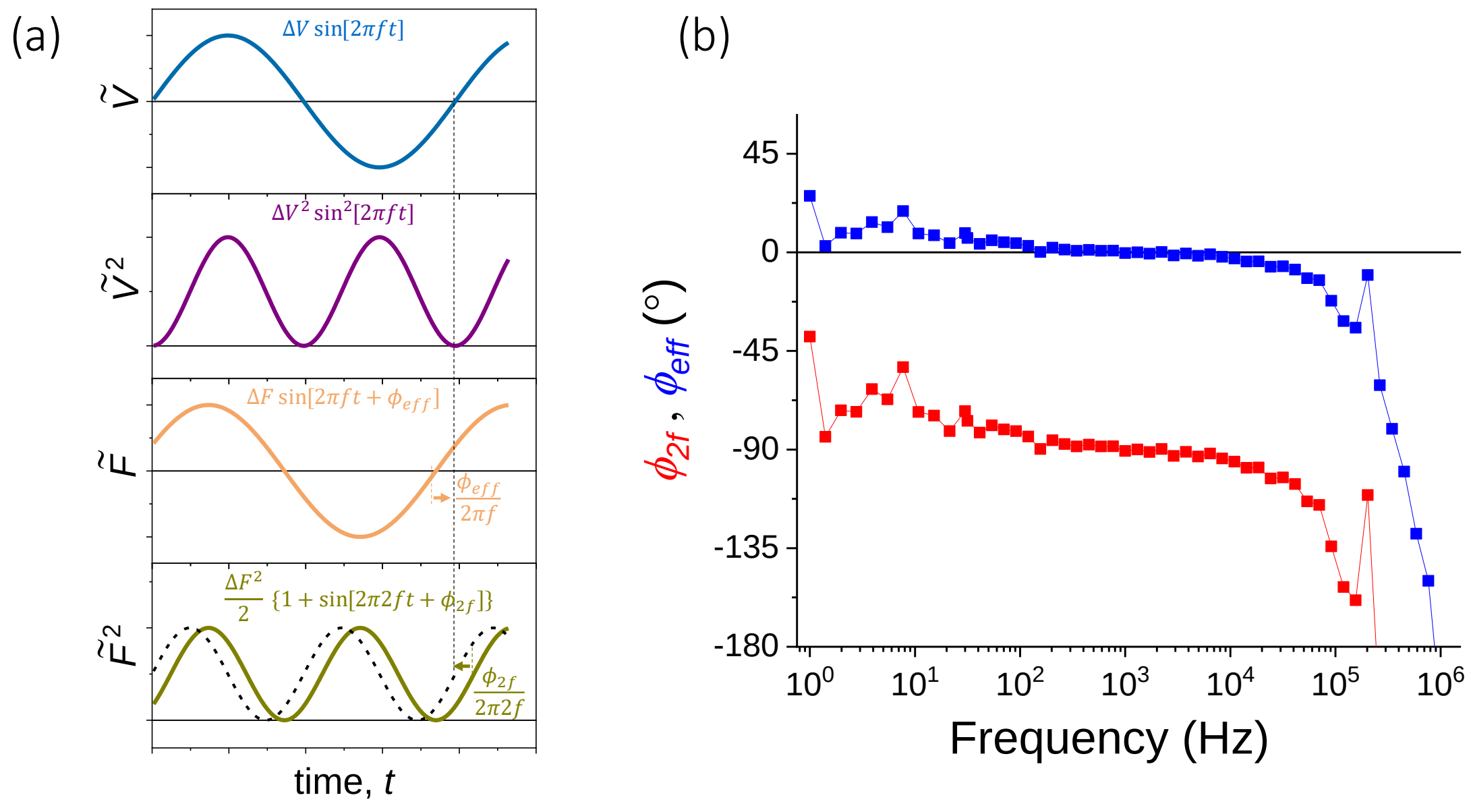

**Figure S1. Considerations on phase for 2nd harmonic EA measurements.** (a) Relative phase between the applied voltage in an electroabsorption experiment and the other relevant components: square change of the applied voltage $\tilde{V}^2$, change in electric field $\Delta F$ resulting from the applied voltage and from some screening in the device, square of the change in electric field $\Delta F^2$. The diagram shows the definition of $\phi_{eff}$ with respect to the measured phase $\phi_{2f}$ during the 2nd harmonic measurement. $\phi_{eff}$ is representative of the phase lag between the change in field and the applied voltage. (b) Comparison between the $\phi_{2f}$ phase obtained from the 2nd harmonic electroabsorption measurement and the calculated $\phi_{eff} = \phi_{2f}/2 + \pi/4$.

Figure S1a shows the definition of $\phi_{eff}$ based on the measured phase ($\phi_{2f}$) of the 2nd harmonic electroabsorption of solar cells. Figure S1b displays the comparison between these two quantities as a function of frequency. While $\phi_{2f}$ has a value of -90° in the kHz frequency range where negligible electric field screening is expected to occur in the active layer, $\phi_{eff} \approx 0°$, consistent with the applied voltage and the electric field in the active layer being in phase. As mentioned in the main text, $\phi_{eff}$ corresponds to $\phi_f$ for the case of $\tilde{F}$ having a sinusoidal profile (Figure S1a). Because in our experiments large values of $\Delta V$ are used, the $\tilde{F}$and $\tilde{F}^2$ profiles are not sinusoidal. This means that when taking the square of the Fourier series describing at each point (see next section), intermodulation terms that vary with $2f$ frequency are present and can ultimately change the value of $\phi_{eff}$ calculated from 2nd harmonic measurements compared with $\phi_f$.

*Interpretation of frequency dependent electroabsorption measurements*

We now comment on the interpretation of the changes in 2nd harmonic electroabsorption as a function of frequency. We expect the 2nd harmonic electroabsorption signal to be proportional to the square of the changes in electric field in the active layer of the solar cell (see Equation 2 in the main text). Considering a 1D approximation of the solar cell stack, we define the scalar function $F(x,t)$ as the electric field as a function of position and time in the perovskite layer of thickness *d*. In general the function $F(x,t)$ resulting from the applied oscillating voltage $V = \tilde{V} = \Delta V \sin[2\pi f t]$ can be expressed as the Fourier series

$$F(x,t) = \bar{F}(x) + \tilde{F}(x,t) = \bar{F}(x) + \sum_{n=1}^{\infty} \Delta F_{nf}(x) \sin[2\pi n f t + \phi_{nf}(x)] \qquad \text{Eq. S1}$$

We define $F_{ms}(t)$ as the square change in electric field averaged across the perovskite layer thickness as follows

$$F_{ms}(t) = \frac{1}{d}\int_0^d \left\{\sum_{n=1}^{\infty} \Delta F_{nf}(x) \sin\left[2\pi n f t + \phi_{nf}(x)\right]\right\}^2 dx \qquad \text{Eq. S2}$$

The 2nd harmonic electroabsorption signal amplitude is proportional to the amplitude of the component of $F_{ms}(t)$ oscillating at $2f$

$$\frac{\Delta T_{2f}}{\bar{T}} \propto \sqrt{\left\{2f \int_0^{\frac{1}{f}} F_{ms}(t) \sin[2\pi 2 f t]\, dt\right\}^2 + \left\{2f \int_0^{\frac{1}{f}} F_{ms}(t) \cos[2\pi 2 f t]\, dt\right\}^2} \qquad \text{Eq. S3}$$

To provide a simplified picture of the complex dependence of $\frac{\Delta T_{2f}}{\bar{T}}$ on the field distribution at different frequencies, let us consider two extreme scenarios:

(i) *In a first scenario, we assume that the entire change in applied potential drops within the perovskite layer*. This implies that the integrated value of the change in electric field over the active layer thickness equals the change in applied potential at all times (and at all frequencies):

$$\int_0^d \tilde{F}(x,t) dx = \Delta V \sin[2\pi f t] \qquad \text{Eq. S4}$$

We can now analyze the expected magnitude of the EA signal for both the high and the low frequency cases (indicated below as HF and LF). For high frequencies that are lower than the $R_sC_g$ limited regime but higher than the SR2 regimes (see Figure 5 and 6), we consider the simplified case where no screening within the active layer of the solar cell occurs. Thus, a sinusoidal change in electric field that is independent on position across the perovskite and that is in phase with the applied oscillating voltage is expected. We can then write $\tilde{F}_{HF}(x,t) = \tilde{F}_{HF}(t) = \Delta F_{f,HF} \sin[2\pi f t]$, where $\Delta F_{f,HF} = \Delta V / d$ is the (position independent) amplitude of the change in electric field in the perovskite layer due to the applied potential at high frequencies. As $F_{ms}(t) = \Delta F_{f,HF}^2 \sin^2[2\pi f t] = \frac{\Delta F_{f,HF}^2}{2}\{1 - \cos[2\pi 2 f t]\}$ it results from Equation S3 that:

$$\frac{\Delta T_{2f,HF}}{\bar{T}} \propto \frac{1}{d}\int_0^d \frac{\Delta F_{f,HF}^2}{2} dx = \frac{1}{2d}\left(\int_0^d \Delta F_{f,HF} dx\right)^2 = \frac{\Delta F_{f,HF}^2}{2} = \frac{\Delta V^2}{2d^2} \qquad \text{Eq. S5}$$

We now consider the effect of an electric field screening process occurring at low frequencies. This varies the electric field profile within the perovskite layer. In such case, we need to consider the explicit dependence of the oscillating change in electric field on position as shown above. For simplicity, we assume that $\tilde{F}(x,t)$ oscillates at the same frequency of the applied voltage ($\Delta F_{nf,LF}(x) = 0$ for $n \neq 1$), similarly to the situation shown in Figure S1a. If we now consider the case of low enough applied frequencies, where all screening processes are able to follow the voltage oscillation we can make a further simplification. As the field is non zero in the perovskite layer at such low frequencies we can use the approximation $\phi_{f,LF}(x) \approx 0$ (see transfer function of EA signal in section 8). Based on Equation S4 and S5, and on the fact that for any function $f(x)$, $\frac{1}{d}\int_0^d f^2(x)dx \geq \frac{1}{d}\left(\int_0^d f(x)dx\right)^2$ , we can write for the 2nd harmonic EA signal at low frequencies $\frac{\Delta T_{2f,LF}}{\bar{T}}$:

$$\frac{\Delta T_{2f,LF}}{\bar{T}} \propto \frac{1}{d}\int_0^d \frac{\Delta F_{f,LF}^2(x)}{2}\, dx \geq \frac{1}{2d}\left(\int_0^d \Delta F_{f,LF}(x)dx\right)^2 = \frac{\Delta V^2}{2d^2}$$

This simplified analysis indicates that, if the change in applied potential were to drop fully within the active layer of the device, a larger EA signal magnitude would be expected at low enough frequencies compared to the signal measured at high frequencies.

(ii) *In the second scenario, we consider the case where most of the change in applied potential at low frequencies drops across the space charges in the contacts and consequently the perovskite layer would be nearly field free*. This is expected to occur when the contact layer capacitance is much lower than the interfacial capacitance on the perovskite side. The EA signal would be expected to drop monotonically when going from high to low frequencies.

Experimentally we observe a drop in the EA magnitude at low frequencies. We conclude that a significant fraction of the applied potential drops across the contact layers, as described in (ii) above. On the other hand, high values of capacitance are measured at low frequencies (see Figure 5a). If we interpret the low frequency behavior of the device in terms of electrode polarization and assume a lower contribution from stoichiometric polarization in the bulk of the device, this would imply that the capacitance associated with the space charge regions in the contact layers is relatively large (> 6 $\mu$F cm$^{-2}$). In section 3 of this document, we discuss drift-diffusion simulations using representative input parameter sets that approximately reproduce the capacitance spectra of the devices. The data show a non-monotonic drop in the simulated electroabsorption signal at low frequencies (see also Figure 5a), a behavior described by scenario (i). Experimentally we are not able to resolve the very low frequency region due to signal-to-noise limitations. While we do see an increase in the signal magnitude in the 1 Hz region for some measurements, this is comparable to the background low frequency noise due to the probe lamp used in this study (see Figure S2). We therefore cannot comment precisely on the changes in signal in the 1 Hz region and below. Based on the analysis of impedance spectra, which point towards similar values of capacitance on the contact and on the perovskite side, a mixed situation between scenario of (i) and (ii) is in fact most likely.

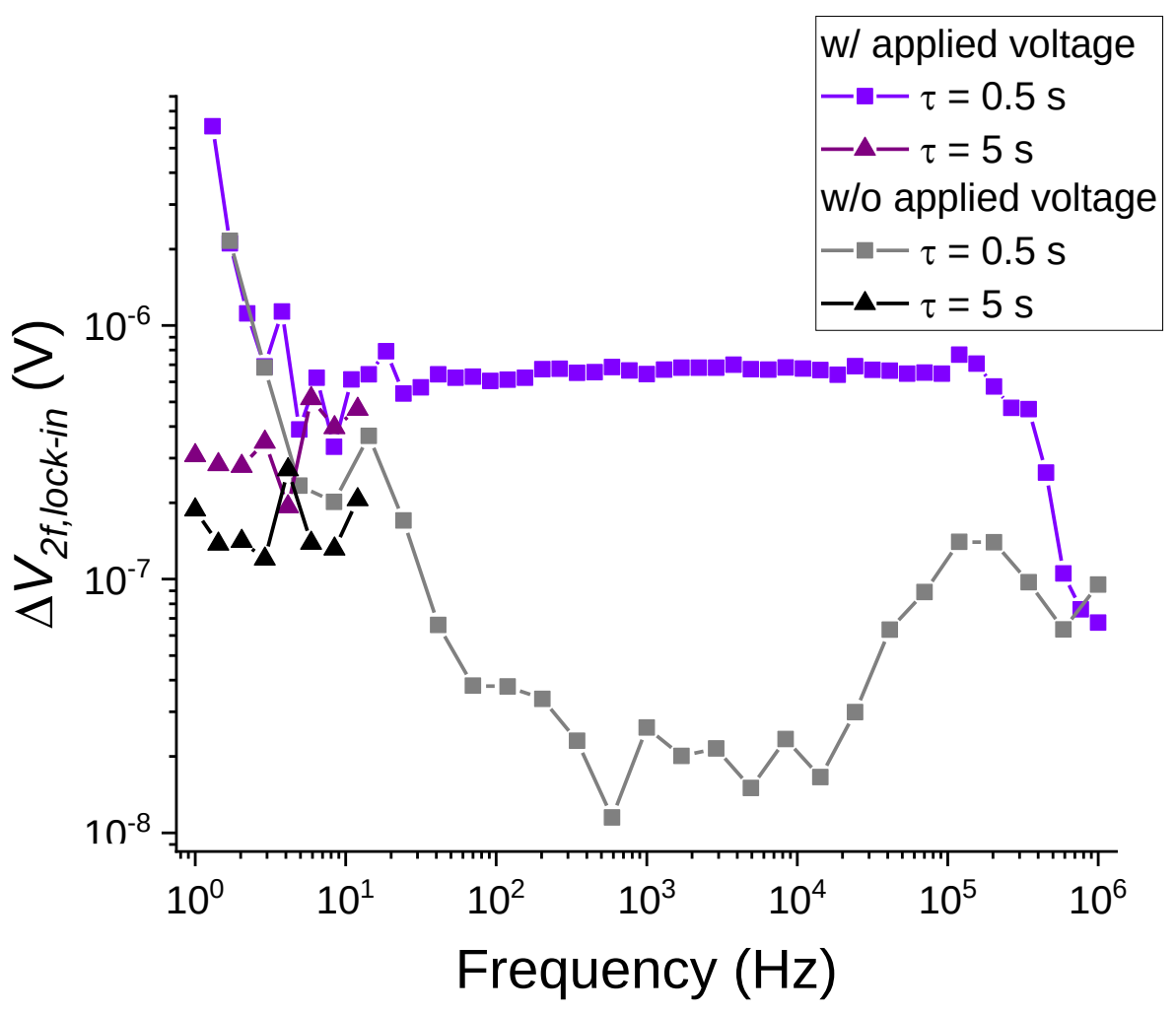

**Figure S2. Signal-to-Noise test for electroabsorption measurements**. Output voltage magnitude from the lock-in amplifier, used to evaluate the electroabsorption signal. The measurements were conducted with and without the applied voltage to quantify the background noise level. A longer integration time constant $\tau$ was set in the lock-in amplifier for detecting the low frequency regime.

In the discussion above and in the main text, we have focused on the effect of ion redistribution and accumulation at the interfaces between the perovskite layer and ion-blocking contacts. Another possibility is that, not only ions accumulate at such interface to screen the applied potential, but they also reversibly penetrate in the interlayer materials. If charged ions were to penetrate the contacts, one could expect a change in (compensating) electronic charge concentrations. While we cannot rule out this possibility for the solar cells investigated in this work, Figure S3 shows that no significant increase in the electron concentration in $TiO_2$ (which would be visible in the 1st harmonic electroabsorption signal in the 1–2 eV photon energy range [2]) is observed when changing the modulation frequency of the measurement. We cannot comment on possible changes in hole concentration increase in Spiro OMeTAD from the data in Figure S3. The oxidized Spiro OMeTAD signal is expected in the energy range 2–3 eV or below 1 eV [3]. The former photon energy range overlaps with the EA signal from the absorber layer in this case.

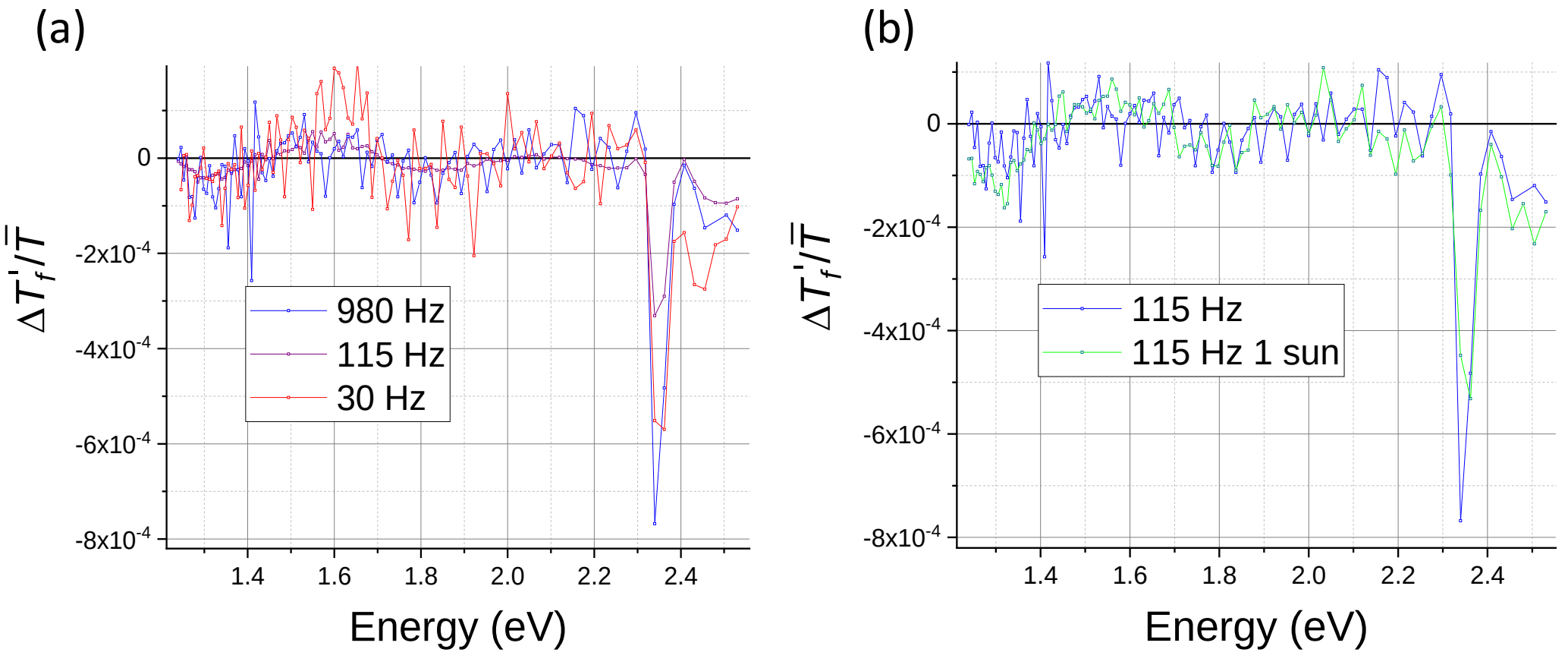

**Figure S3. Frequency and illumination dependent 1st harmonic EA.** 1st harmonic electroabsorption spectrum measured for a $TiO_2$/$MAPbBr_3$/Spiro OMeTAD solar cell at (a) different frequencies and (b) different bias light conditions.

## 3. Drift-diffusion simulations

We performed drift-diffusion simulations of perovskite solar cells using the open-source software DriftFusion (https://github.com/barnesgroupICL/Driftfusion). The code uses MATLAB's Partial Differential Equation solver for Parabolic and Elliptic equations (PDEPE) to solve the continuity equations and Poisson's equation for electron and hole densities (*n, p*), a positively charged mobile ionic charge density (*c*), and the electrostatic potential (*V*) as a function of position *x* and time *t*. While the software can also simulate the case with two ions, here we focus on the case where only cations are mobile. More details on the software are available in Ref. [4], [5], [6]

The input parameters for the simulations were selected based on literature and varied to obtain a similar response to the experimental observations. We used electron and hole transporting contacts with energy levels and work functions that are representative of the materials ($TiO_2$, Spiro OMeTAD). The complete parameter set used in the simulations is provided in Table S1.

**Table S1. Drift-diffusion simulation parameters.** These parameters were used for the simulated data displayed in this section, unless stated otherwise.

| Parameter name | Symbol | HTM | Active layer | ETM | Unit |
|---|---|---|---|---|---|
| Layer thickness | $d$ | 200 | 360 | 200 | nm |
| Band gap | $E_g$ | 2.7 | 1.6 | 3.2 | eV |
| Built in voltage | $V_{bi}$ | 0.7 | 0.7 | 0.7 | V |
| Relative dielectric constant | $\varepsilon_s$ | 4 | 32 | 30 | |
| Mobile ionic defect density | $N_{ion}$ | 0 | $10^{19}$ | 0 | $cm^{-3}$ |
| Ion mobility | $\mu_a$ | - | $10^{-10}$ | - | $cm^2 V^{-1} s^{-1}$ |
| Electron mobility | $\mu_e$ | 0.02 | 20 | 0.1 | $cm^2 V^{-1} s^{-1}$ |
| Hole mobility | $\mu_h$ | 0.02 | 20 | 0.1 | $cm^2 V^{-1} s^{-1}$ |
| Donor doping density | $N_A$ | $5.8 \times 10^{19}$ | - | - | $cm^{-3}$ |
| Acceptor doping density | $N_D$ | - | - | $5.8 \times 10^{19}$ | $cm^{-3}$ |
| Work function | $\phi_W$ | -4.85 | -4.6 | -4.15 | eV |
| Ionization potential | $IP$ | -4.9 | -5.4 | -7.3 | eV |
| Electron affinity | $EAff$ | -2.2 | -3.8 | -4.1 | eV |
| Effective density of states | $N_c$, $N_V$ | $4\ 10^{20}$ | $10^{19}$ | $4\ 10^{20}$ | $cm^{-3}$ |
| Band-to-band recombination rate coefficient | $k_{btb}$ | $10^{-12}$ | $10^{-12}$ | $10^{-12}$ | $cm^{-3} s^{-1}$ |
| SRH trap energy | $E_t$ | $E_{VB}$+0.15 | $E_{CB}$-0.8 | $E_{CB}$-0.15 | eV |
| SRH time constants | $\tau_n$, $\tau_p$ | $10^{-9}$ | - | $10^{-12}$ | s |
| Generation rate | $G$ | - | $2.64 \times 10^{21}$ | - | $cm^{-3} s^{-1}$ |

***Impedance and electroabsorption simulations***

For simulations of impedance and electroabsorption, a sinusoidal perturbation of the form

$$V_{app} = \bar{V} + \tilde{V} = \bar{V} + \Delta V \cdot \sin(\omega t) \qquad \text{Eq. S6}$$

was applied to the device stack (ω = 2π × frequency). For the impedance simulations, a small voltage amplitude was used ($\Delta V_{sim} =$ 20 mV) as the experiment was also carried out as a small perturbation from the steady state ($\Delta V_{exp} =$ 20 mV). The small signal perturbation was superimposed on a d.c. applied voltage, $\bar{V}$ =$V_{OC}$. The open circuit voltage was obtained for each light intensity by running the simulation until the device reached a steady-state. For the frequency dependent electroabsorption simulations the same a.c. and d.c. applied voltage amplitudes were used as in the practical experiment ($\Delta V_{sim} = \Delta V_{exp} =$ 800 mV, $\bar{V} = 0\ V$). In every case, the number of simulated periods has been extended until when no variations on a larger-than-one-period scale could be observed.

For the impedance simulations, the in-phase and out-of-phase current was evaluated via demodulation as previously described [6] to determine the simulated real and imaginary part of the impedance.

For electroabsorption measurements, the same procedure was used to evaluate the electric field in the active layer of the device, the square of which is related to the changes in absorption of the device detected via the 2nd harmonic electroabsorption measurements according to Eq. 2 in the main text. The resulting real and imaginary part of the electric field are function of position in the perovskite layer such that, at a given frequency, the electric field distribution in the perovskite can be expresses as (see also section S2 and methods section):

$$F_{sim}(x,t) = \bar{F}_{sim}(x) + \tilde{F}_{sim}(x,t) = \bar{F}_{sim}(x) + \sum_{n=1}^{\infty} \Delta F_{sim,nf}(x)\sin[2\pi nft + \phi_{sim,nf}(x)] \quad \text{Eq. S7}$$

We express the square of the change in electric field averaged across the perovskite film's thickness as $F_{ms}(t)$:

$$F_{ms}(t) = \frac{1}{d}\int_0^d \left(\sum_{n=1}^{\infty} \Delta F_{sim,nf}(x)\sin[2\pi nft + \phi_{sim,nf}(x)]\right)^2 dx \quad \text{Eq. S8}$$

where *d* is the thickness of the perovskite layer.

By extracting the 2nd harmonic component $F_{ms}|_{2f}(t)$ from $F_{ms}(t)$, one obtains:

$$F_{ms}|_{2f}(t) = (\Delta F_{sim})^2 \sin(2\pi 2ft + \phi_{sim,2f}) \quad \text{Eq. S9}$$

The amplitude $(\Delta F_{sim})^2$ of this harmonic is the quantity of interest, in that $\Delta T_{2f}/\bar{T} \propto (\Delta F_{sim})^2$. $\Delta F_{sim}$ is used in the main text and in this document as a quantity that can be compared with the experimental EA signal $\left(\Delta T_{2f}/\bar{T}\right)^{1/2}$.

***Step-dwell-probe simulations***

Simulations of step-dwell-probe measurements were performed by following a perturbation protocol consistent with the experimental procedure:

1. The device was equilibrated in the dark at short circuit
2. At $t$ = 0 s, a forward bias voltage step of 0.6 V was applied to the stack. For SDP under light, a bias light was simultaneously included at this point.
3. The device was kept at 0.6 V for a variable dwell time $\Delta t$.

4 A monochromatic light pulse (wavelength 638 nm) with intensity of 1 sun equivalent was applied. The pulse duration was long enough to resolve the electronic photocurrent value after the initial transient, similarly to the experiment.

5 From the resulting transient photocurrent (function of $\Delta t$), the value of $\Delta J_{sim}$ was extracted as the final value during the application of the pulse.

***Simulation results***

Below we discuss results obtained from the simulations described above. In Figure S4 we show data obtained using the input parameters in Table S1 and for two different thicknesses of the active layer.

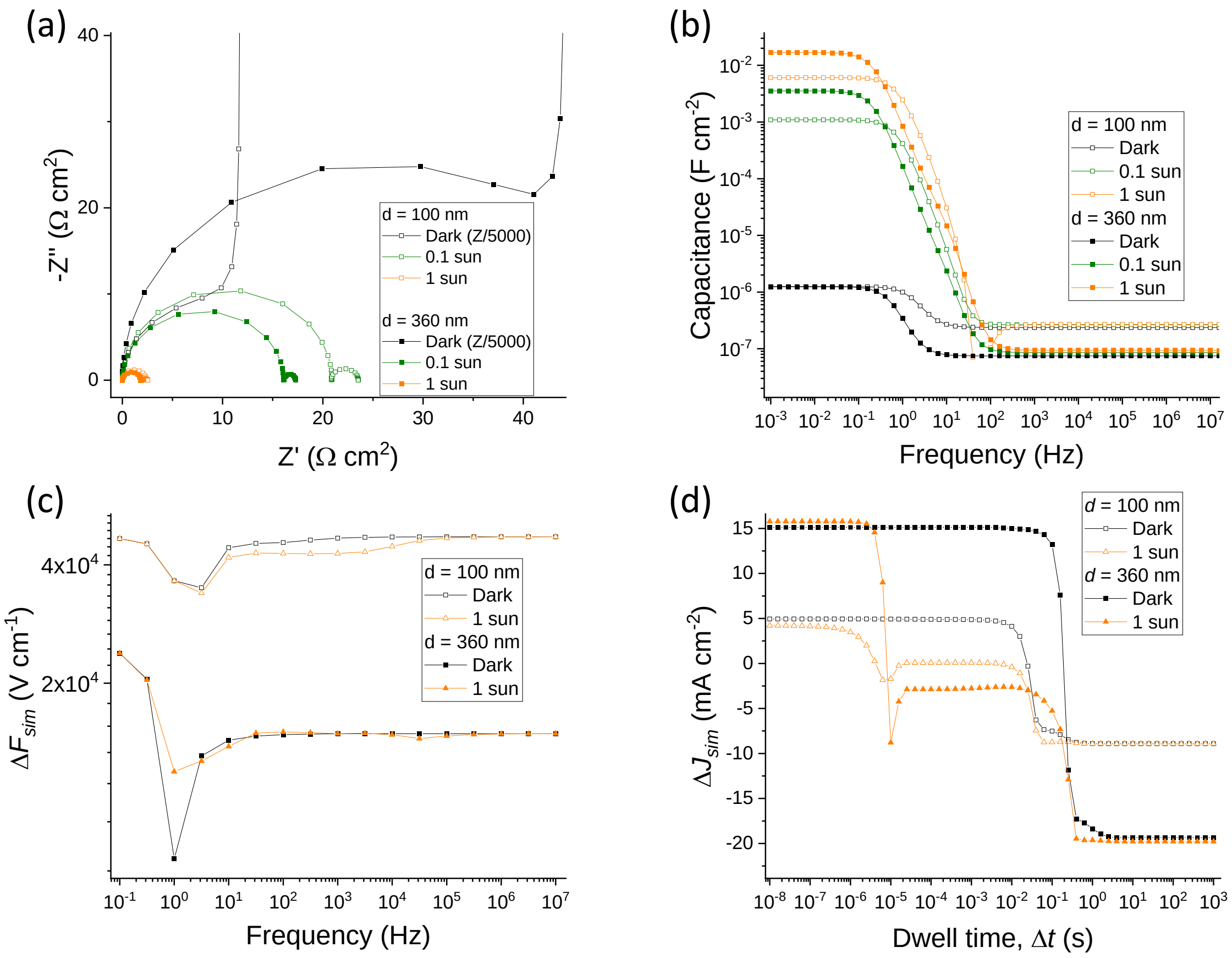

**Figure S4. Drift-diffusion simulations of $TiO_2$/MAPI/Spiro OMeTAD solar cell structures**. Simulations comparing two different active layer thicknesses. Data corresponding to the cell under dark and under light are included. (a) and (b) show impedance simulations (Nyquist and capacitance plots respectively), (c) shows the simulated rms of the electric field in the active layer and (d) displays the simulated SDP response of the cells.

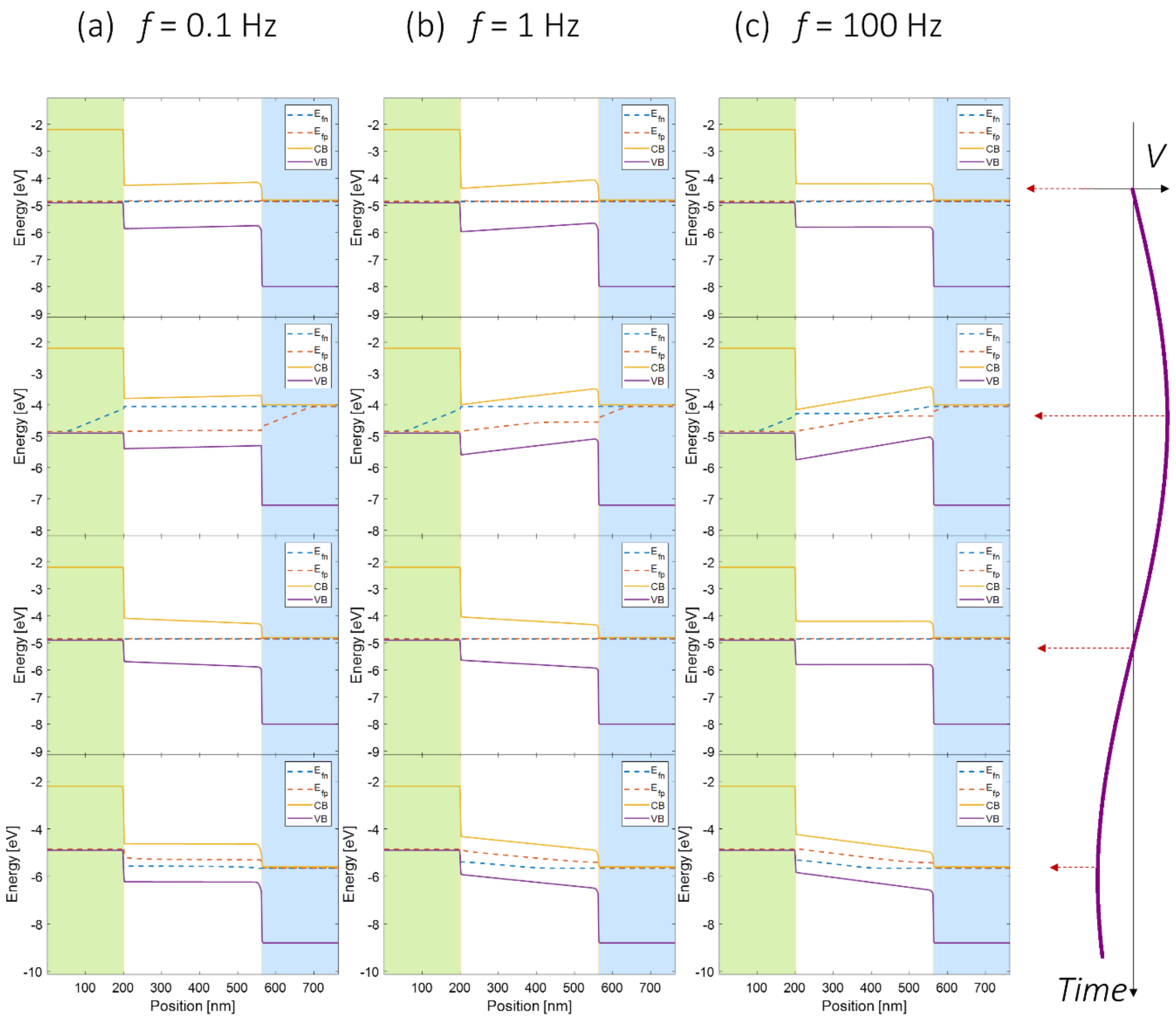

**Figure S5. Simulated energy level diagram corresponding to the electroabsorption simulation for a device with the properties shown in Table S1**. An applied voltage with $\Delta V = 0.8\ V$ is considered and the cases of applied frequencies of (a) 0.1 Hz, (b) 1 Hz and (c) 100 Hz are shown here. For each column, four different diagrams are shown, corresponding to the four values of applied voltage indicated by the sinusoidal profile on the right.

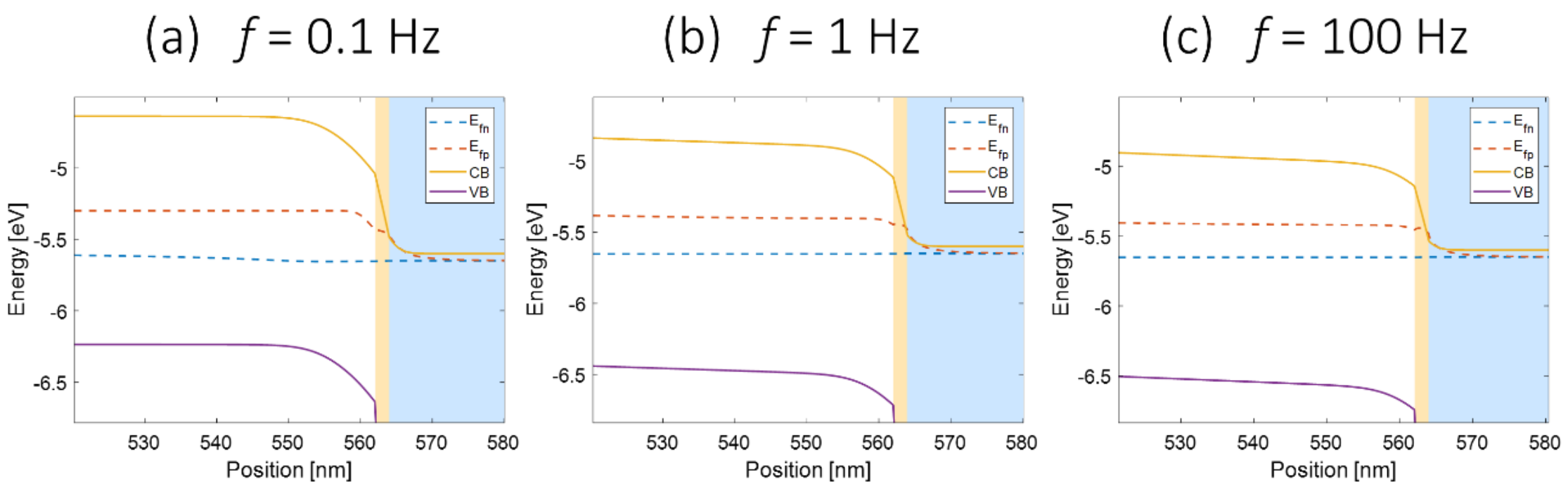

**Figure S6. Interfacial potential profile obtained from EA simulations.** The graphs emphasize the potential drop in the MAPI (white background) at the interface with the electron transport material (blue background) when $V$ = -0.8 V (see also Figure S5), for different applied frequencies. Such drop increases the lower the applied frequency.

Below we discuss some key observations. Regarding the simulations *in the dark*:

- The impedance simulations show values of capacitance going from ~ 800 nF cm$^{-2}$ at high frequencies to about 1 µF cm$^{-2}$ at low frequencies. Both these values are slightly lower than the experimental results. For the former (high frequency), roughness in the real device would contribute towards a larger geometric capacitance. For the latter (low frequency), the large experimental value could be due to even higher doping in the contacts and ionic defect concentration in the perovskite layer, than the values used in Table S1. We also cannot exclude contribution from stoichiometric polarization at very low frequencies. [7]
- From the simulations of the three measurement techniques considered in this study, we observe a clear change in the screening time constant when varying the active layer thickness. Thinner devices respond faster to the applied potential perturbation, consistent with the lower ionic resistance.
- The electroabsorption simulations highlight a drop in signal magnitude when going to low frequencies (~1-10Hz). This is due to the ionic screening, which reduces the electric field in the bulk of the device (see Figure S5). This drop in magnitude is followed by an increase in signal at very low frequencies. As discussed in section 2, this is an effect due to the input parameters chosen here, which result in a non-negligible fraction of the electric field dropping within the perovskite, giving rise to an overall increase of $\Delta F_{sim}$ at very low frequencies. An example of the significant potential drops in the perovskite at the interface with the contacts at low frequencies is shown in Figure S6 for the case of reverse bias. This behavior cannot be easily resolved experimentally, due to the limitations of our setup. We cannot comment on whether such effect would be present in the experiments too.

The behavior *under light* highlighted by the experiments is only partially reproduced:

- The impedance simulations qualitatively correspond to the experimental data. We note that the time constant from capacitance simulations (related to the frequency at which capacitance starts to drop from the low frequency value) scales with thickness as expected. However, the value of $f_{0,C}$, as defined in the main text, does not scale in the same way, due to the fact that the geometric capacitance also varies with thickness. This result is consistent with the experimental observation (see data in Figure S4 and Figure 6 in the main text).
- On the other hand, both the electroabsorption and the SDP simulations show some clear differences from the results obtained experimentally under light. For the former, we do not observe a sharp drop in signal in the kHz frequency range as it is the case for the $TiO_2$/MAPI/Spiro OMeTAD cells discussed in the main text. Only in some cases (see for example the results for the thin devices in Figure S4 but also below) we do observe a slight decrease in magnitude at high frequencies, which we ascribe to the contribution of the photo-generated electronic charges within the perovskite layer. As for the SDP simulations, we reproduce the observation of two regimes of electric field screening discussed in the main text, although the drop in the simulated transient photocurrent amplitude associated with SR1 occurs at much faster time scales compared to the experiments. The differences in transient photo-current magnitude between the two simulated situations is due to the different optical thickness of the two cases considered.

Next, we discuss the effect of varying the capacitance of the contact materials. This is expected to influence the overall time constant of ionic redistribution, but also the relative fraction of the applied potential dropping across the perovskite and the contact layers. In order to investigate the effect of the interfacial contact capacitance, while keeping the situation as unchanged as possible, we vary the effective density of states of the contacts and maintain the work function and band edge positions constant for both contact materials. Note that varying the contact doping density would also influence the depletion layer capacitance, but a change in built-in voltage of the cell would be expected in that case. The results are displayed in Figure S7. We highlight the following points:

- The impedance simulations show that by increasing the contact capacitance (larger value of $N_C$ and $N_V$ in the contacts result in larger capacitance on equal position of the Fermi level) the low frequency capacitance in the dark is increased. For the extreme case of very large values of the contact capacitance (e.g., very high doping density, metal contacts etc. not shown here) the low frequency capacitance would be dominated by the ionic double layer or depletion capacitance on the perovskite side.
- The trend in the low frequency capacitance under light shows the opposite dependence on the contact capacitance. Based on the concept of ionic-to-electronic current amplification [6], this can be explained in terms of a lower fraction of the applied potential driving changes in recombination at the interfaces when the contact capacitance increases. This results in a reduced value of the apparent capacitance. The effect is also evident from the relative size of the low frequency semicircle in the Nyquist plot: the change in recombination resistance between the high frequency and the low frequency situation is smaller the larger the contact capacitance.
- The electric field in the solar cell in the electroabsorption simulations shows that with larger contact capacitance, the increase in $\Delta F_{sim}$ at very low frequency is more pronounced, consistently with the discussion in section 2 in this document. The simulated behavior under light is rather different from the experimental observation, although it is interesting to see that, for large values of $N_C$ and $N_V$, the electroabsorption signal would be expected to be larger under light than in the dark, a result that is observed for MAPI solar cells with $SnO_2$ contact layers (see section S6).

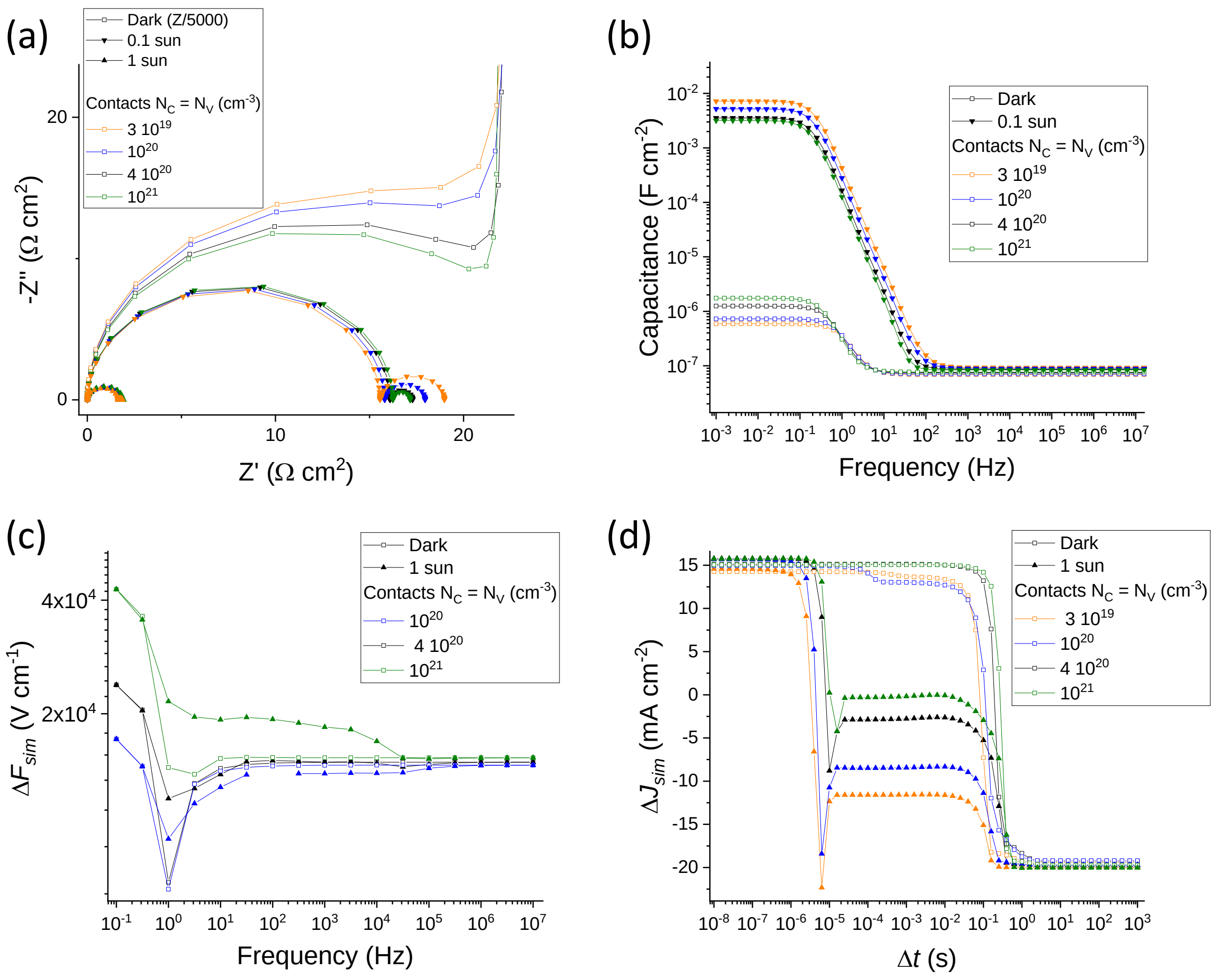

**Figure S7. Drift-diffusion simulations of $TiO_2$/MAPI/Spiro OMeTAD solar cell structures**. Simulations comparing different values of the effective density of states in the contact layers. Data corresponding to the cell under dark and under light are included. (a) and (b) show impedance simulations (Nyquist and capacitance plots respectively), (c) shows the simulated rms of the electric field in the active layer and (d) displays the simulated SDP response of the cells.

# 4. 1st Harmonic electroabsorption study

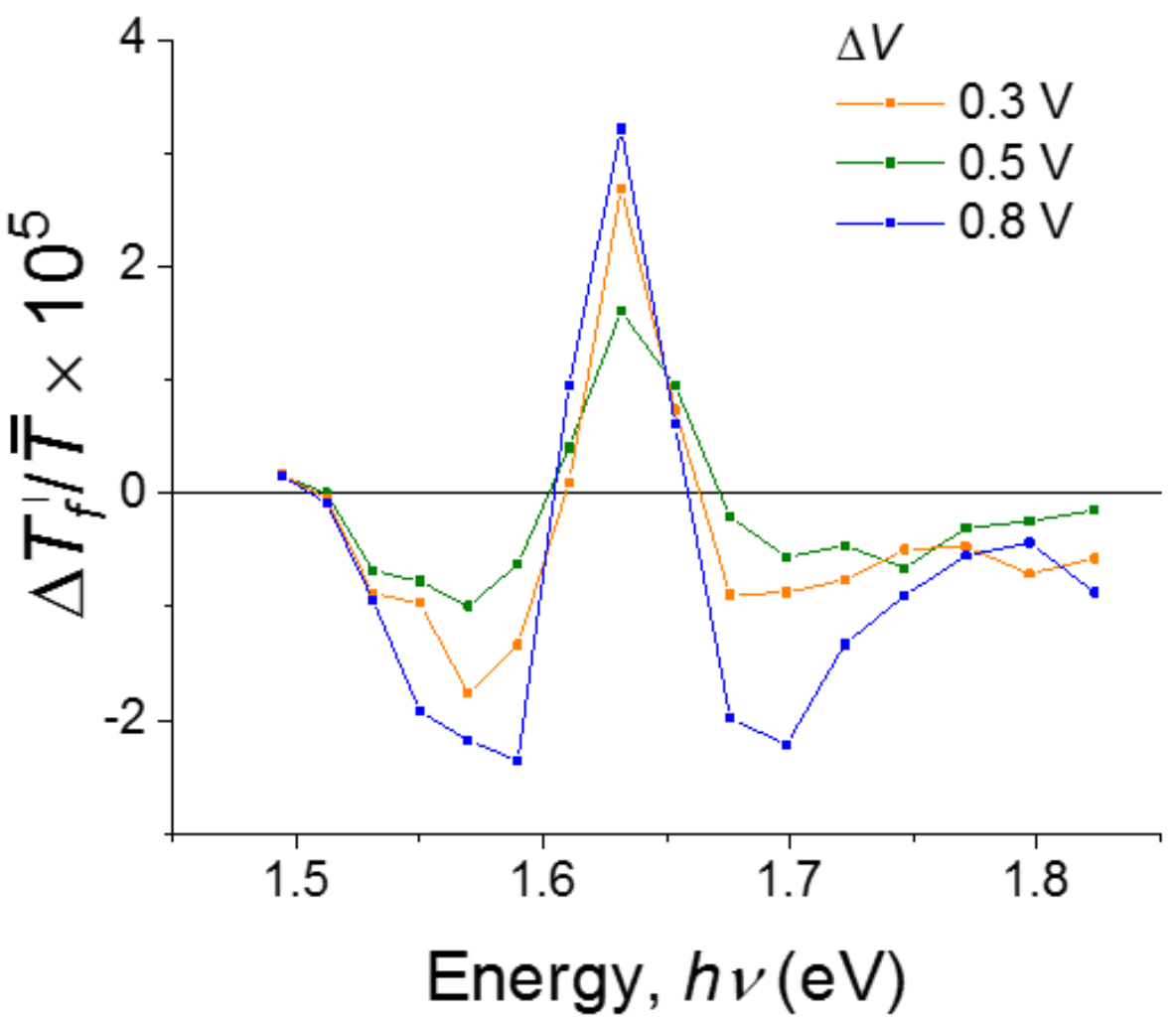

**Figure S8. 1st harmonic electroabsorption measurements on a $TiO_2$/MAPI/Spiro OMeTAD device.** The data show the EA measurement run for different values of the AC voltage amplitude. The scaling of the spectra magnitude does not follow the expected linear dependence. For MAPI cells, we find that the shape and magnitude of the 1st harmonic spectrum under nominally identical conditions can vary depending on the history of the sample, despite stabilization times before the measurements of over 200 seconds. This is ascribed to changes in the static electric field $\overline{F}$, which this signal depends on (see equation 1 in main text).

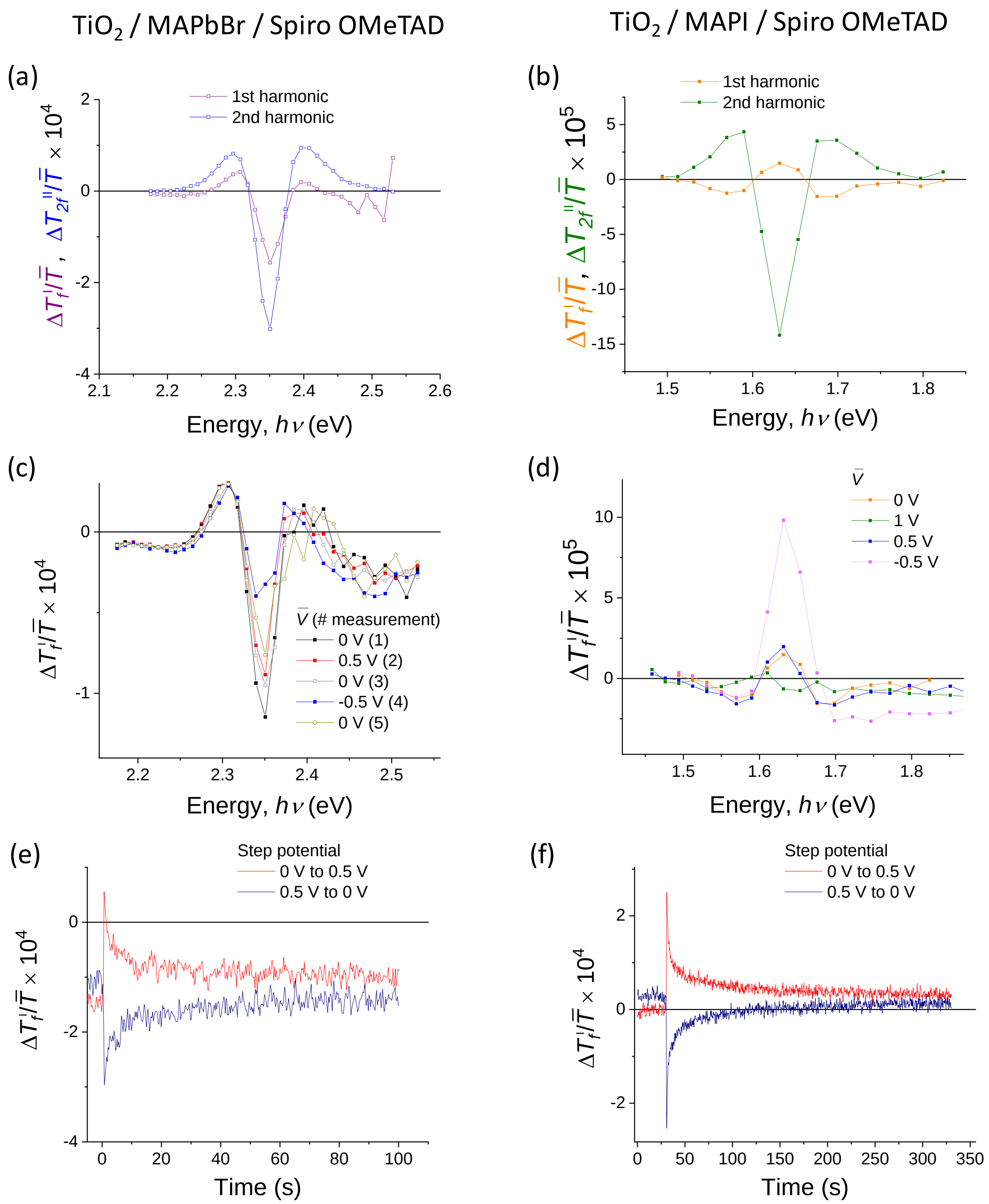

**Figure S9. 1st harmonic electroabsorption measurements on $MAPbBr_3$ and MAPI cells**. Left column refers to a $TiO_2$/$MAPbBr_3$/Spiro OMeTAD device while the right column to a $TiO_2$/MAPI/Spiro device. (a) (b) Comparison between 1st and 2nd harmonic signal. (c) (d) DC voltage dependence on the 1st harmonic EA spectrum. (e) (f) Transient Electroabsorption of the 1st harmonic signal for a $\bar{V} = 0\ V \rightarrow 0.5\ V$ step and for a $\bar{V} = 0.5\ V \rightarrow 0\ V$ step probed at (e) 527.5 nm and (f) 760 nm.

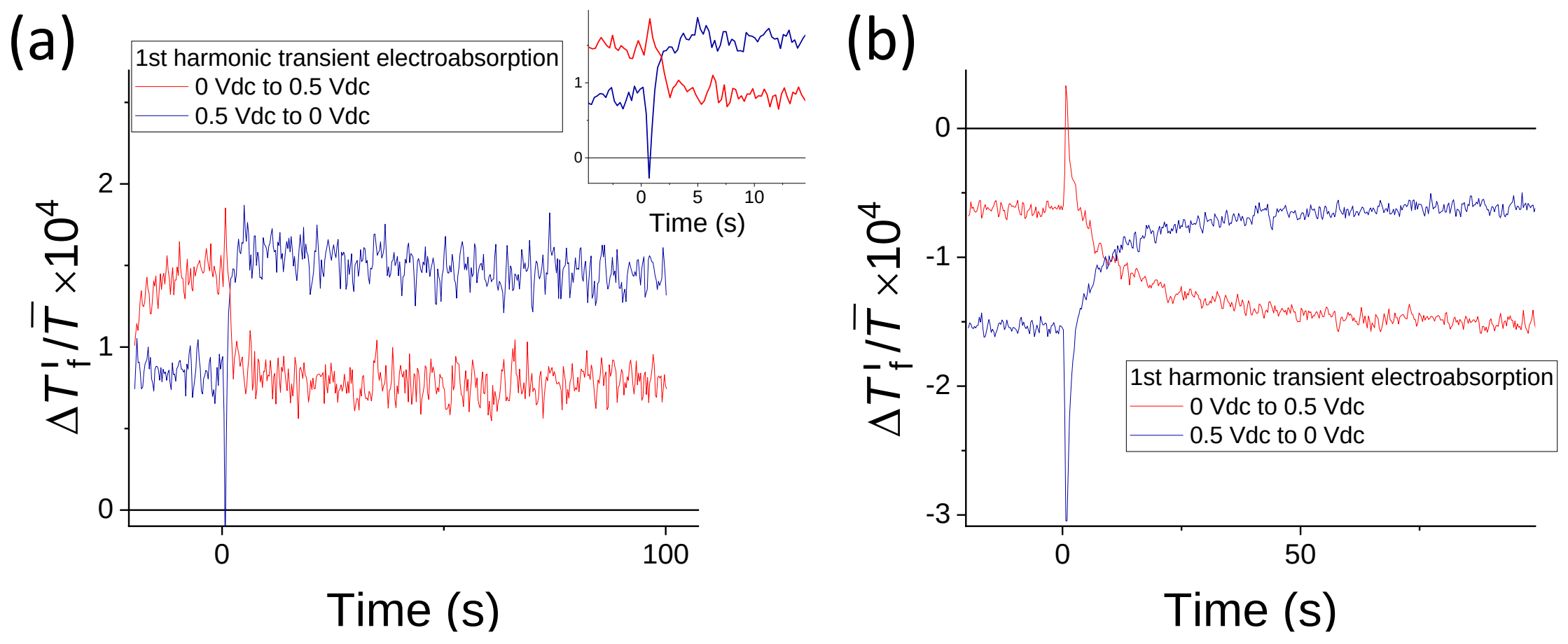

**Figure S10. 1st harmonic transient electroabsorption on $SnO_2$/MAPI/Spiro OMeTAD devices**. Measurements run for two solar cells with MAPI as active layer, Spiro OMeTAD as HTM and $SnO_2$ as ETL. The transients are obtained from applying potential steps $\bar{V}$= 0 V -> 0.5 V and $\bar{V}$= 0.5 V -> 0 V superimposed to the oscillating voltage ($\Delta V = 0.5\ V$ f = 1 kHz). (a) MAPI thinkness 360 nm, (b) MAPI thickness 100 nm. The inset of (a) shows the early time scale peaks.

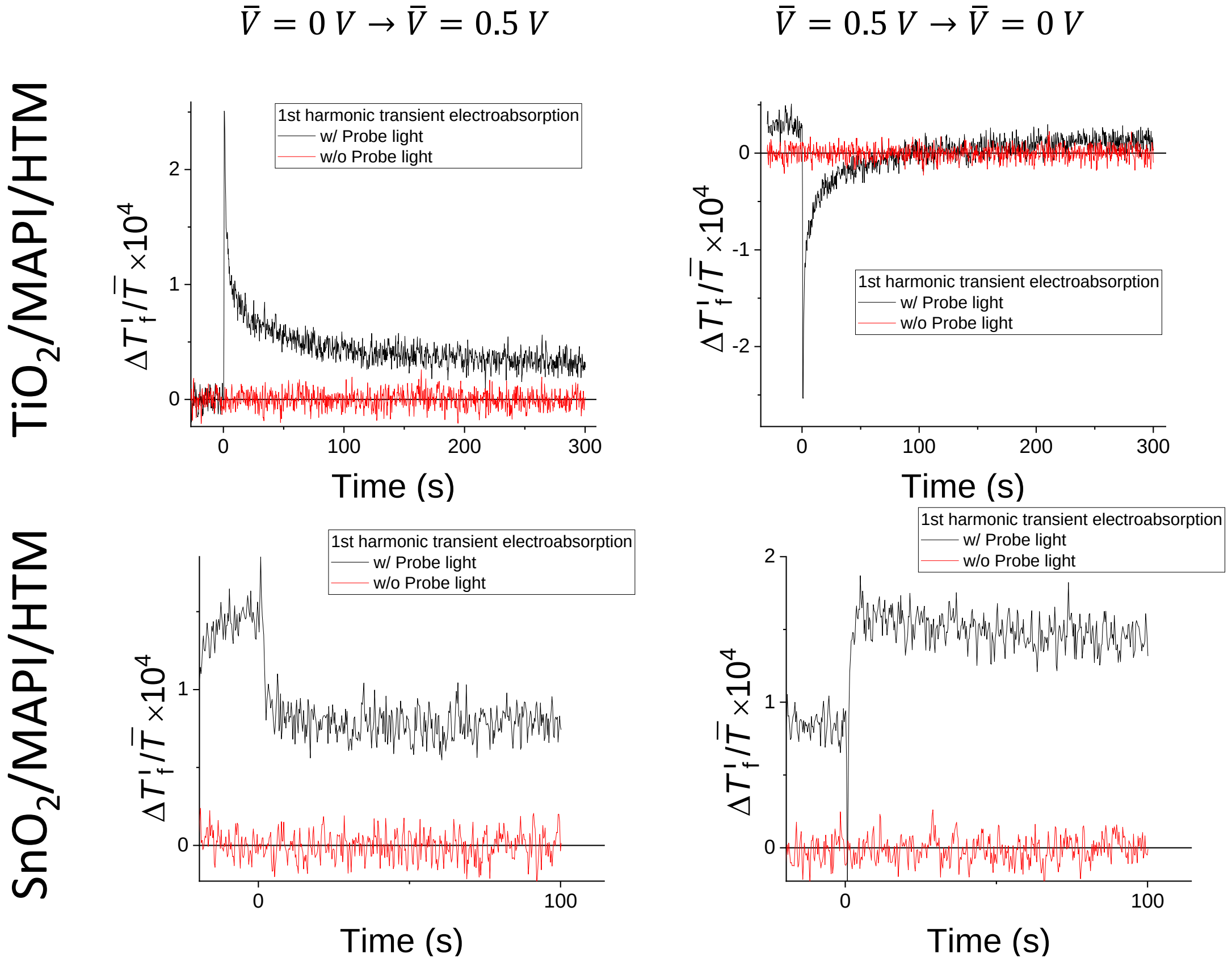

**Figure S11. 1st harmonic transient electroabsorption control experiment**. Measurements run for two solar cells with MAPI as active layer, Spiro OMeTAD as HTM and either (top graph) $TiO_2$ or (bottom graphs) $SnO_2$ as ETL. Transients related to potential steps (left graphs) $\bar{V}$= 0 V -> 0.5 V and (right graphs) $\bar{V}$= 0.5 V -> 0 V are shown. For each experiment, an additional measurement was run without the probe light, to check that the measured signal is not influenced by electroluminescence induced by the oscillating voltage ($\Delta V = 0.5\ V$ f = 1 kHz).

# 5. Step-dwell-probe measurements

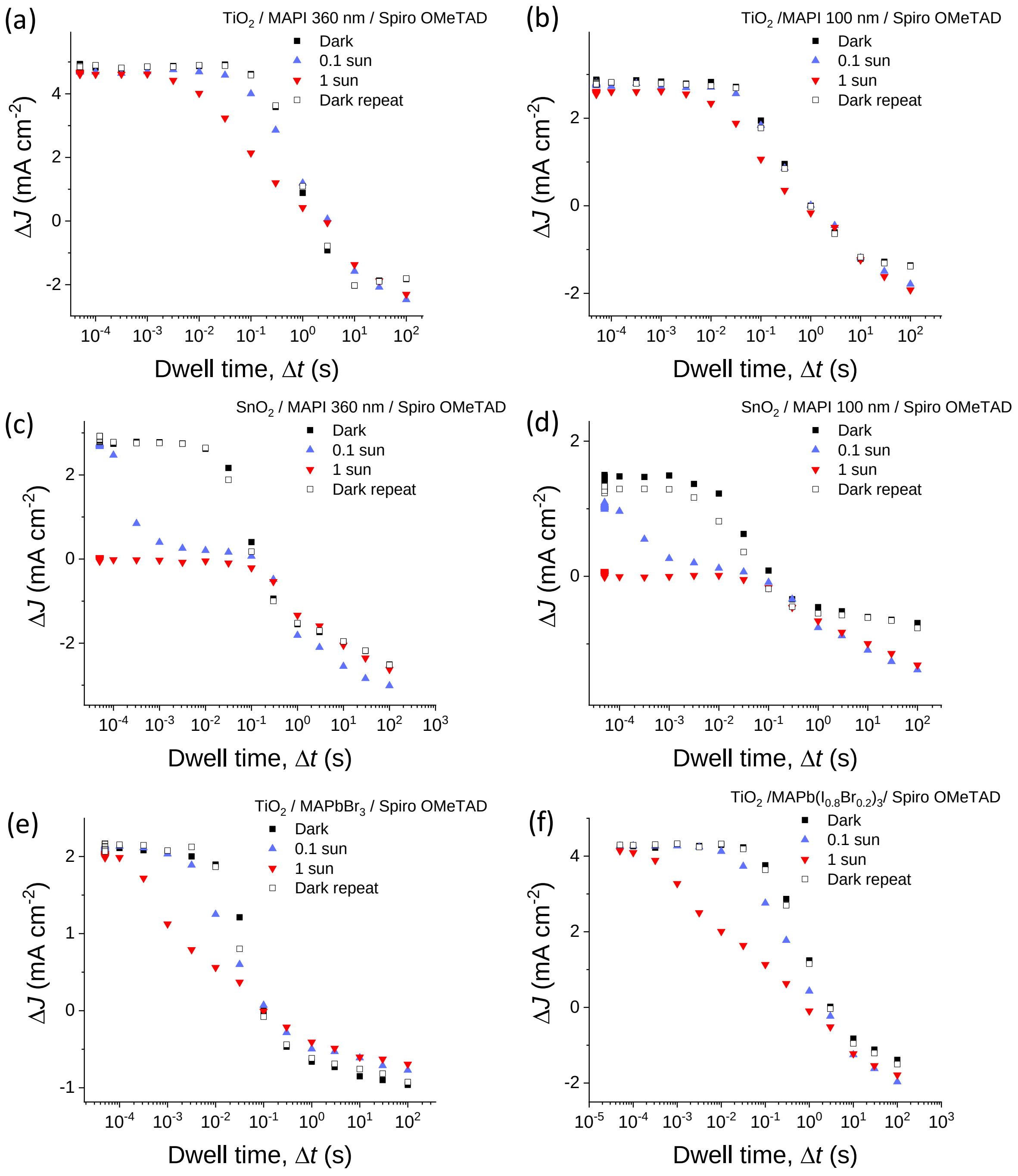

**Figure S12. Step-dwell-probe measurements.** Measurements run for solar cells with different active layer thicknesses, ETL interlayer materials and active layer compositions. Results obtained in the dark and under 0.1 and 1 sun equivalent illumination are shown.

# 6. Morphology study of MAPI

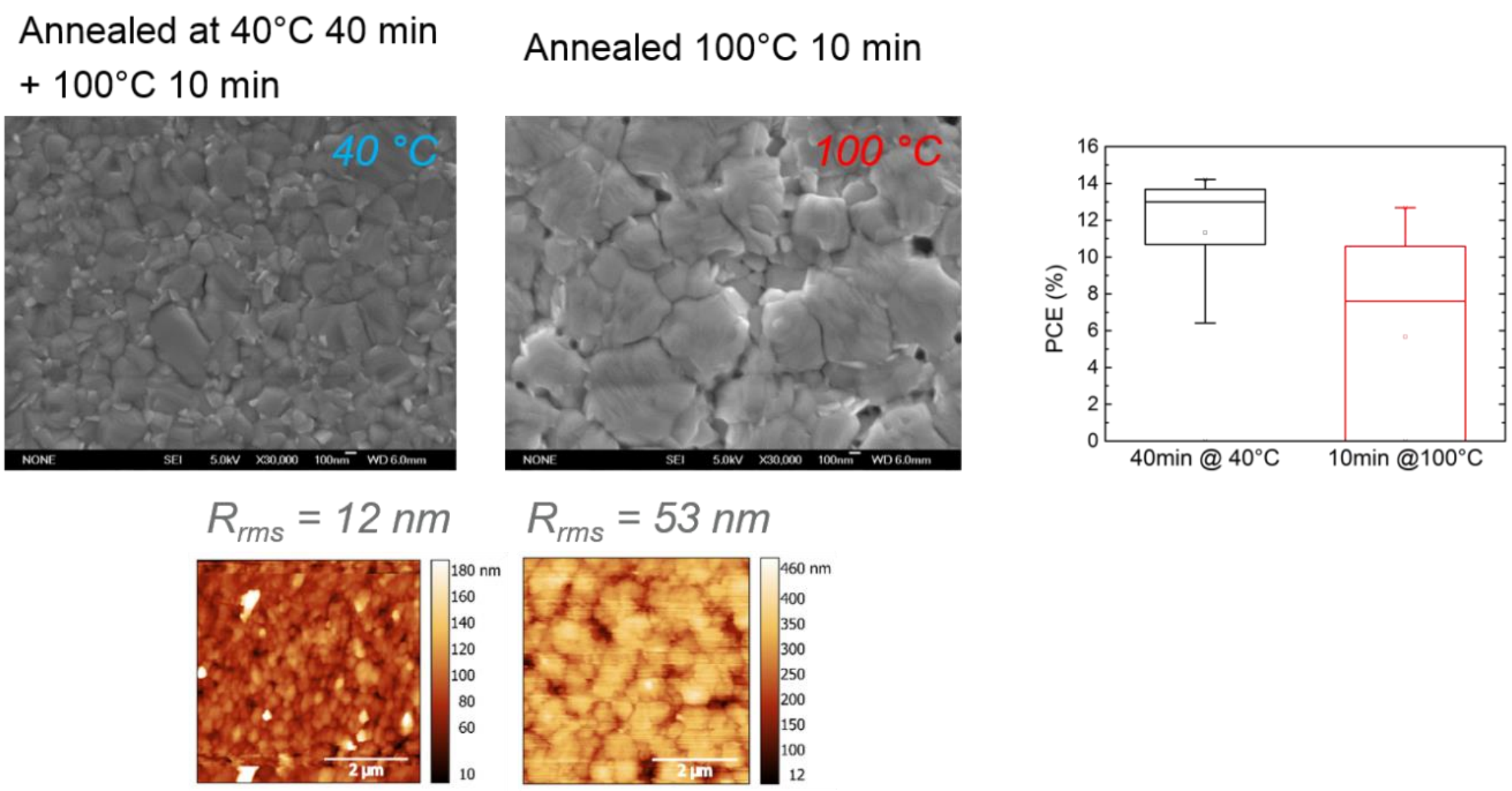

**Figure S13. SEM and AFM measurements performed on MAPI films processed using different annealing recipes**. On the right the statistics on the photoconversion efficiency of solar cell devices fabricated via the two different procedures is shown.

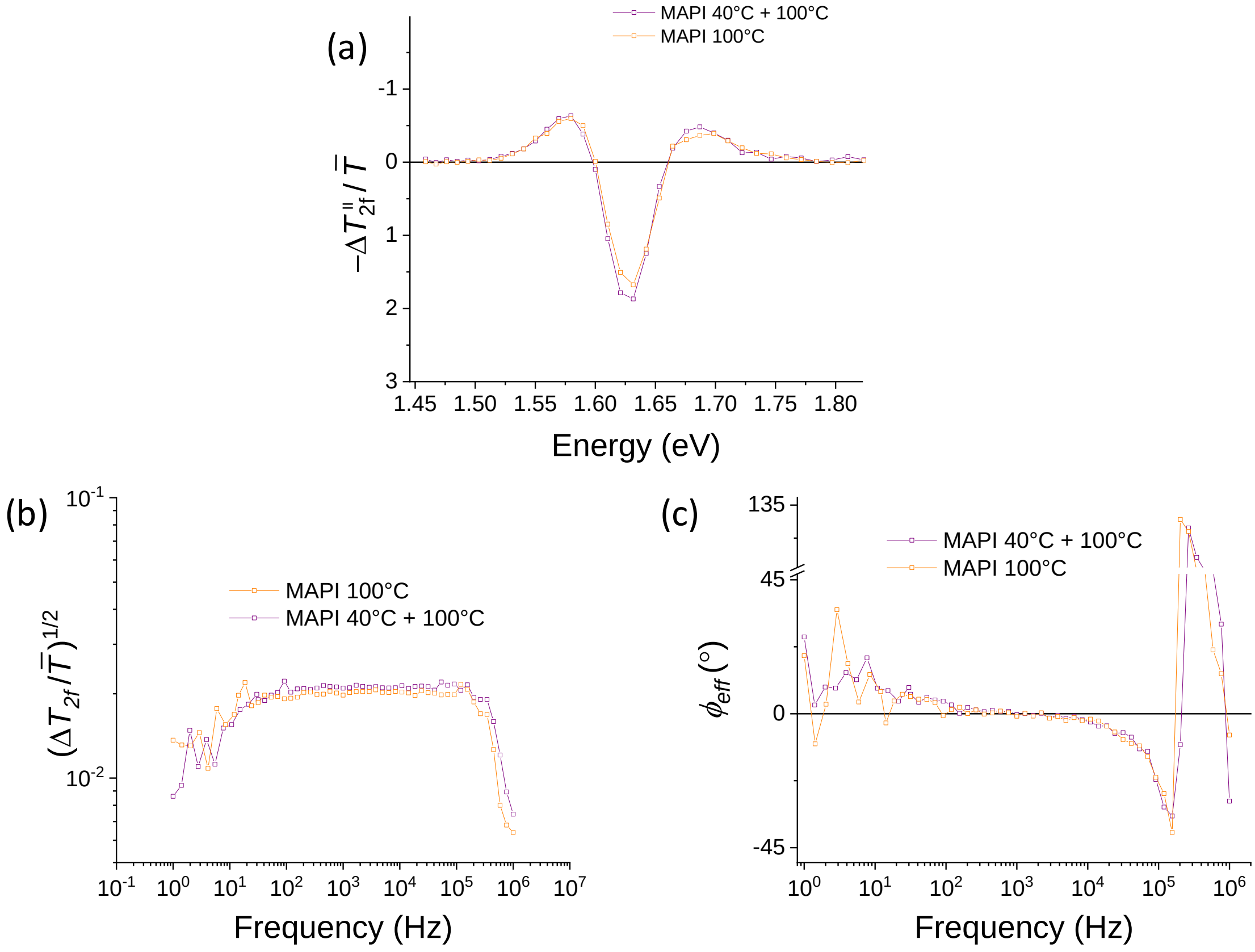

**Figure S14. Electroabsorption study of $TiO_2$/MAPI/Spiro OMeTAD solar cells with different morphology of the active layer**.

# 7. Further electroabsorption measurements

### *Measurements on solar cells with $SnO_2$*

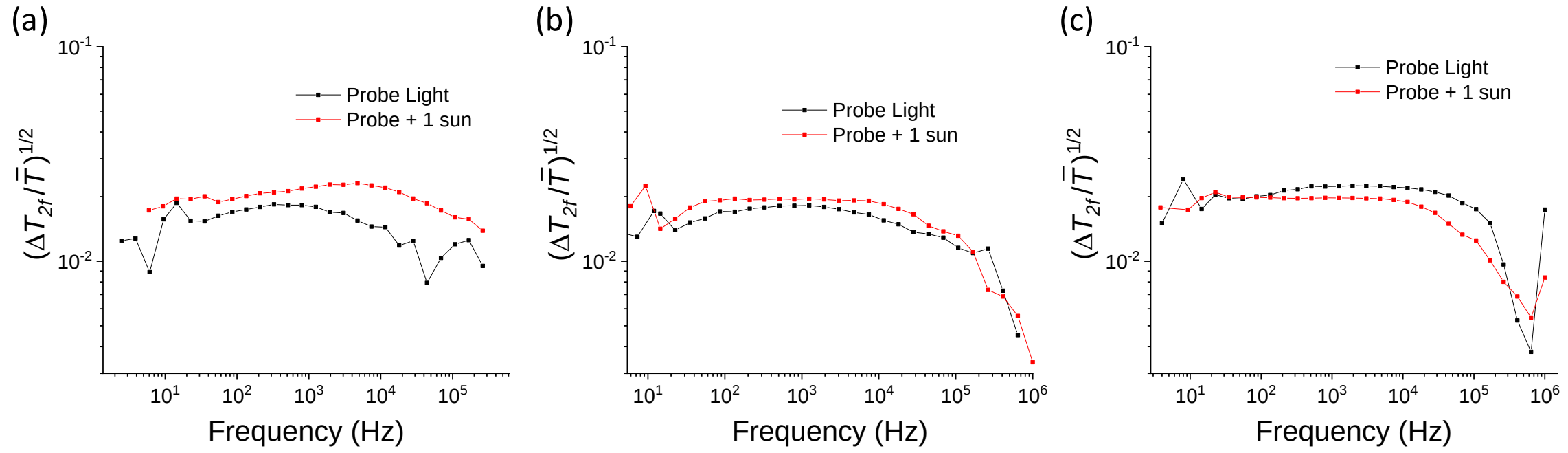

**Figure S15. Thickness dependence of the 2nd harmonic frequency dependence EA measured for $SnO_2$/MAPI/SpiroOMeTAD solar cells**. Devices with different MAPI thickness were considered: (a) 360 nm, (b) 150 nm and (c) 100 nm.

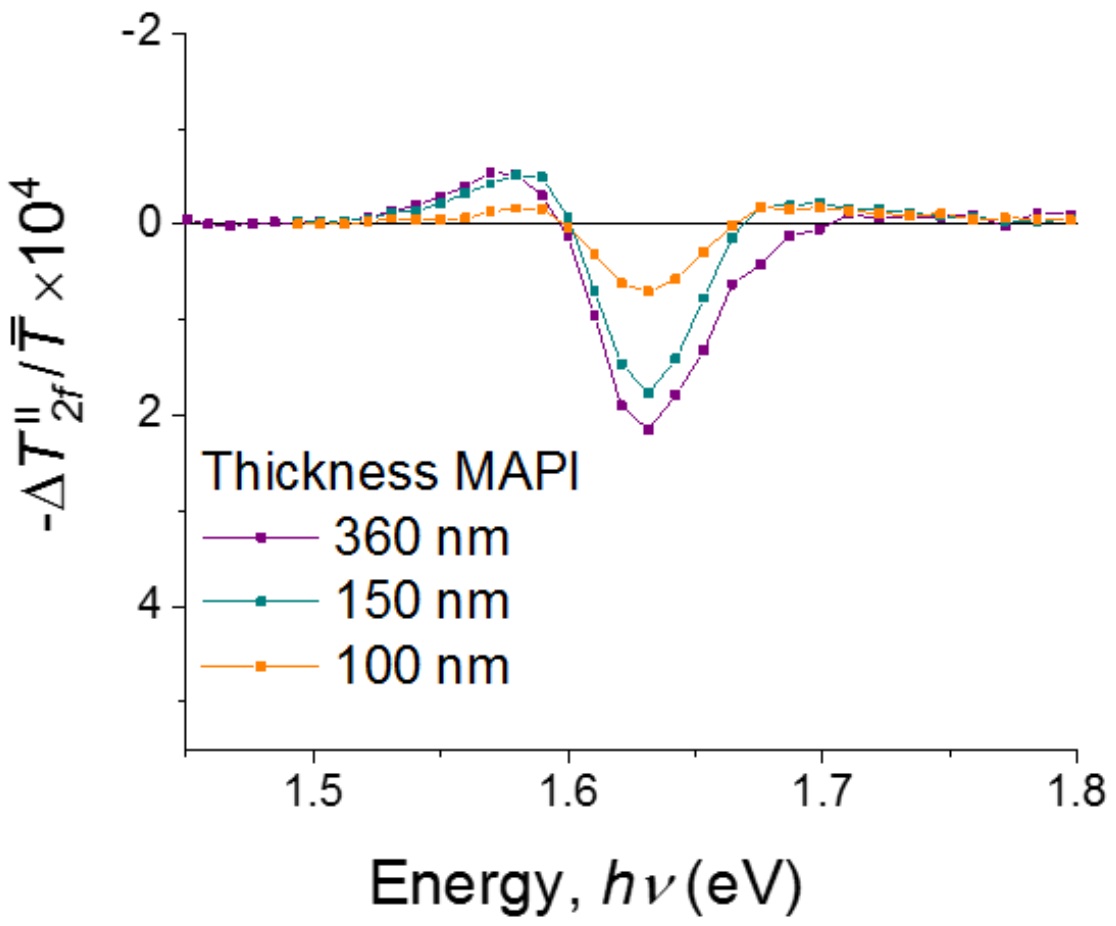

**Figure S16. 2nd harmonic EA spectra for $SnO_2$/MAPI/Spiro OMeTAD solar cells with different active layer thicknesses**.

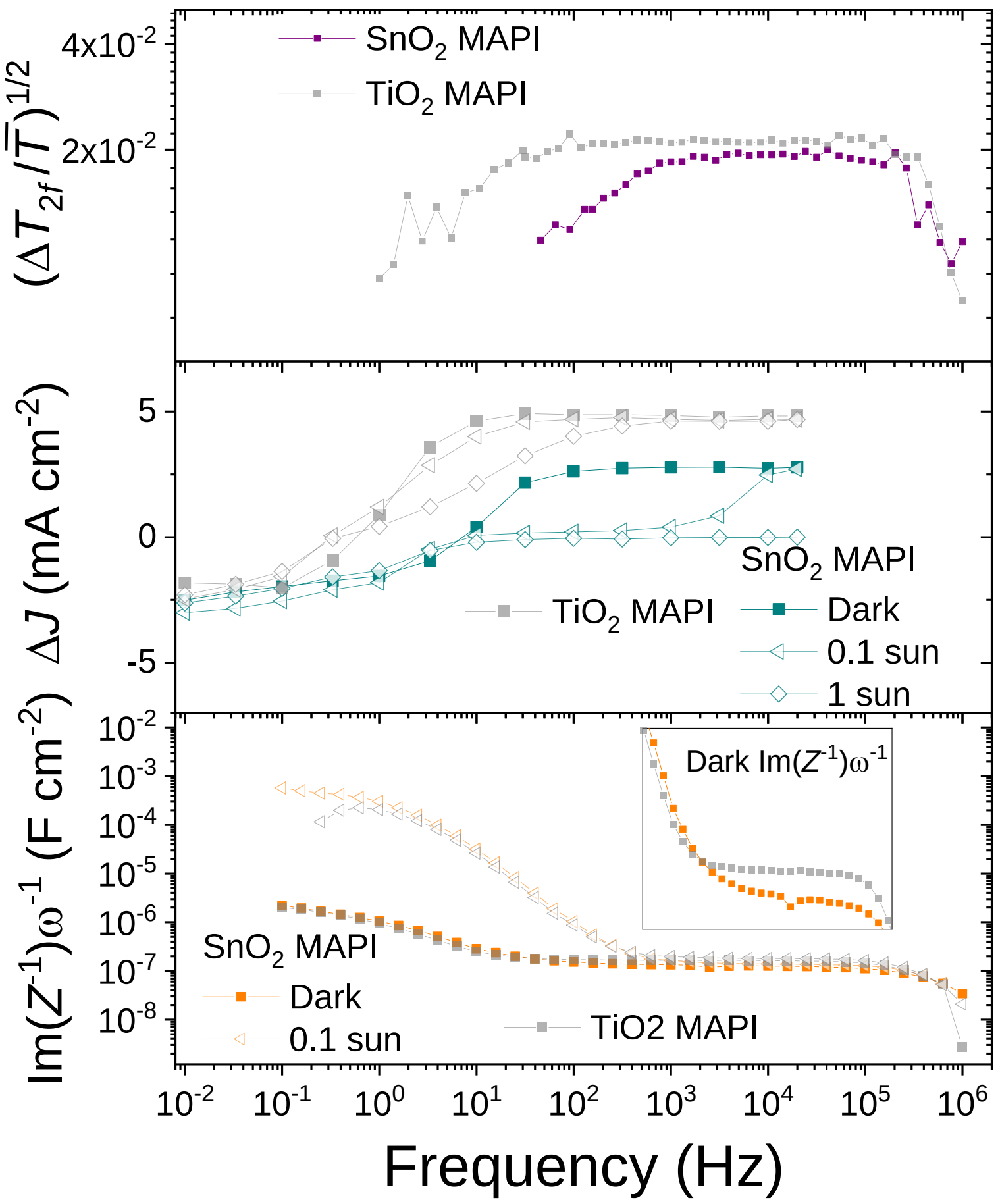

**Figure S17. Comparing (top) frequency dependent EA (middle) Step-Dwell-Probe and (bottom) capacitance measurements for $SnO_2$/MAPI/Spiro OMeTAD and $TiO_2$/MAPI/Spiro OMeTAD solar cells.**

***Light turn-on/off transient electroabsorption measurements***

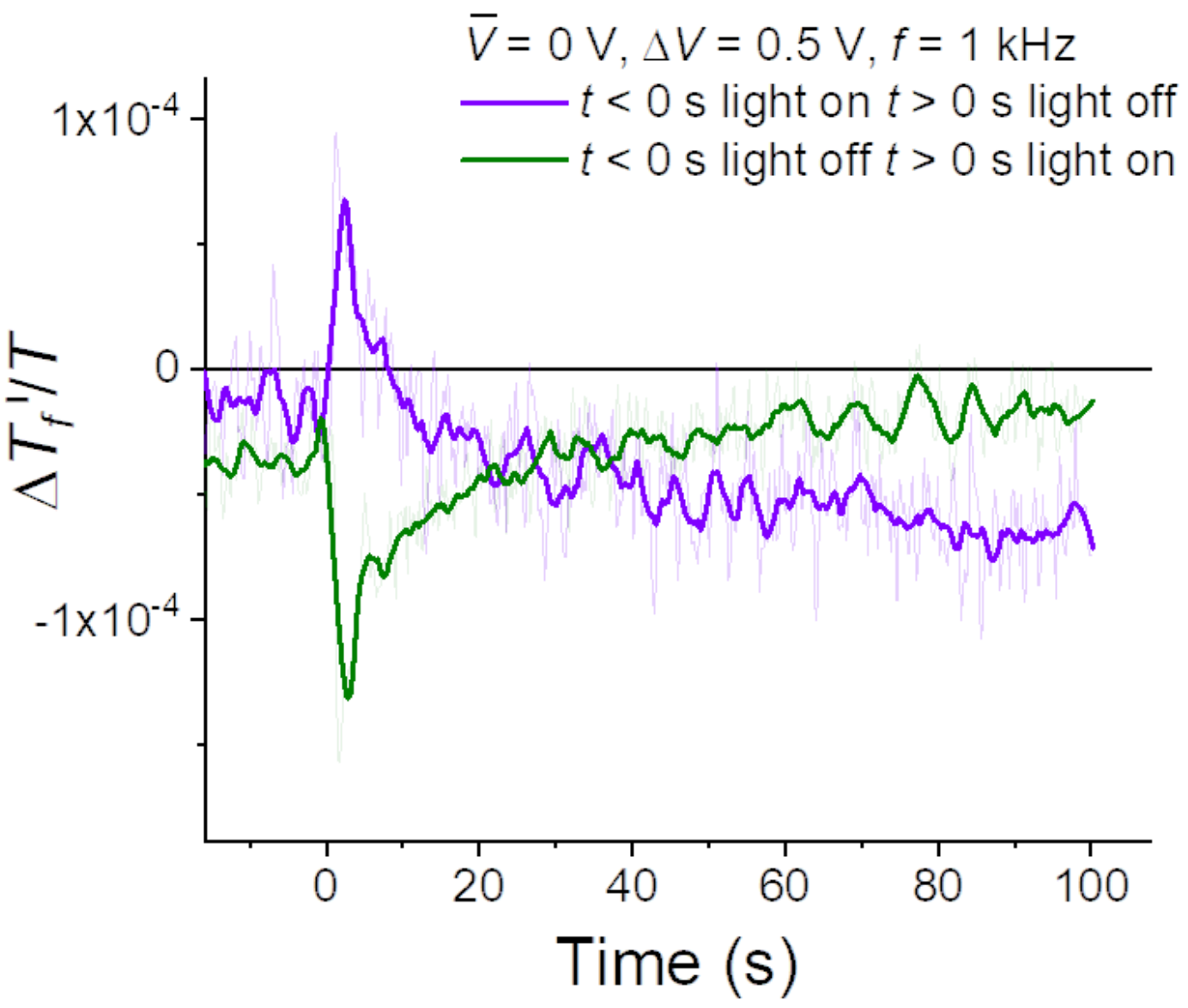

**Figure S18. Light induced transient electroabsorption.** Time resolved transient EA on a $TiO_2$/MAPI/Spiro OMeTAD cell where the bias light was switched either on or off. The applied DC voltage was kept at 0 V and the oscillating voltage had an amplitude of $\Delta V$=0.5 V and a frequency of 1 kHz. Smoothed data (averaging window of 20 data points) are overlaid on the raw data.

**DC Voltage dependent EA measurement**

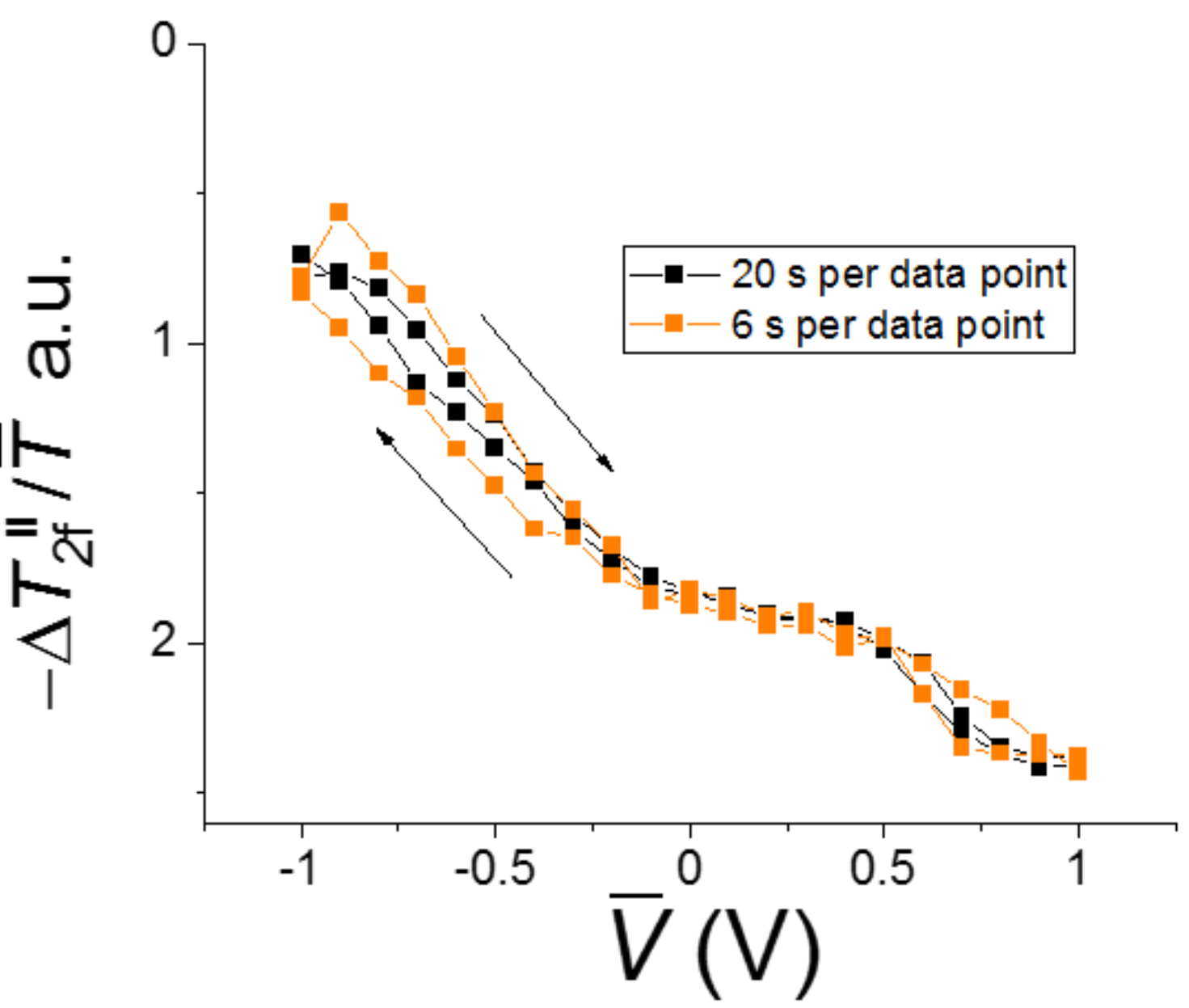

**Figure S19. $\overline{V}$ dependence of the 2nd harmonic electroabsorption signal of a $TiO_2$/MAPI/Spiro OMeTAD device**. The lock-in time constant was set at 0.3 s for the orange data set and at 1 s for the black data set. In both cases the perturbation frequency was 1 kHz and with a voltage amplitude of 0.5 V. At each voltage step the measurement took 6 and 20 s respectively before moving to the next data point. The different voltage scan rates are reflected in slightly different hysteresis in the data. This can be correlated to hysteresis in the electric field.

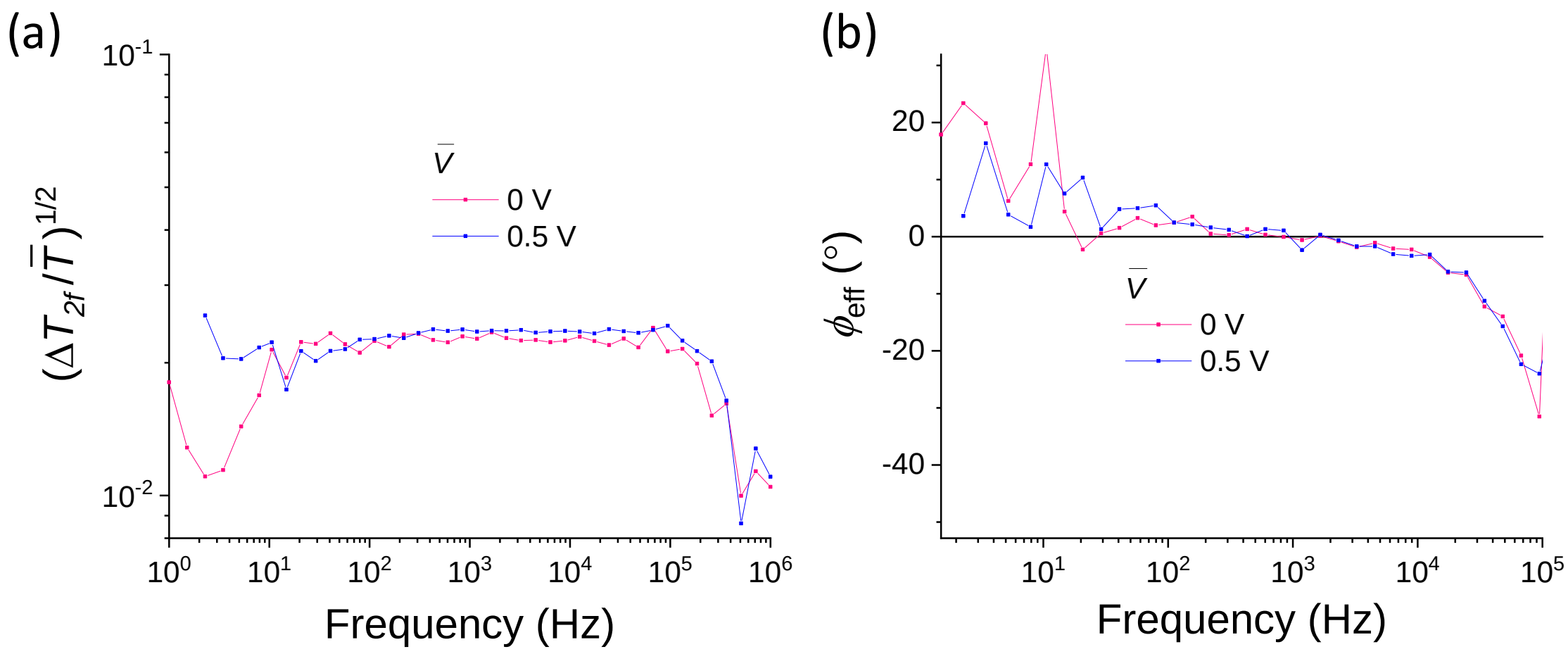

**Figure S20. DC voltage dependence of 2nd harmonic frequency dependent EA**. Frequency dependent (a) magnitude and (b) phase of the 2nd harmonic electroabsorption for a $TiO_2$/MAPI/Spiro OMeTAD solar cell using different values of the DC voltage.

## *Electroabsorption measurements on bromide and mixed iodide/bromide solar cells*

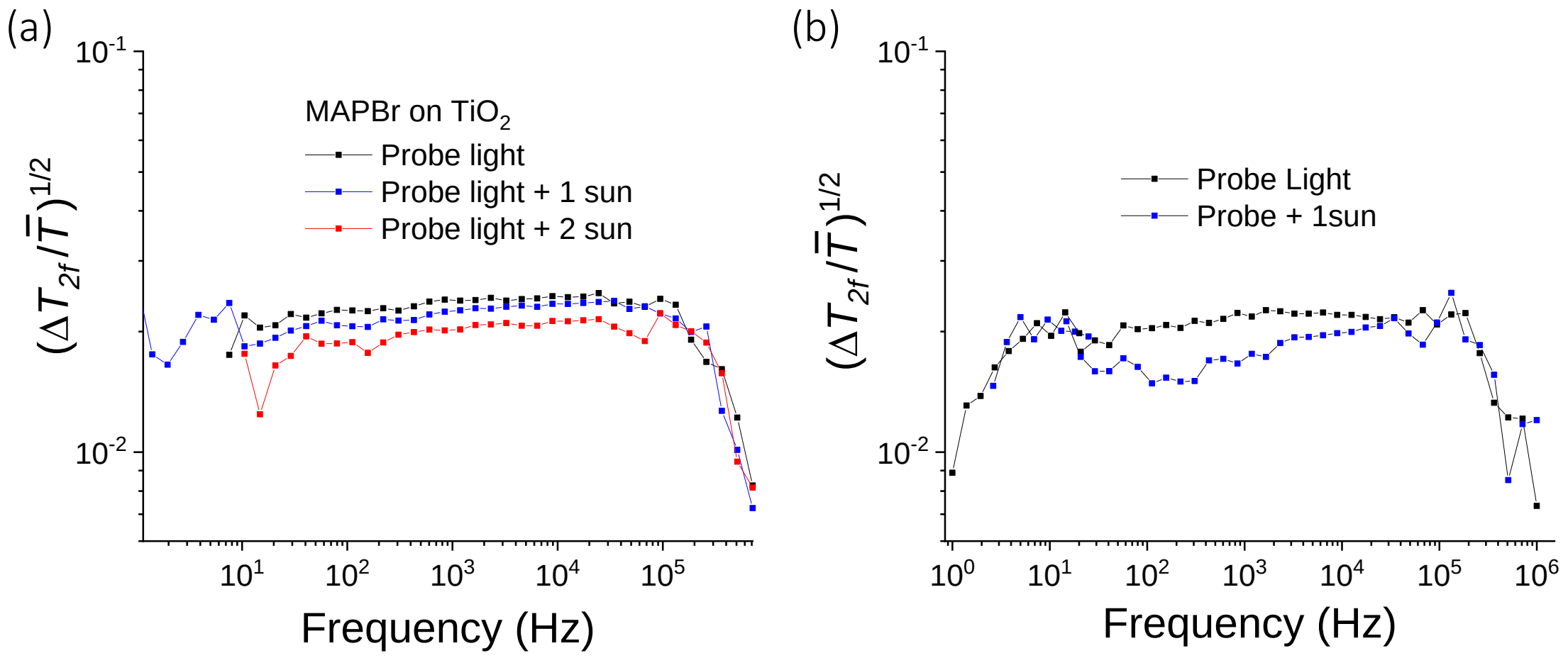

**Figure S21. Frequency dependence 2nd harmonic electroabsorption** on MAPBr3 and $MAPI_{0.8}Br_{0.2}$ devices. Different bias light intensity conditions were considered.(a) $TiO_2$/MAPBr/Spiro OMeTAD; (b) $TiO_2$/$MAPI_{0.8}Br_{0.2}$/Spiro OMeTAD.

***Electroabsorption measurements on $TiO_2$/MAPI/Spiro OMeTAD solar cells: active layer thickness, reproducibility and HTM doping***

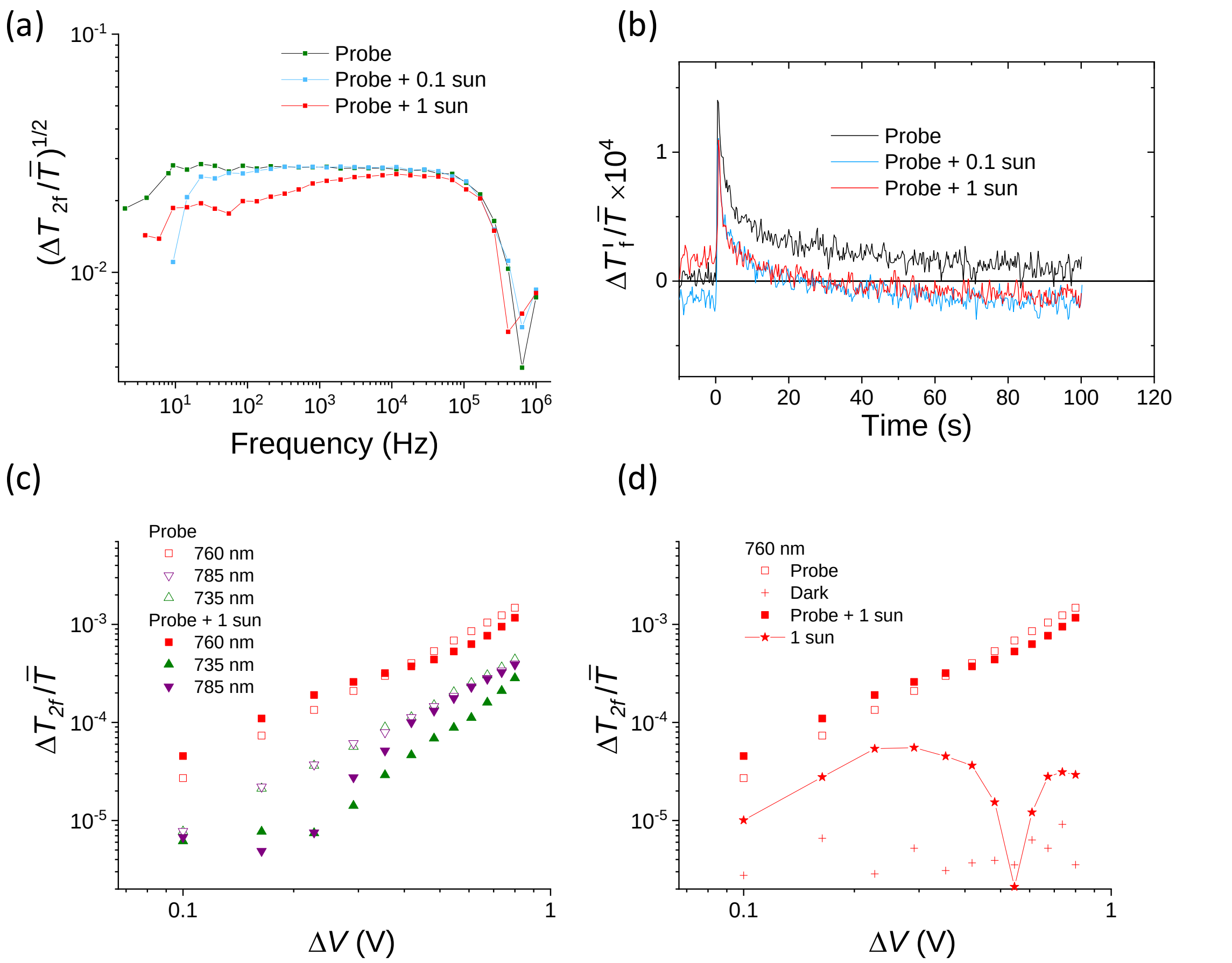

**Figure S22. Electroabsorption measurement on a thin MAPI solar cell.** (a) Frequency dependence 2nd harmonic electroabsorption (b) Time resolved transient 1st harmonic Electroabsorption and (c)(d) applied voltage amplitude dependence electroabsorption for a $TiO_2$/MAPI/Spiro OMeTAD (MAPI thickness 106 nm) solar cell at different bias light intensity conditions. Note that in (d), some signal is detected for the case of 1 sun illumination without probe light, suggesting the possibility of voltage modulated 2nd harmonic photoluminescence.

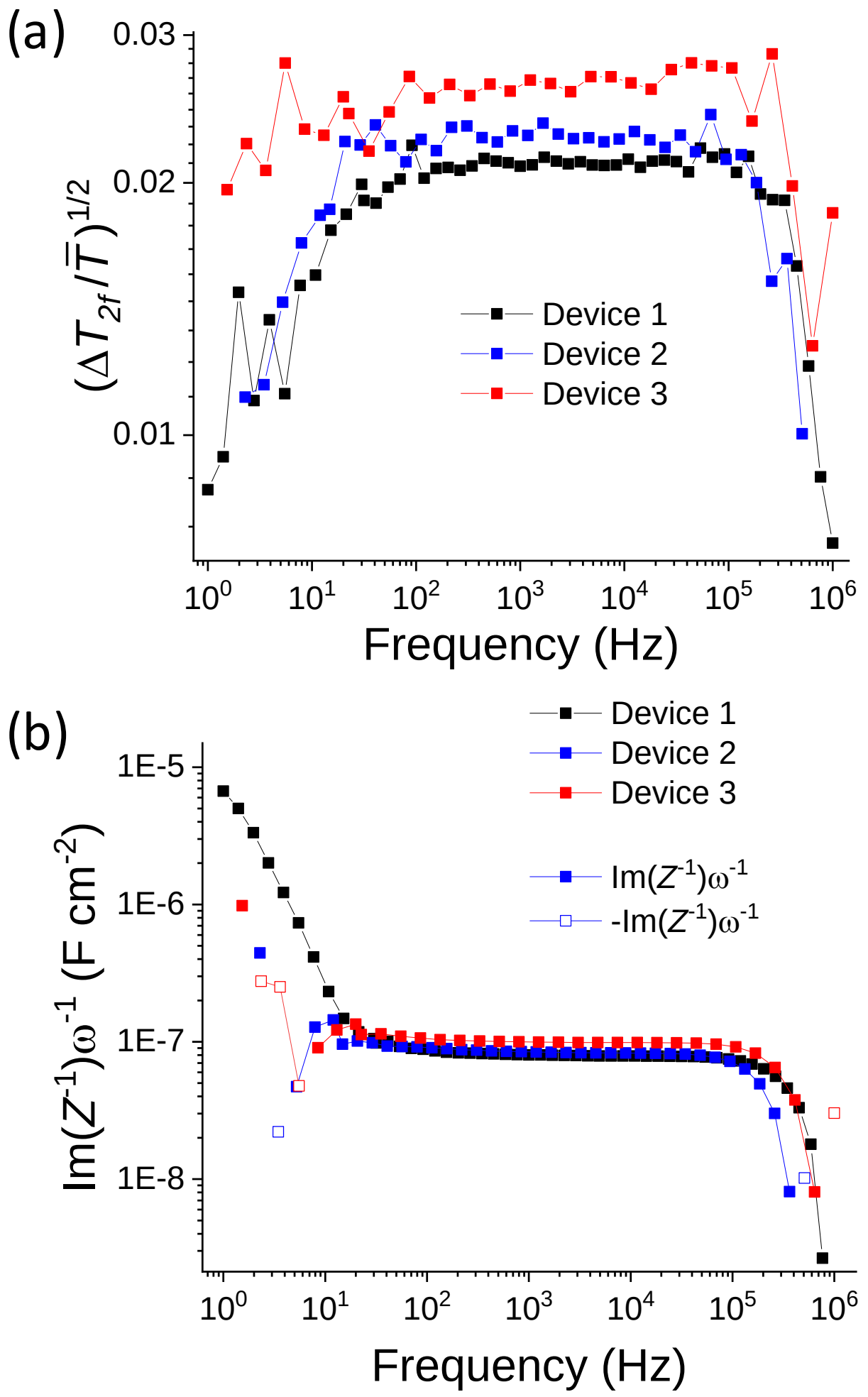

**Figure S23. Simultaneous frequency dependent EA and impedance measurements**. Data are shown for three different $TiO_2$/MAPI/Spiro OMeTAD solar cells. (a) 2nd harmonic EA signal and (b) apparent capacitance of the devices, to which a voltage with $\Delta V = 0.8\ V$ was applied. In (b) we show both the positive and the negative values of negative capacitance [6]. We observe that for Device 3 (and to a lower extent for Device 2), a significant inductive (negative capacitance) behavior is observed between 1 and 10 Hz. This also correlates with the larger EA signal and milder screening observed at low frequencies for this device compared to Device 1 and 2.

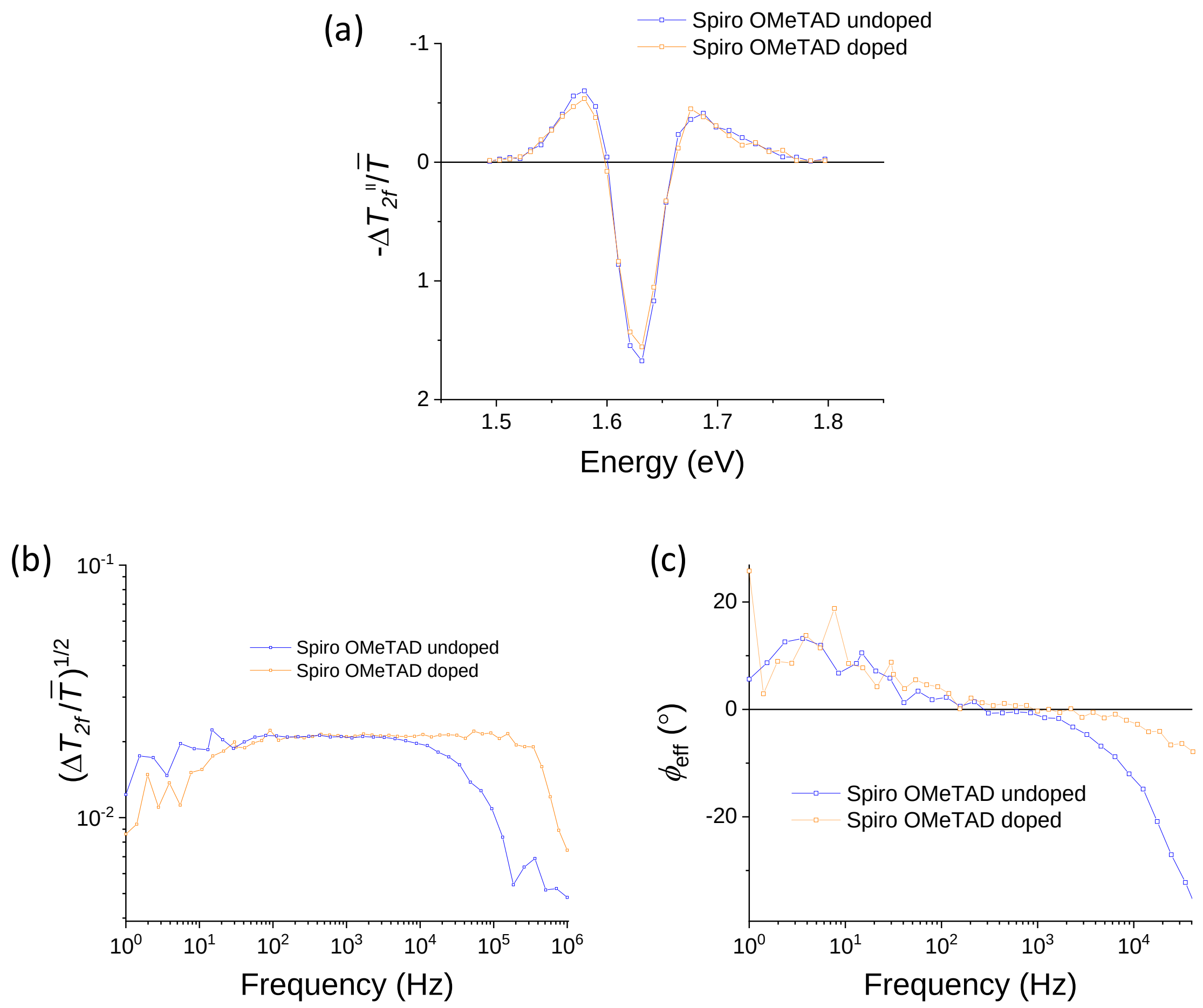

**Figure S24. Effect of additives (LiTFSI and tBP) in the HTM on the EA of MAPI solar cells.** (a) Spectrum and (b)(c) frequency dependence of the 2nd harmonic electroabsorption for $TiO_2$/MAPI/Spiro OMeTAD solar cell for the cases of Spiro OMeTAD processed with (doped) or without (undoped) additives (LiTFSI and tBP).

*Electroabsorption for mixed cation perovskite solar cells*

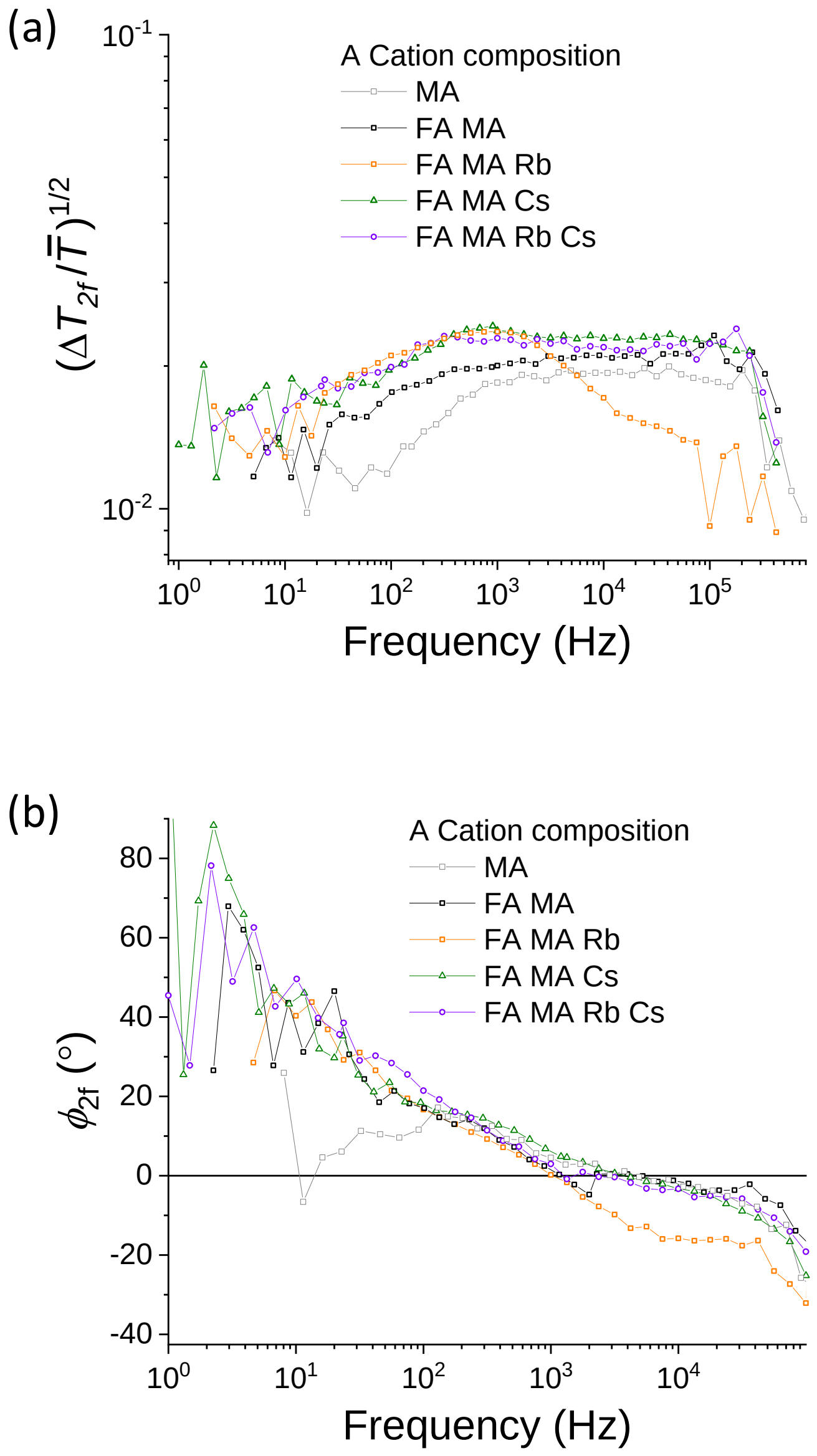

**Figure S25. Frequency dependent EA measurements on solar cells with different A-cation compositions**. Frequency dependent electroabsorption of $SnO_2$ / $APb(I_{0.83}Br_{0.17})_3$ / Spiro OMeTAD solar cells, where different A cation compositions are tested. For the detail regarding the composition, see section S1 of this document.

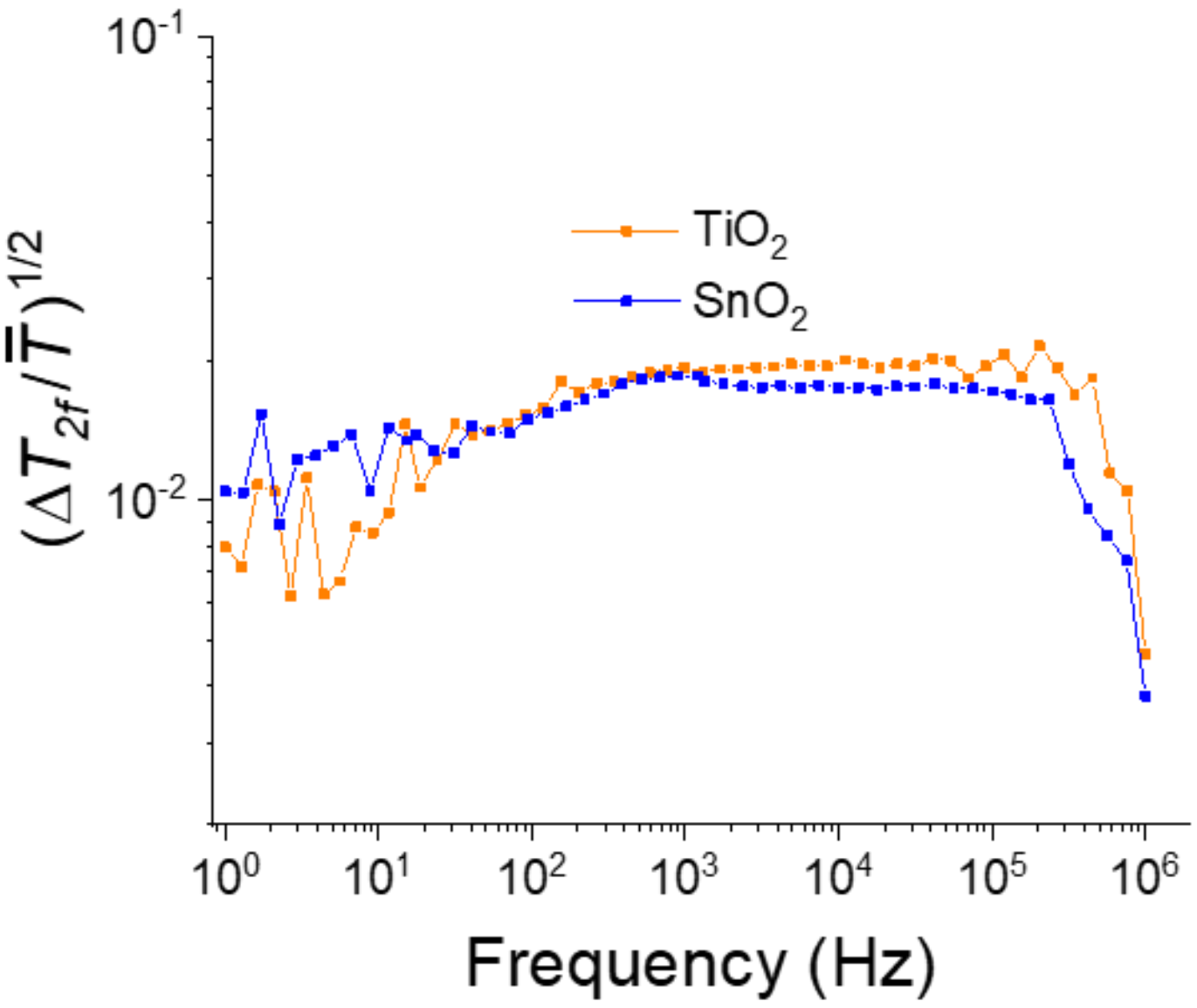

**Figure S26. ETL dependence of the EA measurements on triple cation solar cells**. Frequency dependence 2nd harmonic electroabsorption of $Cs_{0.05}[(FA_{0.83}MA_{0.17})]_{0.95}Pb(I_{0.83}Br_{0.17})_3$ / Spiro OMeTAD solar cells with either $TiO_2$ or $SnO_2$ electron transporting materials.

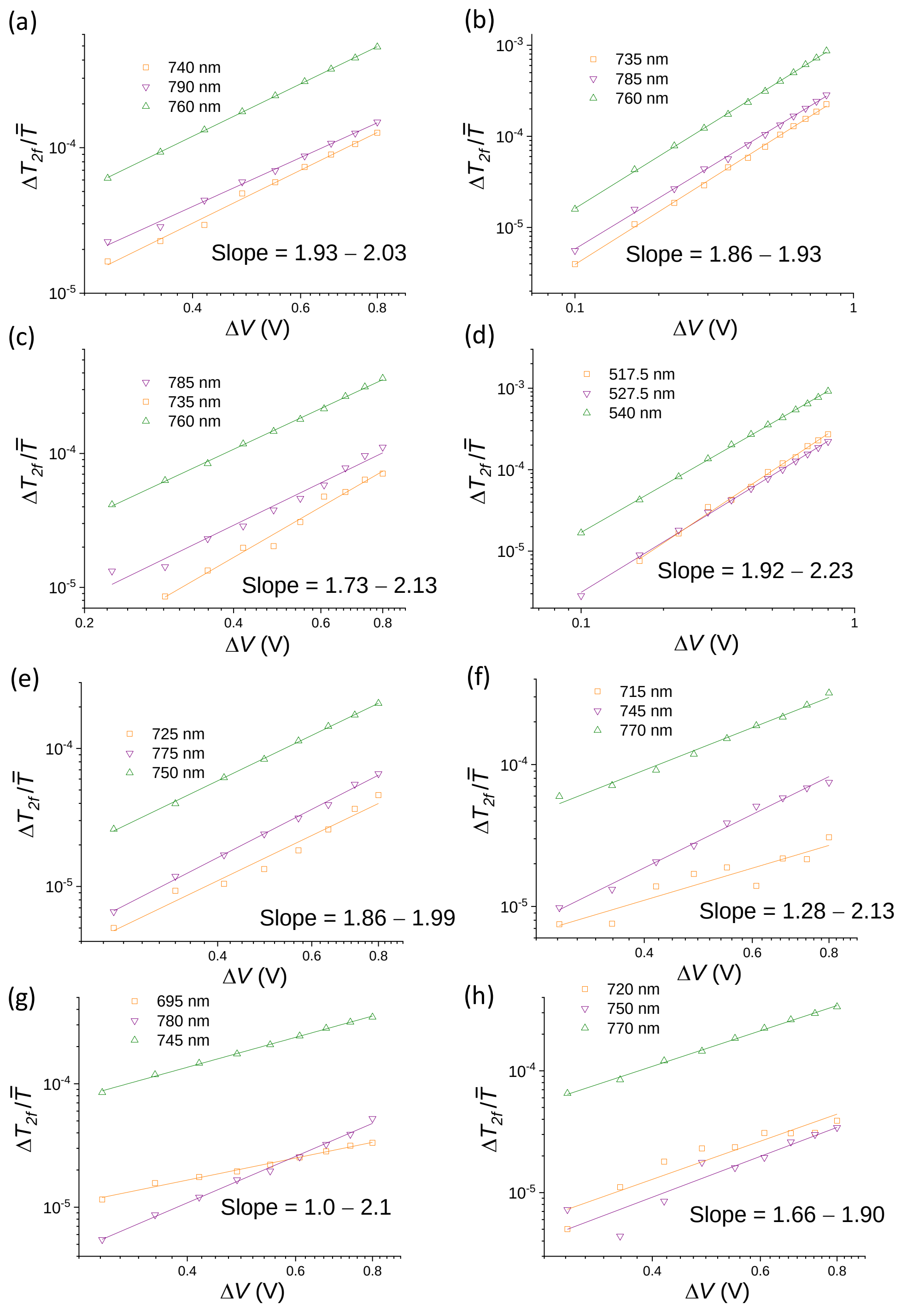

**Figure S27. AC voltage dependence of the 2nd harmonic EA signal.** The signal at the peak wavelengths were monitored as a function of the applied voltage amplitude for different hybrid perovskite solar cell architectures (the lines correspond to fits to the data in log-log scale):

(a) $TiO_2$/MAPI (360 nm)/Spiro OMeTAD, (b) $TiO_2$/MAPI (100 nm)/Spiro OMeTAD,

(c) $SnO_2$/MAPI (360 nm)/Spiro OMeTAD, (d) $TiO_2$/$MAPBr_3$/Spiro OMeTAD,

(e) $SnO_2$/ $(FA_{0.83}MA_{0.17})Pb(I_{0.83}Br_{0.17})_3$/Spiro OMeTAD,

(f) $SnO_2$/ $Cs_{0.05}(FA_{0.83}MA_{0.17})_{0.95}Pb(I_{0.83}Br_{0.17})_3$/Spiro OMeTAD,

(g) $SnO_2$/ $Rb_{0.05}(FA_{0.83}MA_{0.17})_{0.95}Pb(I_{0.83}Br_{0.17})_3$/Spiro OMeTAD and

(h) $SnO_2$/ $Rb_{0.05}Cs_{0.05}[(FA_{0.83}MA_{0.17})]_{0.9}Pb(I_{0.83}Br_{0.17})_3$/Spiro OMeTAD.

# 8. Impedance characterization

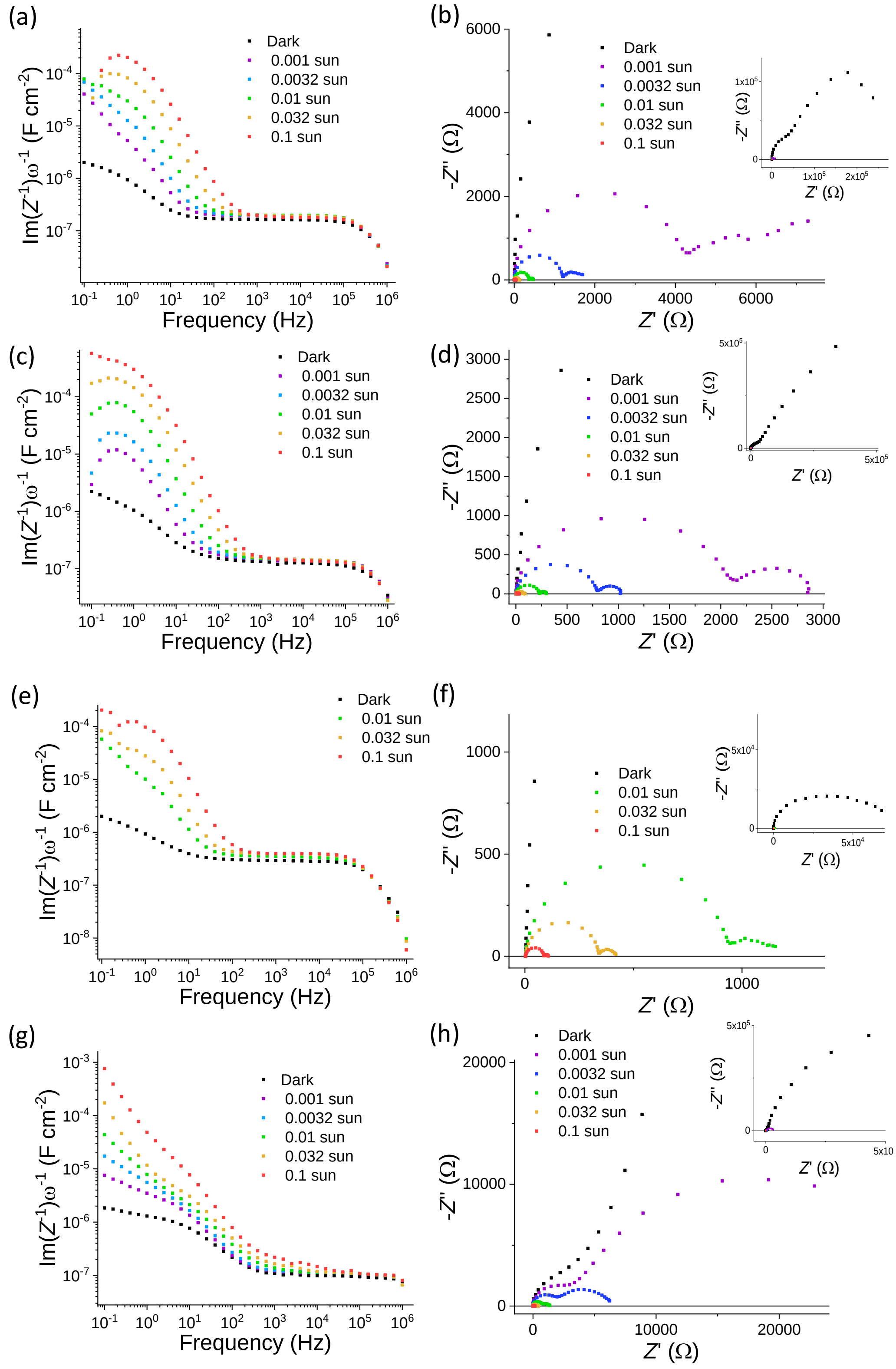

**Figure S28. Impedance measurements performed on the solar cell architectures investigated in this study**. (a) (b) $TiO_2$/MAPI (360 nm)/Spiro OMeTAD; (c) (d) $SnO_2$/MAPI (360 nm)/Spiro OMeTAD; (e) (f) $TiO_2$/MAPI (100 nm)/Spiro OMeTAD; (g) (h) $TiO_2$/$MAPbBr_3$/Spiro OMeTAD.

In figure 6d in the main text, we only show data related to $f_{0,C}$ defined as the intersection between a linear fit in log-log scale of the main low frequency capacitance feature and the geometric capacitance plateau observed at higher frequencies. We note that additional increase in capacitance is superimposed to this plateau at high light intensities for some of the devices. This is consistent with an increasing contribution of electronic chemical capacitance at high light intensity. [6]

The low frequency features of the impedance spectra shown in Figure S28 highlight complex long time scale behavior of the devices (including inductive behavior or multiple apparent capacitance features) which may also be related to incomplete stabilization of the devices at the applied bias conditions. Despite for each solar cell the open circuit voltage was monitored for 100 seconds and its final value was then applied for further 100 s prior to the impedance analysis, this might not be enough to reach a stable enough state for low frequency measurements. [8] As a result, extracting useful information from the low frequency feature, such as a time constant for electric field screening, would not give reliable results. In the main text we focus on the position of $f_{0,C}$ which, despite not being expected to scale with the low frequency time-constant, allows to track the changes in capacitance spectra for different bias conditions, as shown in Figure 5. Here we also show the $V_{OC}$ evolution measured before impedance at 0.1 sun (Figure S29) Interestingly, the time constant at long time scales (10–100 s) is quite similar for MAPI devices with different thickness and oxide contact, as well as for a $MAPbBr_3$ device. The short time scale (0.1–10 s) points towards a similar trend as discussed in the main text as far as the role of the oxide is concerned. The thin MAPI device and the $MAPbBr_3$ device also display faster change in $V_{OC}$ than the reference sample in this time scale.

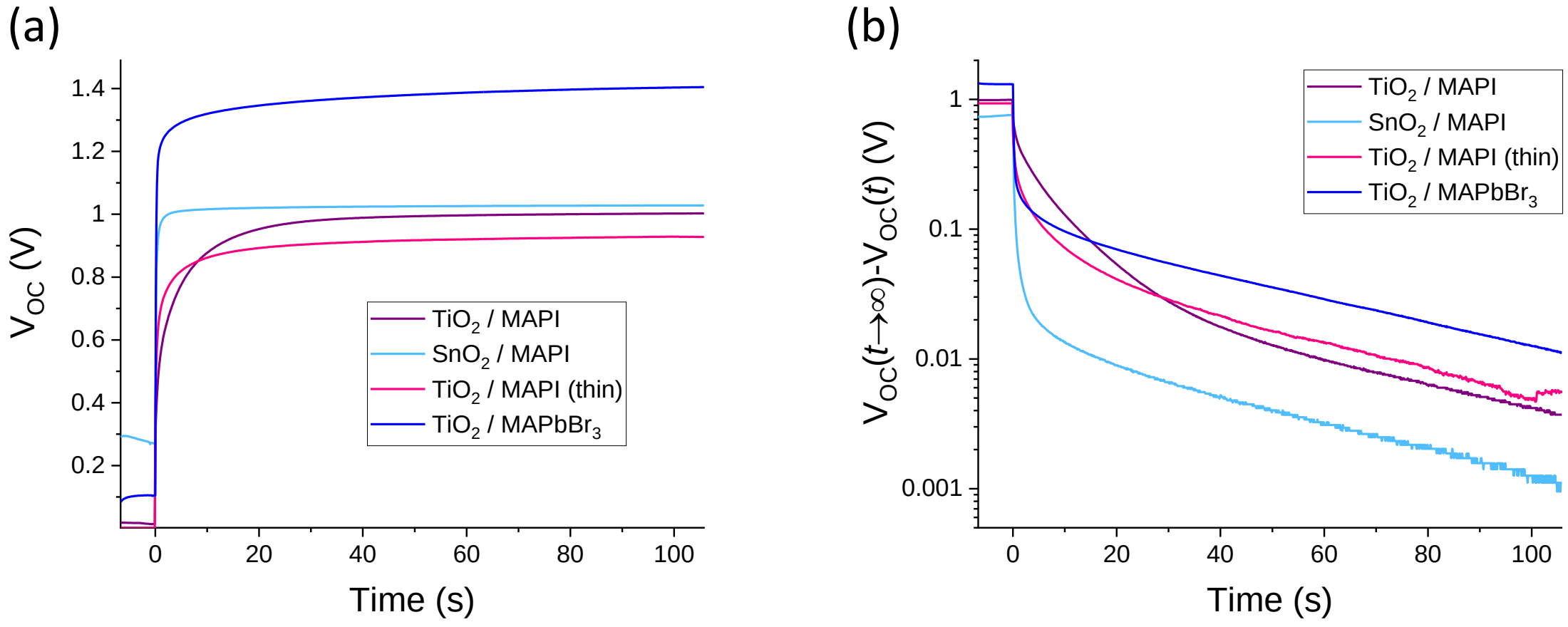

**Figure S29. $V_{oc}$ transients.** Open circuit voltage measurement performed on different cells (see Figure S28) before impedance measurements run at 0.1 sun equivalent illumination. Light is switched on at Time = 0 s and the voltage of the cell is monitored as a function of time (see methods section in the main text).

We performed fitting of impedance measurements on the reference device, also as a consistency check with the simulations discussed in section 3 of this document. In Figure S30 we show the data corresponding to the $TiO_2$/MAPI/Spiro solar cells measured at different light intensity, as well as the fitting to the data using the equivalent circuit model displayed in Figure S30c, where 6 parameters were varied manually in the model to reproduce the experimental data. The circuit shows a slight modification to the version presented in ref. [6] regarding the treatment of the geometric capacitance, which is explicitly shown here. The circuit is an approximation, which we expect to be valid as long as that the concentration of mobile ionic defects in the devices is much larger than the concentration of electronic carriers.

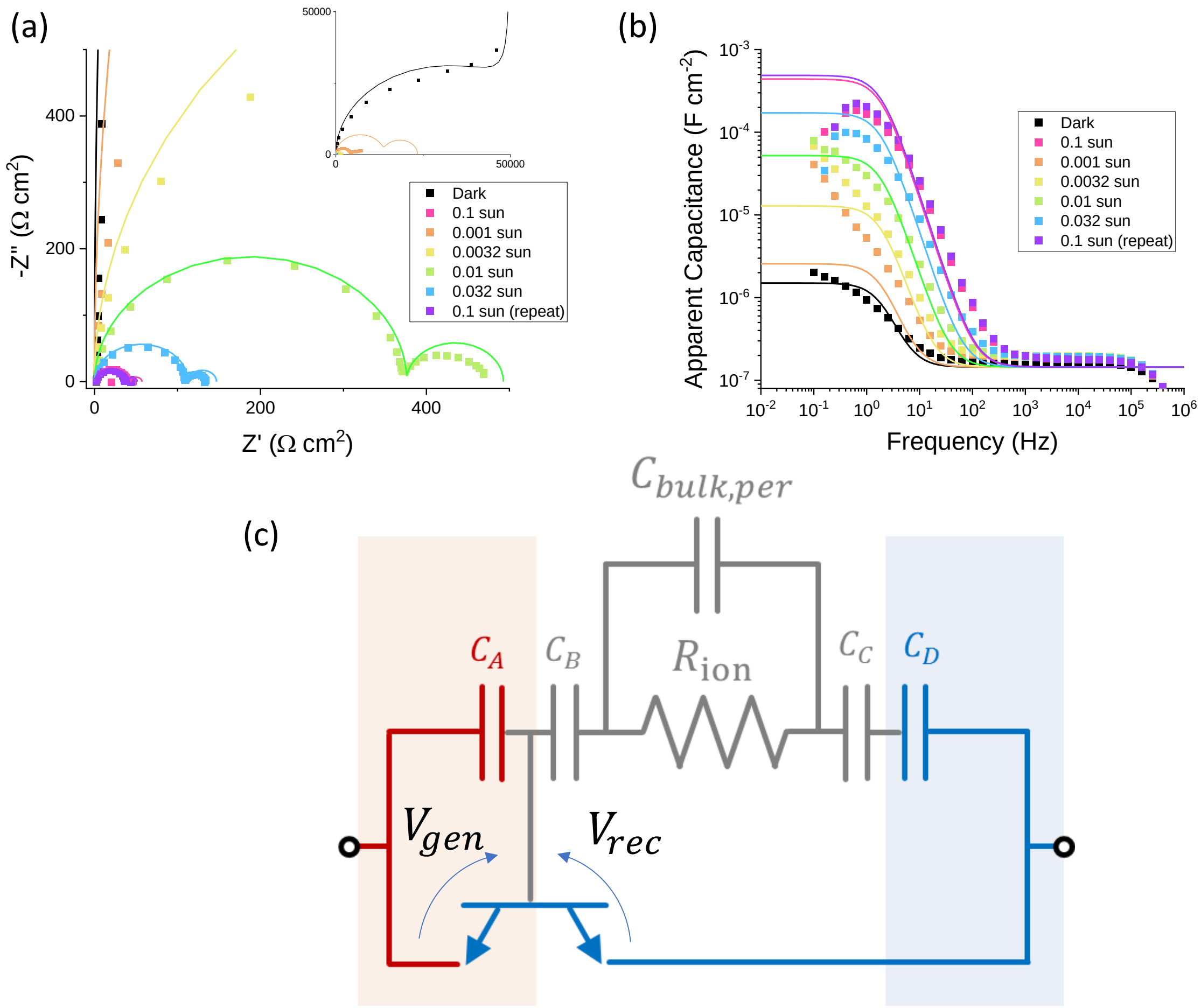

**Figure S30. Fitting of impedance spectra for a $TiO_2$/MAPI/Spiro OMeTAD solar cell**. (a) and (b) show the Nyquist plot and capacitance spectra for measurements at different light intensity. (c) shows the equivalent circuit model used for the fits (solid lines in a and b).

**Table S2. Fitting parameters values associated with the data in Figure S30.**

| Parameter (units) | Fitted value |
|---|---|
| $R_{ion}$ $(\Omega\ cm^2)$ | 5 x $10^4$ |
| $C_{con}(F\ cm^{-2})$ | 6 x $10^{-6}$ |
| $C_{per}(F\ cm^{-2})$ | 6 x $10^{-6}$ |
| $j_{0,rec}(A\ cm^{-2})$ | 9 x $10^{-13}$ |
| $C_{bulk,per}(F\ cm^{-2})$ | 1.6 x $10^{-7}$ |
| $m$ | 1.4 |

For the fit, we assumed the same values for the contact capacitances and of the capacitances associated with the perovskite space charge regions ($C_A = C_D = C_{con}$ and $C_B = C_C = C_{per}$)

The parameters $j_{0,rec}$ and $m$ are used to describe the current flowing through the recombination transistor at interface 1 as follows:

$$J_1 = J_{\text{rec}} - J_{\text{gen}} = J_{0,\text{rec}} e^{\frac{qV_{\text{rec}}}{mk_B T}} - J_{0,\text{rec}} e^{\frac{qV_{\text{gen}}}{mk_B T}} \qquad \text{Eq. S10}$$

The voltages $V_{rec}$ and $V_{gen}$ represent the potentials driving the recombination and generation processes at interface 1. [6]

We note that the value of $R_{ion}$ used for the drift-diffusion simulations was in the order of 2 $10^5$ $\Omega\, cm^2$. This higher resistance was compensated by the lower value of interfacial capacitance used for such simulations (very high capacitance for the contact layers requires high doping levels which complicate solving the numerical problem) to yield comparable time constant to the experiments and the fits shown above.

**Analytical expressions for *f₀***

Based on the circuit model in Figure S30c, it is possible to extract an analytical expression for the values of the $f_0$ parameters used in this study:

- $f_{0,EA}$: this can be associated to the pole in the expression of the transfer function relating the potential dropping on the series $C_B, R_{ion}, C_C$ and the applied potential.
- $f_{0,C}$: this is extracted from the intersection of the capacitance at high frequencies ($C_{bulk,per}$) and the low frequency capacitance behavior.

For simplicity, we consider the case of $C_A = C_D = C_{con}$ and $C_B = C_C = C_{per}$. The transfer function $G_{EA}(\omega)$ relating the change in potential dropping across the perovskite layer $\tilde{V}_{per}$ and the applied voltage $\tilde{V}$ can be expressed as follows:

$$G_{EA}(\omega) = \frac{\tilde{V}_{per}}{\tilde{V}}(\omega) = \frac{C_{con}}{C_{con}+C_{per}} \frac{1+i\omega R_{ion}\left(\frac{C_{per}}{2}+C_{bulk,per}\right)}{1+i\omega R_{ion}\left(\frac{1}{2}\frac{C_{per}C_{con}}{C_{con}+C_{per}}+C_{bulk,per}\right)} \quad \text{Eq. S11}$$

From Equation S11, we note the following:

a) $G_{EA}(\omega \to 0) \to \frac{C_{con}}{C_{con}+C_{per}}$, which is a constant real value and therefore has a phase of 0°.
b) For large values of $C_{per}$ (corresponding to scenario (ii) discussed in section 2)

$G_{EA}(\omega) \approx \frac{i\omega R_{ion}\frac{C_{con}}{2}}{1+i\omega R_{ion}\left(\frac{C_{con}}{2}+C_{bulk,per}\right)}$. In such case, $G_{EA}(\omega \to 0) \approx i\omega R_{ion}\frac{C_{con}}{2}$ (phase 90°).

We also identify $f_{0,EA}$ as:

$$f_{0,EA} = \frac{1}{2\pi}\frac{1}{R_{ion}\left(\frac{1}{2}\frac{C_{per}C_{con}}{C_{con}+C_{per}}+C_{bulk,per}\right)} \quad \text{Eq. S12}$$

For $C_{g,per} \ll C_{con}$ and $C_{g,per} \ll C_{per}$

$$f_{0,EA} = \frac{1}{2\pi}\frac{1}{R_{ion}\frac{1}{2}\frac{C_{per}C_{con}}{C_{con}+C_{per}}} = \frac{1}{2\pi R_{ion}C_{ion}^{\perp}/2} \quad \text{Eq. S13}$$

For the characteristic frequency associated with capacitance measurements, $f_{0,C}$:

$$f_{0,C} = \frac{1}{2\pi R_{ion}C_{ion}^{\perp}/2}\sqrt{\frac{C_{ion}^{\perp}/2}{C_{g,per}}} \quad \text{Eq. S14}$$

Note that the values of $C_{per}$ and $C_{con}$ are functions of the applied voltage. Here such dependence is not explicitly considered. While for small perturbation measurement (e.g. impedance) a constant value of the capacitances can be used for any given steady-state condition, the treatment for large perturbation measurements (e.g. EA) is more complex.

The expressions above refer to the case of a cell in the dark and have been extracted by considering the ionic branch of the circuit model in Figure S30. Under light, the circuit model is able to explain the behavior observed from impedance, where an increase in the low frequency apparent capacitance occurs due to ionic-to-electronic current amplification. One could express the value of $f_{0,C}$ analytically in that case too. As noted in the main text, such value is expected to increase with light intensity, despite the unchanged time constant of the ionic branch in the circuit model. Changes in $C_{per}$, $C_{con}$ and $R_{ion}$ as a function of bias could explain the changes in $f_{0,EA}$ observed experimentally when applying light to the solar cells investigated in this study.

# 9. Current voltage curves of perovskite solar cells

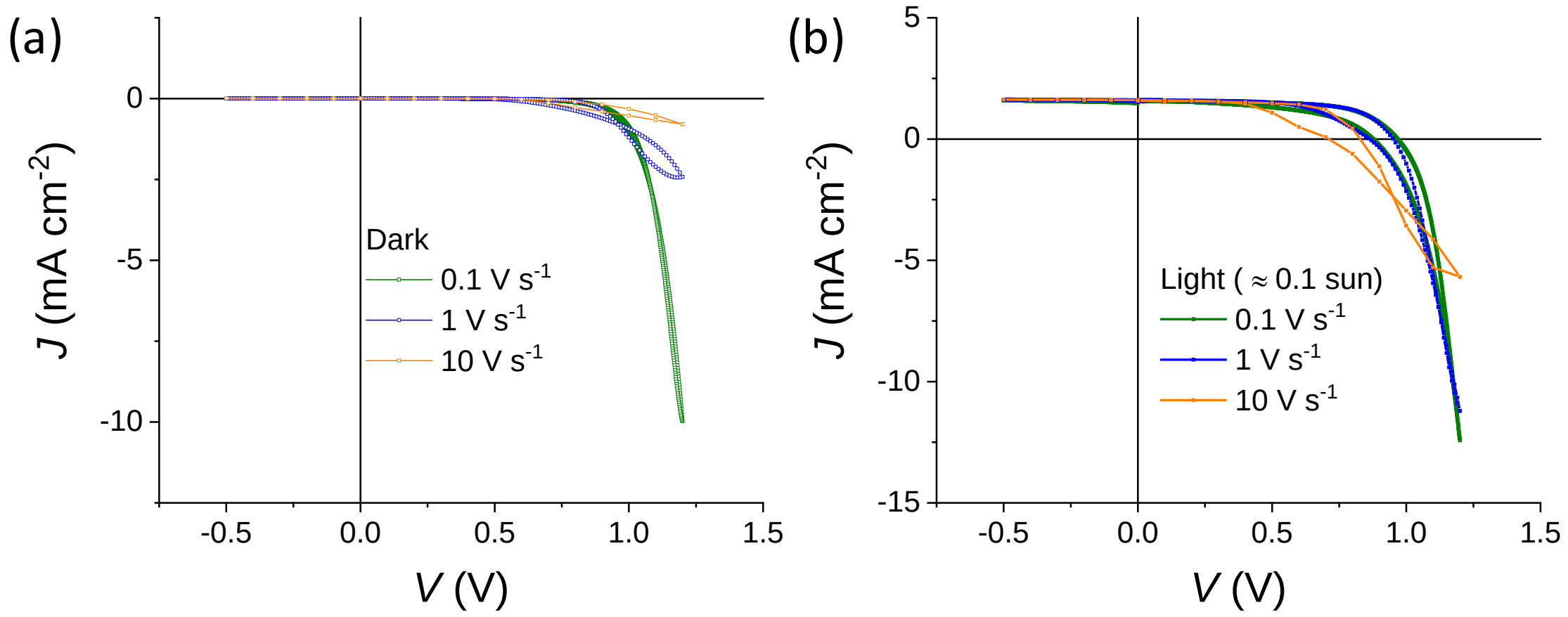

**Figure S31. Current voltage curves for a $TiO_2$/MAPI(360 nm thick)/Spiro OMeTAD solar cell performed at different scan rates under (a) dark and (b) light conditions**.

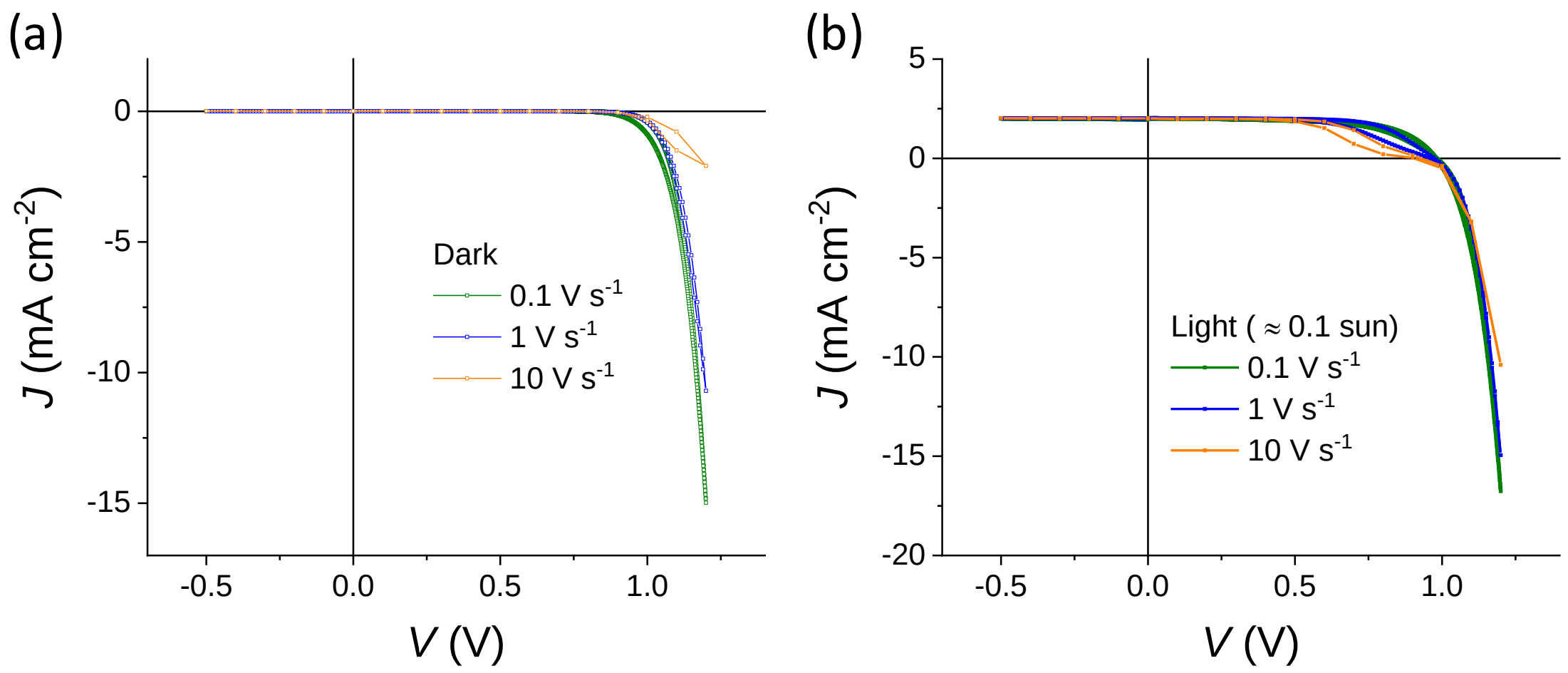

**Figure S32. Current voltage curves for a $SnO_2$/MAPI(360 nm thick)/Spiro OMeTAD solar cell performed at different scan rates under (a) dark and (b) light conditions**.

Here we comment on the potential contribution of electronic charges in the electric field screening observed experimentally and discussed in the manuscript. Electronic transport across the devices investigated in this study is expected to be fast, such that it would not be a rate limiting factor for the field screening occurring in the frequency range tested in our EA measurements (f < 200 kHz). Also, the transient behavior before reaching quasi equilibrium between generation and recombination of carriers should also be fast compared to the dynamics observed here (electronic charge carrier lifetime in similar hybrid perovskite devices is typically below 10 µs at steady state under 1 sun equivalent illumination). [9] This suggests that equilibration of charge density due to the generation/recombination rate at any given condition should be fast enough compared to the frequency range used here.

In Figure S31, we show current voltage characteristics at different scan rates of a $TiO_2$/MAPI/Spiro OMeTAD cell under dark and light. The measured forward bias current suggests recombination limitation at low voltages (current increases with increasing scan rates) and injection limitations at larger voltages (current decreases with increasing scan rates). The same experiment performed on a device with a $SnO_2$ interlayer (Figure S32) shows low hysteresis and relatively constant $V_{oc}$ values at different scan rates, consistent with a lower surface recombination than the case of $TiO_2$. [5] We can correlate the current voltage hysteresis measured for the devices under light with their surface recombination. It follows that electric field screening in the dark occurring below about 1 kHz observed in samples with $SnO_2$ where charges may have longer lifetime may have an electronic contribution. As the amount of charge injected in the dark at forward bias for each voltage cycle in the EA experiment decreases with frequency this could also explain the relatively low frequency at which such contribution occurs. In Figure S33, we show the J-V curves of the devices discussed in Figure 6.

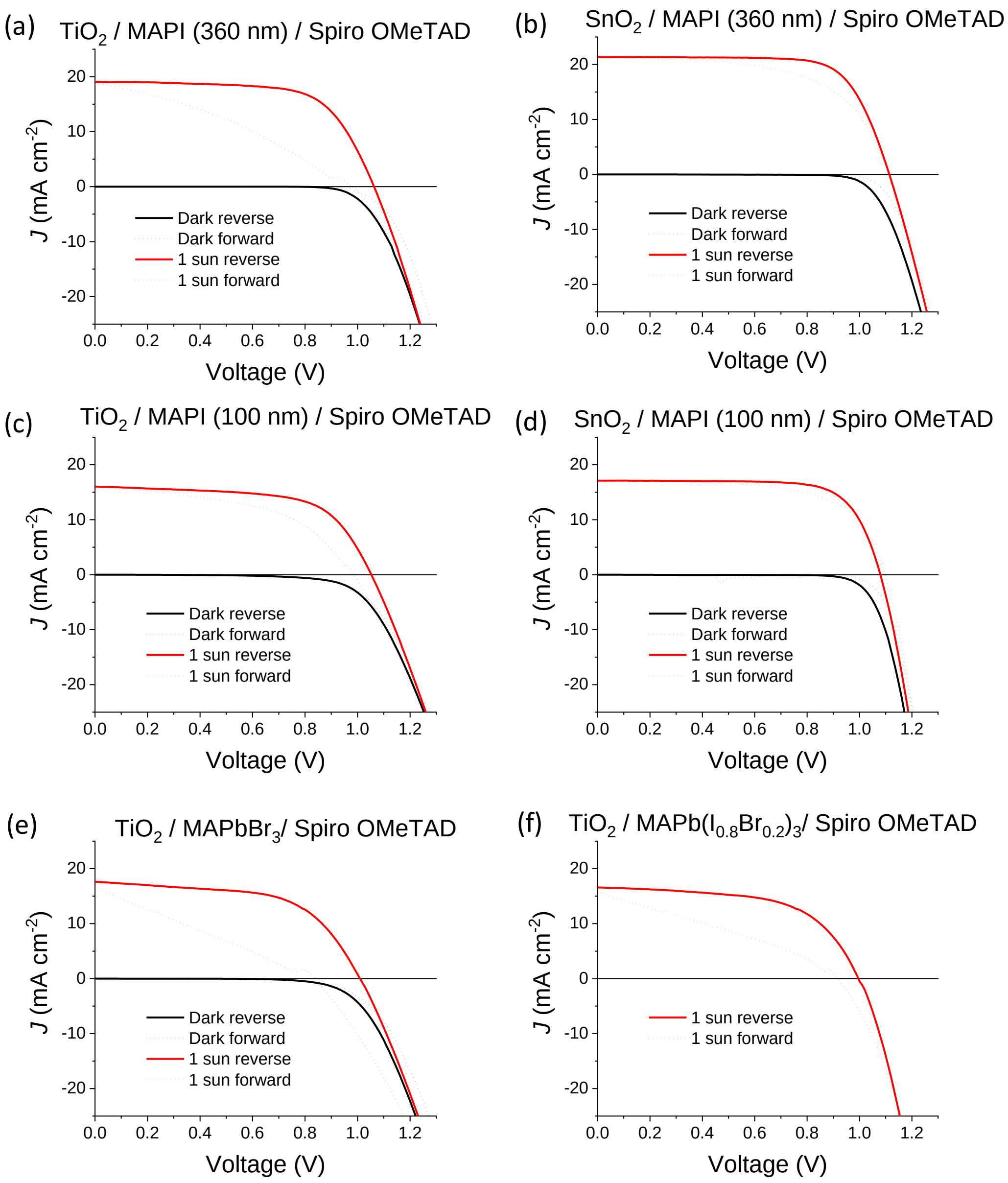

**Figure S33. Current voltage curves under dark and 1 sun equivalent illumination of the solar cell architectures investigated in this study**. The small kink in forward J-V measurements is assigned to the automatic change in the current range of the instrument used [10]

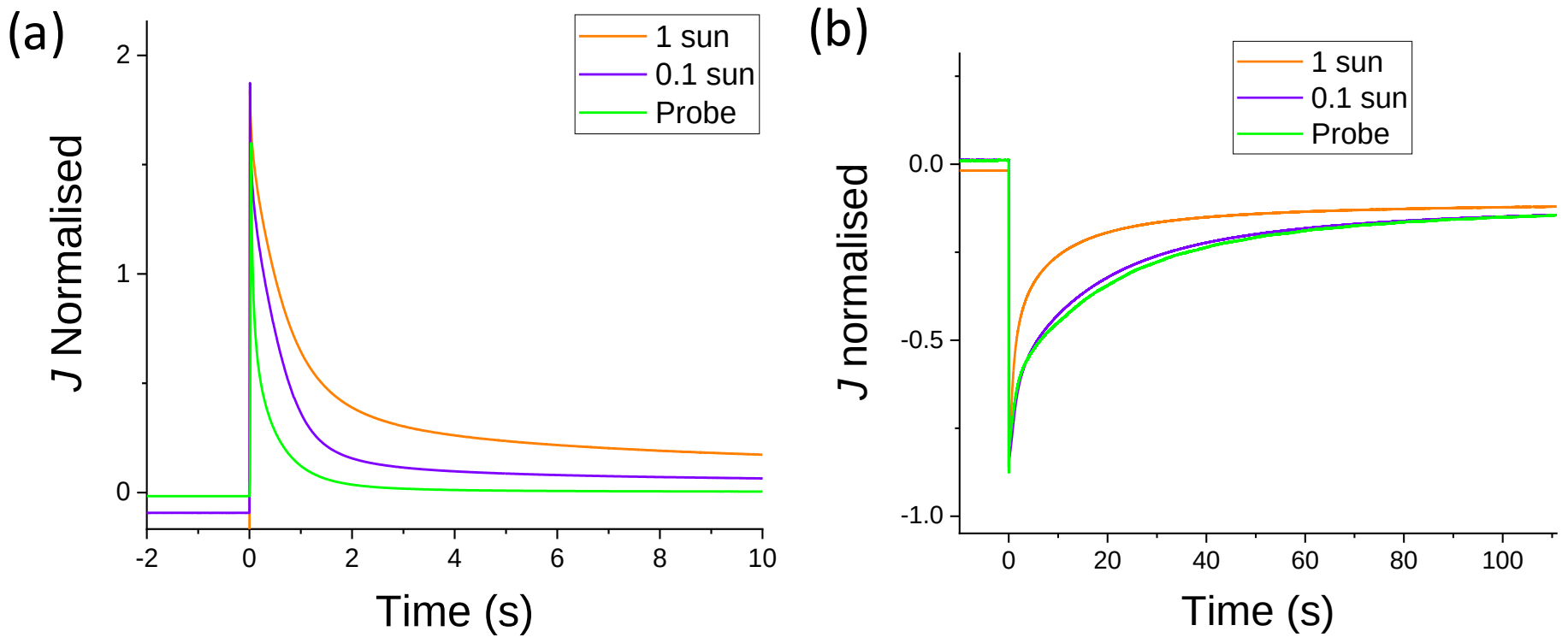

**Figure S34. Normalized transient photocurrent on a $TiO_2$/MAPI/Spiro OMeTAD solar cell**. The measurements refer to step potential measurements with (a) $\bar{V}$ = 0 -> 0.5 V and (b) $\bar{V}$ = 0.5 -> 0 V measured at different bias light conditions.

## 10. Impedance study of contact layers

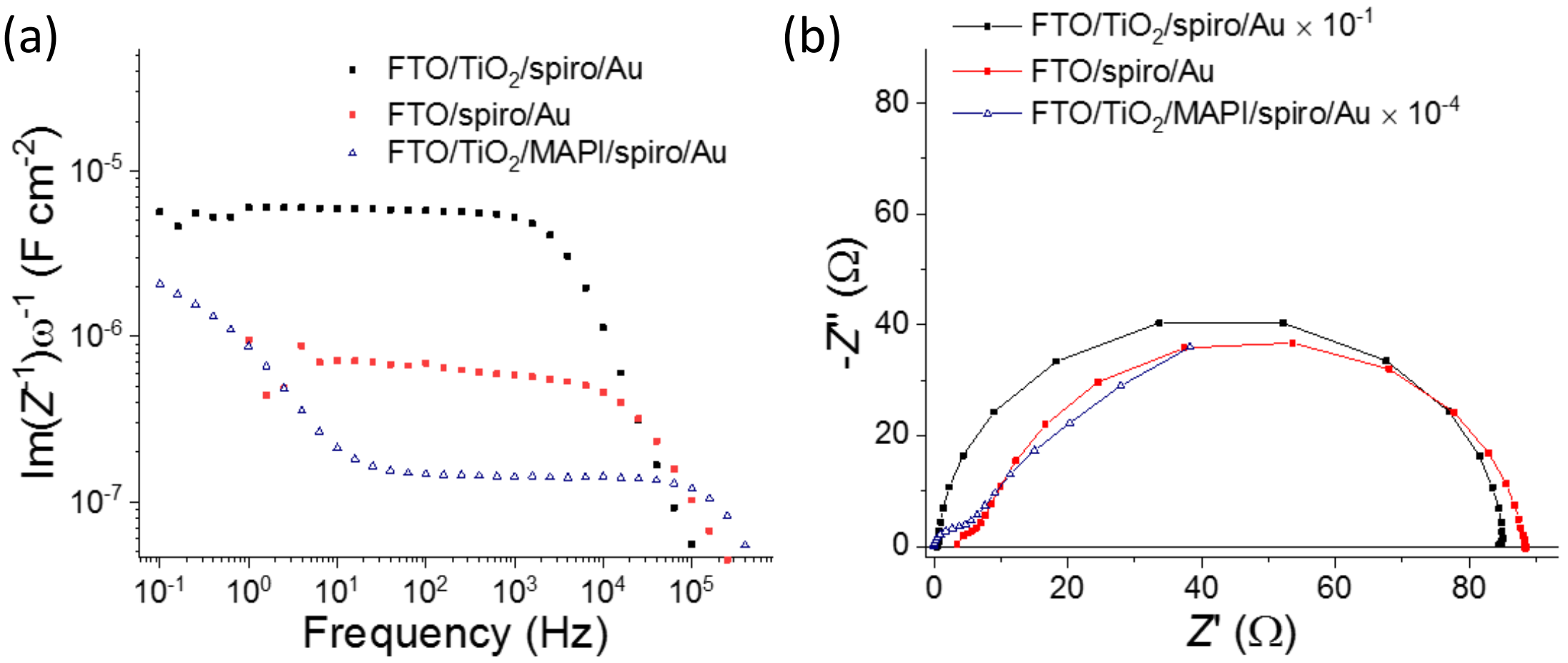

**Figure S35. Impedance measurements performed on devices with different architectures in the dark to test the properties of the contact layers**. (a) Capacitance and (b) Nyquist plots.

The impedance measurement run on a solar cell structure without active layer (black square data points) in Figure S35 shows that values of capacitance up to ~ 6 μF $cm^{-2}$ can be observed for the interfacial capacitance associated to the $TiO_2$/Spiro OMeTAD interface. A smaller space charge potential would be expected to drop in each of the contacts' depletion layer once the perovskite layer is introduced between them. This would also result in a narrower space charge width and larger value of capacitance for $C_A$ and $C_D$. It follows that the capacitance value measured for the $TiO_2$/Spiro OMeTAD interface is a lower limit to the series of the contact capacitance $(C_A^{-1} + C_D^{-1})^{-1}$ in the full solar stack. This indicates that even the large values for the low frequency capacitance measured for solar cells in the dark (see empty blue triangles) could still be explained based on interfacial capacitance arguments (see also discussion in section 3 of this document).

**Supporting references**